\documentclass{article}
\usepackage[
    style=apa,
    sortcites=true,
    backend=biber,natbib,
    labeldate=year
]{biblatex}

\usepackage{lineno}

\usepackage{graphicx, booktabs, rotating, subfig, authblk, amsmath, placeins, tabularx, csquotes, amsfonts}
\usepackage[breaklinks=true]{hyperref}

\AtEveryBibitem{\clearfield{month}\clearfield{day}\clearfield{date}}
\AtEveryCitekey{\clearfield{month}\clearfield{day}\clearfield{date}}

\title{Deconstructing Jazz Piano Style Using Machine Learning}
\author[1]{Huw Cheston\thanks{\href{mailto:hwc31@cam.ac.uk}{\texttt{hwc31@cam.ac.uk}}}}
\author[1]{Reuben Bance}
\author[1]{Peter M. C. Harrison\thanks{\href{mailto:pmch2@cam.ac.uk}{\texttt{pmch2@cam.ac.uk}}}}
\affil[1]{Centre for Music and Science, University of Cambridge}

\date{\today}

\begin{document}

\maketitle

\begin{abstract}
Artistic style has been studied for centuries, and recent advances in machine learning create new possibilities for understanding it computationally. However, ensuring that machine-learning models produce insights aligned with the interests of practitioners and critics remains a significant challenge. Here, we focus on musical style, which benefits from a rich theoretical and mathematical analysis tradition. We train a variety of supervised-learning models to identify 20 iconic jazz musicians across a carefully curated dataset of 84 hours of recordings, and interpret their decision-making processes. Our models include a novel multi-input architecture that enables four musical domains (melody, harmony, rhythm, and dynamics) to be analysed separately. These models enable us to address fundamental questions in music theory and also advance the state-of-the-art in music performer identification (94\% accuracy across 20 classes). We release open-source implementations of our models and an accompanying web application for exploring musical styles.\footnote{\url{https://cms.mus.cam.ac.uk/jazz-piano-style-ml}\label{note:rsi_webapp}} 
\end{abstract}

\providecommand{\keywords}[1]{\textbf{\textit{Keywords: }} #1}
\keywords{music information retrieval, machine learning, explainable artificial intelligence, music performer identification, jazz}\clearpage

\section{Introduction}\label{sec:rsi_introduction}

A fundamental task in the study of the arts is to deconstruct the ``style'' of individual artists. For a visual artist, their style might include aspects such as subject choice, colour choice, and brush techniques; for a writer, it might include vocabulary, syntactic constructions, and narrative archetypes; for a composer, it might include harmonic progressions, rhythmic patterns, and melodic motifs. Individual differences across all these parameters, and more, come together to define each artist's unique style.

Most of these stylistic parameters can theoretically be assessed by human experts. However, such assessments are necessarily slow and hence hard to apply at scale. Subjectivity is also a problem, since every human analyst comes with their own history of artistic exposure that will inevitably affect how they interpret artworks.

Computational methods promise a more scalable and objective approach to this problem. Once a researcher has crafted an algorithm that captures a particular stylistic parameter --- for example, using entropy to capture vocabulary complexity --- then a computer can easily apply the algorithm to large datasets, and hence compare different artists using this parameter \citep{Cheston2024-RSOS, Li2012, Abry2013, Deepaisarn2023}.

Such hand-crafted algorithms are generally limited to relatively low-level aspects of artistic style that are easy to represent with mathematical formulae \citep[e.g.,][]{Cheston2024-RSOS, Falomir2015, Saunders2004, Li2012, Abry2013, Ens2020, Ramirez2010}. However, a more recent possibility is to avoid hand-crafting features, and instead train machine-learning models (in particular, deep neural networks) to capture relevant stylistic parameters in their own way \citep{Setzu2024, Theophilo2022, Kim2020, VanNoord2015, Mahmudrafee2023, Tang2023, Yang2021, Zhang2023-Horowitz, Kong2020, Abbasi2022}.

Can we use such algorithms to deliver insights that are relevant both to art critics and educational for artists? This remains an ongoing challenge. Various studies have shown that deep neural networks can indeed learn important aspects of artistic style \citep{Foscarin2022, Zhang2023-Horowitz}, and have investigated what kinds of features the models look at across a variety of artistic domains \citep{Dervakos2022, VanNoord2015, Chirosca2024}. Nevertheless, it has proved relatively hard to work out just what these networks have learned about individual artists. 

In this paper, we consider the particular case of musical style. Though music is primarily an aural phenomenon, it benefits from a rich array of theoretical literature describing the structural principles of musical styles. Importantly, these theoretical concepts (e.g. intervals, chord roots, patterns and ``licks'') are also commonly used by musicians, both in educational and collaborative contexts \citep{Monson1996}. This rich theoretical tradition makes it a natural domain for interpretable machine learning.

We build on an existing tradition of music composer \& performer identification research that originated with the work of \citet{Widmer2004} and \citet{Saunders2004}. This research typically involves training supervised-learning models to identify the composer or performer of a given musical excerpt. A model that performs well at this task must, in theory, have learned the characteristic style of the composer or performer it identifies. Deep-learning models have recently achieved impressive accuracy levels in such tasks \citep[see][]{Tang2023, Kim2020, Edwards2023, Mahmudrafee2023, Chou2024, Yang2021, Zhang2023-Symbolic}; however, relatively little progress has been made in extracting interpretable insights from these models \citep[although, see][for recommendations]{Foscarin2022}.

Such models can theoretically be applied to all kinds of music, including Western classical \citep{Stamatatos2005, Tang2023, Mahmudrafee2023, Zhang2023-Symbolic, Yang2021, Ens2020, Kong2020} and popular music \citep{Chou2024}. Here, we provide a case study of jazz, which is particularly interesting for the freedoms afforded to performers and has been studied in a number of prior works \citep[e.g.,][]{Cheston2024-RSOS, Ramirez2010, Edwards2023, Velenis2023}. Through improvisation, jazz performers manipulate many different aspects of the music they play, such as the harmony, melody, and rhythm of a composition. For a beginner, however, understanding which elements of a jazz performance reflect these personal styles can be difficult. This can act as a barrier to learning this music \citep{Kernfeld1995}: as trumpeter Wynton Marsalis claims, ``when you're just learning jazz, everything is mystical'' \citep[p. 2]{Berliner1994}.

We show that we can use these models to address many issues relevant to music theorists and musicians. In particular, we ask:

\begin{itemize}
\item[--]{Which high-level musical domains (e.g., harmony, melody, rhythm) are the most important for defining musical style?}
\item[--]{Which domains best distinguish particular performers?}
\item[--]{Which local musical features (e.g., individual melodic patterns or harmonic progressions) are associated with particular performers?}
\item[--]{How does musical style differ between performance contexts, such as playing in an ensemble versus unaccompanied?}
\item[--]{How are the styles of particular performers related?}
\item[--]{What is the most ``characteristic'' example of a particular performer's playing style?}
\item[--]{Who is the performer of an unknown musical piece? (i.e., the archetypical task considered in most prior work on this topic.)}
\end{itemize}

In order to answer these questions, we consider a succession of different machine-learning models that represent both the `handcrafted' and the `representation-learning' approach, including a new deep-learning model with an interpretable architecture that allows its predictions to be explained in terms of four fundamental musical dimensions --- melody, harmony, rhythm, and dynamics.

\section{Dataset}\label{sec:rsi_dataset}

Our dataset consists of recordings of jazz piano improvisation by twenty famous performers, transcribed using an automatic system into MIDI ``piano roll'' format. We study both ensemble and solo performance styles, which are known to differ in systematic ways. For instance, in jazz the piano shares responsibility with the bass and drums for defining the harmonic and rhythmic movement of a performance \citep{Monson1996}. When a pianist performs unaccompanied, they may need to compensate for the absence of the other instruments, such as by emphasising harmonically ``fuller'' chords or bass lines \citep{Berliner1994, Levine2011-1}.

We use transcriptions from two existing open source datasets: (1) the Jazz Trio Database \citep[JTD:][]{Cheston2024} and (2) the Piano Jazz with Automatic MIDI Annotations \citep[PiJAMA:][]{Edwards2023} dataset. JTD contains transcriptions of 1,294 improvised solos by 34 different jazz pianists performing in a trio with a bassist and drummer. Suitable performers were identified from online listening data and pedagogical textbooks, with all recordings from their discographies that met a predefined inclusion criteria (e.g., instrumentation, musical attributes) included in the dataset. PiJAMA contains transcriptions of 2,777 full-length performances by 120 different pianists, without accompaniment. Suitable performers were identified both from textbooks and records of finalists in international jazz competitions, with all available performances by these pianists included in the dataset. Note that, as JTD and PiJAMA contain recordings of different types of jazz performances (solo and trio) identified manually by the dataset creators, combining both together is highly unlikely to introduce contamination (i.e., same recordings contained in both datasets).

The transcriptions for both datasets were created by applying the automatic music transcription system described by \citet{Kong2021} to an audio signal. For the recordings in PiJAMA, an automatic music tagging system was first used to filter out non-music portions of the audio (e.g., applause, spoken introductions), with the transcription model applied to the remaining sections. For the recordings in JTD, the audio was manually trimmed to the piano solo, a source separation model was applied to isolate the playing of the pianist from the rest of the ensemble in this section, and the transcription model was then applied to the separated source. The transcriptions in both datasets were created at a resolution of 100 frames-per-second, the default setting for the transcription model.

\begin{figure}[]
  \centering
  \includegraphics[width=1\textwidth]{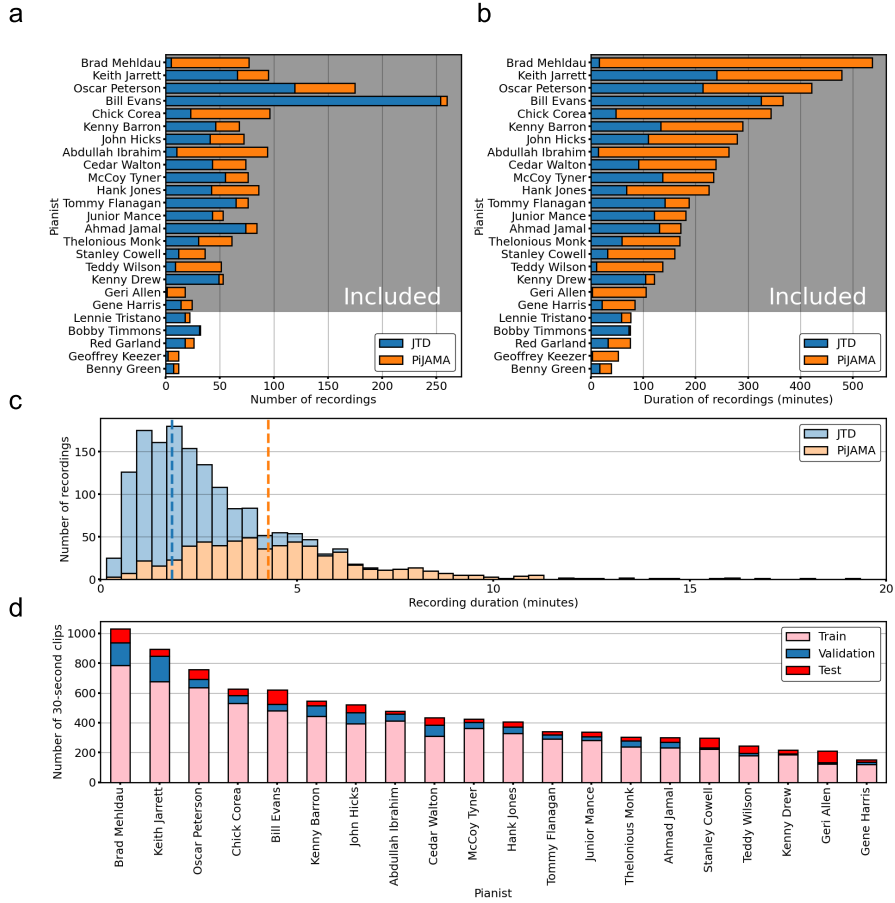}
  \caption{Dataset. The upper row of panels show (a) the number of recordings by each pianist and (b) the total duration of their recordings, stratified by source database. Pianists are sorted in descending order by total duration, with the shaded area indicating when this is greater than 80 minutes. (c) shows the distribution of recording durations across tracks in both databases, with dashed lines representing the median recording duration. (d) shows the number of 30-second clips in each split of the dataset used to train the neural networks.}
\label{fig:rsi_dataset}
\end{figure}

Twenty-five different pianists have at least one recording in both datasets (Figure \ref{fig:rsi_dataset}a). However, the cumulative duration of all recordings varies substantially between performers, from less than half an hour to over nine hours (Figure \ref{fig:rsi_dataset}b). Consequently, we remove pianists with fewer than 80 minutes of recordings across JTD and PiJAMA. This leaves 1,629 performances by twenty pianists, with a total duration of 84 hours. Our dataset includes a greater number of recordings from JTD (1,001) than PiJAMA (628). However, the PiJAMA recordings (which are all full-length performances) are typically longer than those in JTD (which consist only of the piano solo in a recording), with a median duration of 4 minutes 18 seconds versus 1 minute 51 seconds (Figure \ref{fig:rsi_dataset}c).

We randomly split these recordings into training, validation, and test subsets in the ratio $8:1:1$. These splits are stratified by source database, such that a proportional number of solo and ensemble performances are included in each subset. We use these splits to train and evaluate all of the models described in the remainder of the paper. Note that we do not stratify our splits on either a \textit{composition}- or \textit{album}-level, which can minimise over-fitting when training classification models on music data \citep{Zhang2023-Symbolic, Rodriguez-algarra2019}. In the first case (and, unlike Western classical music), different performances of the same composition in jazz are usually very different, due to the emphasis on improvisation; in the second case, symbolic music representations such as MIDI ``piano rolls'' remove almost all acoustic information from a signal, which a model can otherwise learn to rely on \citep{Edwards2023}.

For the models described in section \ref{sec:rsi_handcrafted_features_approach}, features are extracted from an entire recording. For the neural networks in sections \ref{sec:rsi_representation_learning_approach} and \ref{sec:rsi_factorised_inputs_approach}, we first segment each recording into 30-second clips (Figure \ref{fig:rsi_dataset}d), with each clip inheriting the performer label of its parent recording during training. The hop size for each clip is 30 seconds during inference and varies between 15 and 30 seconds during training, as part of our data augmentation pipeline (described in section \ref{sec:rsi:data_augmentation}). During training, the classification loss is calculated using class probabilities estimated separately from each clip. During evaluation, we produce a track-level accuracy metric comparable with the models trained on entire recordings in section \ref{sec:rsi_handcrafted_features_approach} by averaging the class probabilities estimated across all clips taken from a single parent recording \citep[as in][]{Kong2020}. 

\section{Handcrafted Features Approach}\label{sec:rsi_handcrafted_features_approach}

\subsection{Methods}\label{sec:rsi_handcrafted_methods}

We train three classical supervised-learning architectures to identify the pianist playing in each recording using a hand-crafted set of features. These models are: random forests (RF), support vector machines with linear kernel (SVM), and (regularised) multinomial logistic regression (LR). All three architectures have been widely used in previous computational work involving the modelling of musical style \citep{Alvarez2024, Cheston2024-RSOS, Deepaisarn2023, Eppler2014, Ramirez2010, Saunders2004, Widmer2004}. The implementations are taken from the \texttt{scikit-learn} (version \texttt{1.5.1}) Python library \citep{Pedregosa2011}.

\subsubsection{Feature Extraction}\label{sec:rsi_handcrafted_feature_extraction}

While these models could be trained on a wide variety of predictive features, we restrict ourselves to those reflecting the use of harmony and melody by performers, as these can be extracted simply from a musical transcription \citep[e.g.,][]{Ens2020}. For an analogous approach studying rhythmic features obtained from many of the same recordings considered here, see \citet{Cheston2024-RSOS}.

Extracting melody from a symbolic representation of a polyphonic musical performance is a challenging task. Ideally, we would use a sophisticated algorithm that either simulates underlying cognitive procedures involved in melody perception \citep[e.g.,][]{Sauve2018} or that learns to identify melodies from large annotated corpora of polyphonic music. Some deep learning architectures do exist that attempt the latter task \citep[e.g.,][]{Chou2024}, but we found that they performed badly on our data, with very few melody notes identified successfully and the majority of the output being empty (Figure \ref{fig:rsi_sm_melody_extraction_results}). A possible explanation is that the data used to train these models often does not include jazz piano performances.

A simpler method is the ``skyline'' algorithm outlined by \citet{Uitdenbogerd1998}, which extracts the note with the highest pitch among the concurrent notes played at every unique onset time. This approach has drawbacks --- for instance, the extracted notes may ``leap'' between accompaniment and melody. However, in the absence of any more sophisticated methods developed for jazz piano, we nonetheless believe that it represents the best approach to this task. These drawbacks can also be controlled to a certain extent by developing heuristics that filter the ``melodies'' extracted from the skyline (see below).

Before applying the algorithm, we quantise a transcription by ``snapping'' note onsets to the nearest 100 ms (i.e,. 10 frames). This value is roughly equivalent to the duration of a triplet eighth note at the mean tempo of the recordings contained in the JTD \citep{Cheston2024}. We then apply the skyline and obtain a vector of pitch classes, from which we extract $n$-grams (i.e., melodic ``chunks'' obtained over a sliding window). We remove $n$-grams that have either a total span of greater than 12 semitones or that have at least one duration greater than 2 seconds between successive offsets and onsets prior to quantisation. This first heuristic removes cases where the skyline conceivably ``leaps'' between melody and accompaniment, while the second removes cases where $n$-grams might be collected between the boundaries of a musical phrase. Finally, we convert each $n$-gram into a transposition-invariant representation by computing the difference between successive pitch classes.

We use values of $n \in \{3, 4, 5, 6, 7\}$ when extracting melodic patterns from each transcription. In Figure \ref{fig:rsi_sm_lr_extraction_at_n}, we demonstrate empirically that these values of $n$ are sufficient to achieve ceiling accuracy for the LR model when predicting the held-out validation split of the dataset, and that using larger values of either $\text{min}(n)$ or $\text{max}(n)$ does not increase performance \citep{Alvarez2024}. However, this does mean that some of our melodies obtained with lower values of $n$ can more accurately be described as short patterns, rather than full-length phrases. Nonetheless, we note that learning such melodic ``chunks'' does often form a key part of jazz pedagogy \citep{Berliner1994}. For an analogous procedure that solely considers longer melodic patterns in jazz, see \citet{Frieler2018}.

To extract harmony features from the MIDI transcription, we follow a method similar to \citet{Bantula2016}. First, we quantise the transcription to the nearest 100 ms according to the onset time of each note, as before. We keep quantised frames that contain $n \in \{3, 4, 5, 6, 7\}$ notes from each performance --- in other words, chord voicings that contain between three (i.e., triads) and seven notes. We discard chords with two or more leaps of greater than 15 semitones between two adjacent pitches in the chord, since such chords are more-or-less unplayable with an ordinary handspan, and so are likely to correspond to transcription errors instead. Finally, we convert each chord into a transposition-invariant representation by expressing every pitch in terms of the number of semitones it lies above the lowest note in the chord. Note that we choose not to subtract the skylined melody as the highest note of every chord could theoretically have both a harmonic and melodic function, such as in the ``locked hands'' style of jazz piano performance \citep{Levine2011-1}.

We obtain counts for a total of 484,039 features (430,841 $n$-grams, 53,198 voicings) for the 1,629 recordings in the dataset. Similar to how text-based authorship models typically remove both frequent (i.e., ``stop'') and infrequent words \citep{Schonlau2017}, we then discard features contained in fewer than 10 and more than 1,000 recordings in order to reduce the size of the vocabulary. This leaves 21,670 features (17,918 melodic, 3,752 harmonic), with the reduced number of features explainable by a large number of melodic patterns and chord voicings that appear in very few recordings. We transform the matrix of feature counts to a normalised representation using the term frequency-inverse document frequency (TF-IDF) method, previously used for composer classification by \citet{Alvarez2024}. We found that TF-IDF substantially improved predictive accuracy compared with other techniques such as $z$-transformation.

\begin{figure}[]
  \centering
  \includegraphics[width=1\textwidth]{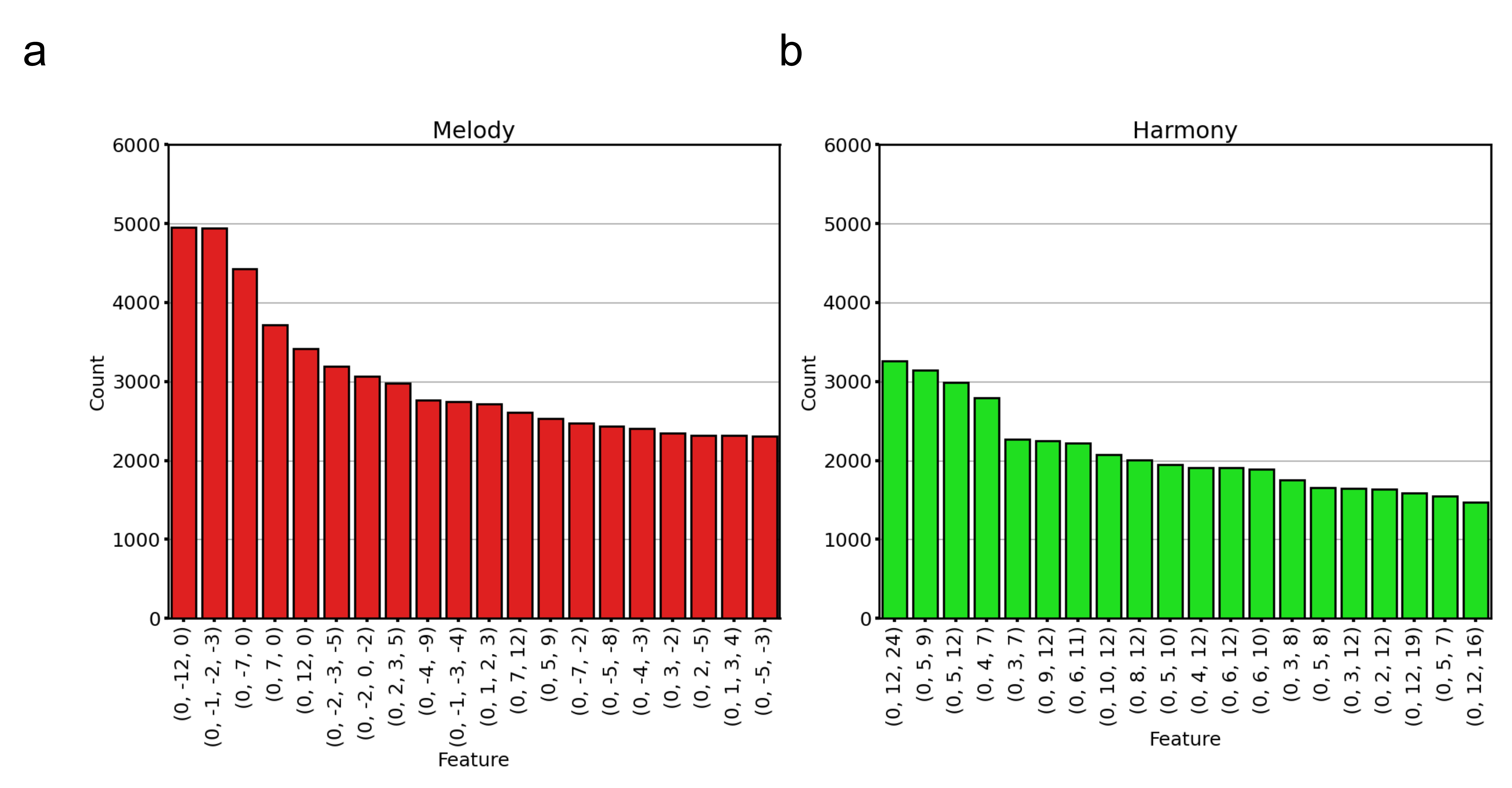}
  \caption{Feature counts. Each panel shows the counts of the twenty most frequently occurring (a) melody and (b) harmony features across all recordings and splits of the dataset. The $x$-axis values can be interpreted as the number of semitones from an initial starting note.}
\label{fig:rsi_feature_counts}
\end{figure}

We show counts for the twenty harmony and melody features that occur most frequently after filtering in Figure \ref{fig:rsi_feature_counts}. Each feature can be interpreted as showing the number of semitones away from an initial starting pitch, such that the feature $(0, -1, -2, -3)$ becomes pitches $(\text{C}, \text{B}, \text{B}\flat, \text{A})$ or alternatively $(\text{G}, \text{G}\flat, \text{F}, \text{E})$. The melody features that occur most often are either alternations of ``perfect'' intervals, such as the octave $(0, -12, 0)$ and fifth $(0, -7, 0)$, or scalar fragments, such as a descending chromatic scale $(0, -1, -2, -3)$ or ascending minor scale $(0, 2, 3, 5)$ As might be expected, shorter patterns appear more frequently than longer ones, with the twenty most common features only containing three or four notes. Several of the harmony features that occur most frequently can be understood with relation to familiar chord types, such as major $(0, 4, 7)$ and minor $(0, 3, 7)$ triads in root position. Both the first $(0, 3, 8)$ and second $(0, 5, 9)$ inversion of the major triad are also common, as are ``stacked'' octave $(0, 12, 24)$ and perfect fourth $(0, 5, 10)$ intervals.

\subsubsection{Training}\label{sec:rsi_handcrafted_training}

The process used to train each of the three models is the same. We generate a two-dimensional hyperparameter space (Tables \ref{tab:rsi_sm_rf_optimised_parameters}--\ref{tab:rsi_sm_lr_optimised_parameters}), randomly sample possible configurations of parameters from this space (with $N = 1,000$ iterations), fit the model to the training split of the dataset, and measure the accuracy when predicting class labels for the validation split. We find that this many iterations is sufficient to achieve ceiling accuracy on the held-out validation data across all classifier types. We then use the hyperparameter configuration that resulted in the highest validation accuracy to predict the class labels of the held-out test split, and report the accuracy for this subset of the data as the overall performance of the model. Following \citet{Cheston2024-RSOS} --- and, for consistency with the neural networks we introduce later --- we do not retrain the model on the training and validation splits after selecting the optimal hyperparameter configuration. A single iteration takes approximately 20 seconds for the LR model, 211 seconds for the RF, and 31 seconds for the SVM. All iterations run in parallel on a machine with 76 CPU cores.

\subsection{Results}\label{sec:rsi_handcrafted_results}

\subsubsection{Who is the Performer of an Unknown Musical Piece?}\label{sec:rsi_handcrafted_evaluation}

We show the accuracy of the three models in predicting the held-out test data in Table \ref{tab:rsi_sm_handcrafted_results}. The LR model (with L2 or ``ridge'' regularisation as the penalty term) performs the best, achieving 0.767 top-1 accuracy when predicting the identity of the twenty jazz pianists considered here. The top-5 accuracy for this model is 0.939 --- meaning that, for nearly 95\% of unseen recordings, the actual pianist is within the five classes with the highest predictive probability estimated by this model, despite it only using melodic n-grams and chord voicings as input features. The RF and SVM models both perform worse, achieving 0.454 and 0.687 top-1 accuracy on the held-out test data, respectively.

\subsubsection{Which Domains are Important for Musical Style?}\label{sec:rsi_feature_importance_by_domain}

One way of ascertaining the relative importance of the input features used by these models is to calculate the decrease in predictive accuracy when the values obtained for a feature (or group of features) are permuted \citep{Breiman2001}. We compute feature importance by permuting either all harmony or all melody features and measuring the loss in held-out test accuracy for the LR versus predicting with the complete feature set. The accuracy loss (averaged over $N = 1,000$ iterations) when permuting $n$-grams is substantially greater than permuting chord voicings (0.403 vs. 0.152). Similar trends are observed for both the optimised RF and SVM models (Figure \ref{fig:rsi_sm_feature_importance_by_handcrafted_model_type}). This might suggest that melodic patterns are more important than chord voicings in differentiating one jazz pianist from another. 

Given the imbalance in the number of features (with over twice as many $n$-grams than voicings), an alternative way to consider the importance of a single feature is to permute random subsets of $K$ melody or harmony features (sampled without replacement from the full feature space) and measure the mean loss in accuracy. With $K = 2,000$ features (approximately 10\% of the total number), the loss in accuracy for the LR is larger when permuting harmony than melody features (0.015 vs. 0.057: mean across $N = 1,000$ iterations). Similar trends are observable for all other model types (Figure \ref{fig:rsi_sm_feature_importance_by_handcrafted_model_type}). This would suggest that the typical chord voicing has over three times the explanatory power of the typical melody $n$-gram; however, as the vocabulary of $n$-grams is much larger than the vocabulary of voicings, permuting all of these features leads to a proportionally greater drop in accuracy. 

\subsubsection{How Does Style Differ Between Performance Contexts?}\label{sec:rsi_feature_importance_by_dataset}

Another question we can ask is whether there are differences in improvisation style between solo and ensemble performances. To do so, we first fit the LR model to the entire dataset. We compute the absolute magnitude of each feature weight for each performer, take the maximum across performers, rank these in decreasing order, and then take the top-$k$ features (with $k = 2,000$, as before). These features correspond to the maximally predictive features across all performers. Further information on this process is provided in the electronic supplementary materials, section \ref{sec:rsi_sm_maximally_predictive_features}.

We then fit separate models to recordings from both datasets using only these features, extract the vector of $k$ weights for every performer, and calculate the correlation coefficients between equivalent vectors from both datasets. We also repeat this process for the top-$K$ harmonic features. Note that, while it would be possible to use all features here, many of these would likely be noise given the high dimensionality of the feature space, which could artificially deflate the correlations.

\begin{figure}[]
  \centering
  \includegraphics[width=1\textwidth]{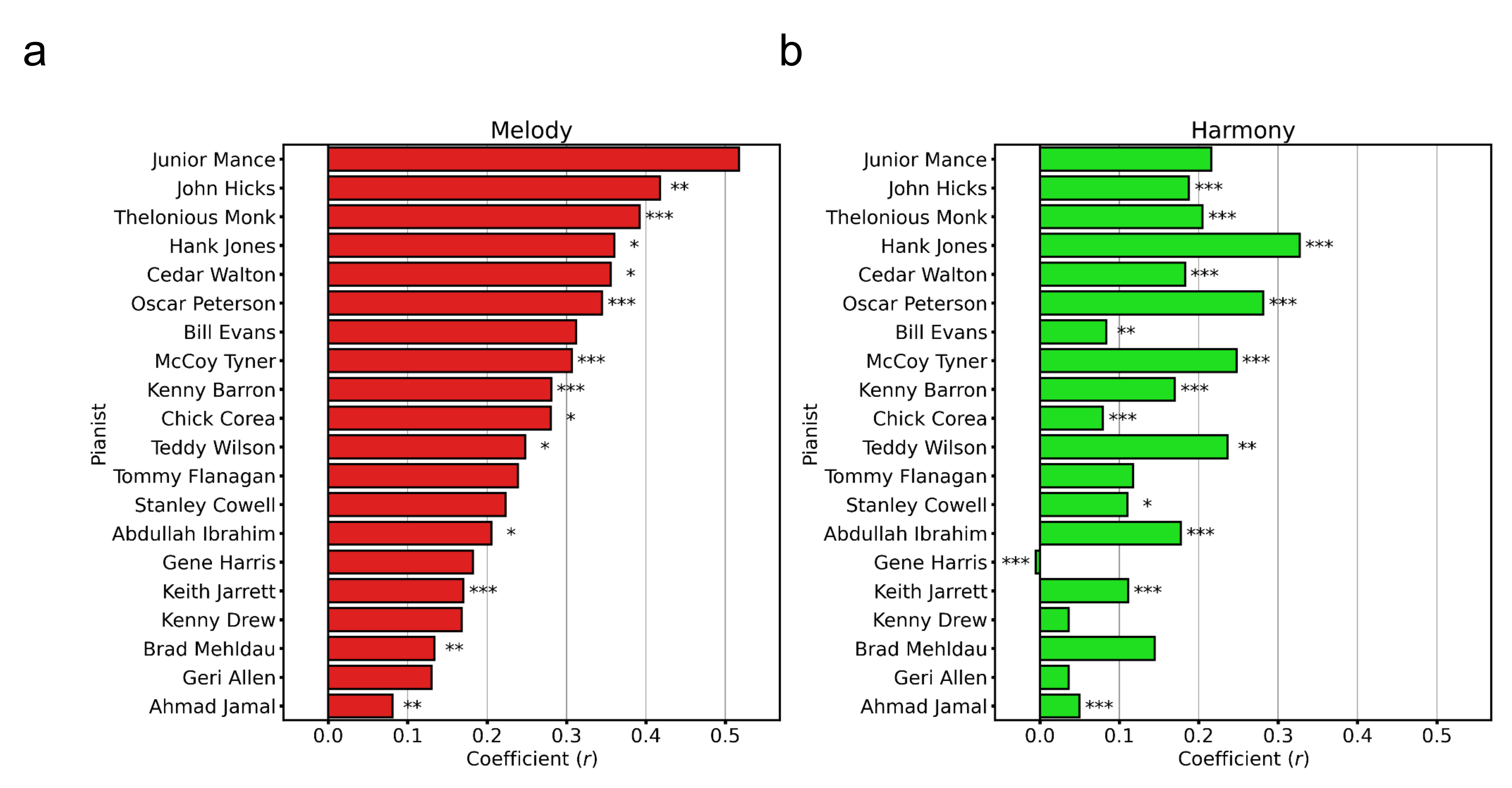}
  \caption{Relationships between databases. Each panel shows correlation coefficients obtained between the weights of every performer for models fitted using (a) melody and (b) harmony features. Performers are sorted in descending order of magnitude according to the correlations in panel (a). Asterisks show the significance of the observed coefficients ($^* \ p < .05, \ ^{**} \ p < .01, \ ^{***} \ p < .001$, with Bonferroni correction applied), calculated using permutation testing ($N = 1,000$ iterations).}
\label{fig:rsi_dataset_correlations}
\end{figure}

In nearly every case, these correlations are positive (melody: $mean(r) = 0.267$, $SD = 0.108$, harmony: $mean(r) = 0.150$, $SD = 0.087$: see Figure \ref{fig:rsi_dataset_correlations}). To the extent that performer style differs between solo and ensemble performances, we would expect correlations to be lower across the datasets (i.e. between solo versus ensemble datasets) than within datasets. To test this hypothesis, we conduct a one-sided (left-tailed) permutation-based significance test, which involves randomly shuffling the dataset label for every recording prior to fitting the model, thereby obtaining a Monte Carlo null distribution of correlation coefficients ($N = 1,000$ iterations) against which we test our observed correlations. We apply Bonferroni correction for the number of feature sets (i.e., 2), thereby controlling false discovery rate on the performer level. These tests indicate that, in the majority of cases, the correlation between feature weights obtained from either dataset are significantly smaller than would be expected were the two datasets to be equivalent. 

This finding is broadly consistent with the literature on jazz improvisation, where musicians frequently comment on differences between playing unaccompanied versus with a group. For pianist John Lewis ``there is very little leeway in transferring your ideas from a group situation to a solo performance. You have a lot more freedom [when playing unaccompanied], but you have to learn how to discipline that freedom'' \citep[p. 81]{Lyons1983}. More specifically, Oscar Peterson noted how playing unaccompanied allowed him to use ``certain harmonic movements ... which [he] couldn't do with the trio'' \citep[p. 141]{Lyons1983}. A note can be made for Junior Mance, who demonstrated the greatest positive correlation between melody feature weights from both datasets ($r = 0.517$, $p = .218$): jazz critic Brian Priestley \citep[p. 319]{Carr1988} describes how ``his trio and solo routines are sometimes predictable ... highly rhythmic, bluesy''.

\subsubsection{Which Features are Associated With Particular Performers?}\label{sec:rsi_feature_importance_by_performer}

While this analysis is interesting in terms of demonstrating the importance of high-level musical dimensions (i.e., melody, harmony) towards improvisation style, it does not tell us which features influenced the model to classify individual performers --- in other words, which melodic patterns or chord voicings best define the style of one particular performer. \citet{Frieler2018} explored this task by ranking how frequently different jazz improvisers used particular melodic patterns. A related method was developed by \citet{Conklin2010}, who defined the distinctiveness of a pattern within a musical corpus as the degree to which it is over-represented with respect to an anti-corpus. 

We take a different approach by instead formalising the ``distinctiveness'' of a given musical feature for a particular performer using the weights it is associated with by our model. This helps mitigate the influence of features that are commonly used by all performers in the dataset (as a result of our TF-IDF pre-processing), while also avoiding the need to explicitly define an anti-corpus.

\begin{figure}[]
  \centering
  \includegraphics[width=1\textwidth]{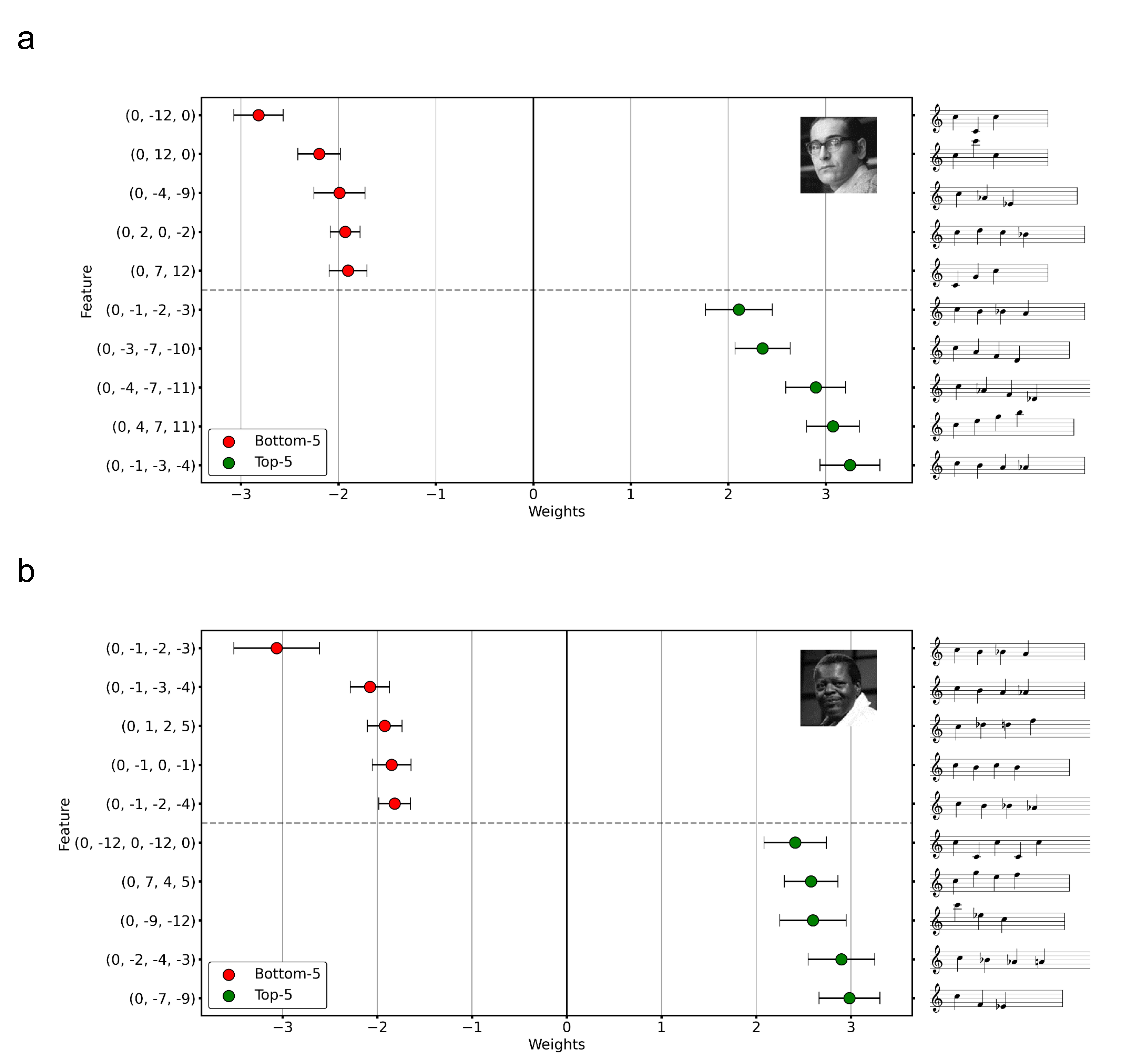}
  \caption{Predictive melody features. Both panels show feature weights obtained for the top and bottom five melody $n$-grams associated with classifications of (a) Bill Evans and (b) Oscar Peterson. Error bars show $SD$, calculated by bootstrapping the dataset used to fit the model ($N = 1,000$ iterations). Corresponding musical notation for every $n$-gram is on the right, transposed such that the first note is C and the mean pitch height is approximately centred around $G_4$. The pitch spelling for each feature is estimated using the algorithm described by \citet{Meredith2006}}
\label{fig:rsi_predictive_melody_features}
\end{figure}

We fit the LR model to the entire dataset and extract the weights obtained for each class and melody feature. Here, a positive weight indicates that the presence of the feature increases the likelihood of predicting a given class, while a negative weight decreases the likelihood. In Figure \ref{fig:rsi_predictive_melody_features}, we show weights for the top and bottom five melody features associated with classifications of Bill Evans and Oscar Peterson --- the two pianists with the greatest number of individual recordings in the dataset (Figure \ref{fig:rsi_dataset}a). We provide similar graphs for all other pianists as part of our supplementary materials (Figures \ref{fig:rsi_sm_ibrahim_melody}--\ref{fig:rsi_sm_flanagan_melody}). The five most distinctive melody features for each pianist can also be listened to within the context of their performances as part of an interactive web application we have developed (see note \ref{note:rsi_webapp}, subheading ``melody'').

For Bill Evans, two of the five patterns with the strongest weighting outline either a descending major $(0, -4, -7, -11)$ or minor $(0, -3, -7, -10)$ seventh arpeggio, beginning on the seventh and falling to the root. Both patterns appear frequently in jazz pianist Jacky Naylor's pedagogical textbook outlining Evans' improvisation style \citep{Naylor2024-BE}. The descending major seventh arpeggio, for instance, is used by Evans to outline a harmonic progression (with the top note descending stepwise through a scale: ``lick 50''), to decorate an extended chord (with the second note of the pattern becoming the top note of the next repetition: ``lick 65''), and over the minor chord in a ``turnaround'' harmonic progression (with the seventh functioning as the ninth above the bass: ``lick 23''). 

Several of the patterns with the strongest weighting for Oscar Peterson can be considered melodic ``enclosures''. For example, the pattern $(0, -2, -4, -3)$ is described by \citet{Naylor2024-OP} in his textbook on Peterson's style as a ``chromatic enclosure around the third'' (``lick 82''), approaching the major third of the underlying harmony from above and then below. Likewise, the $(0, 7, 4, 5)$ pattern acts as a ``scale note enclosure'' (``lick 23''), approaching the note a perfect fourth above the starting pitch diatonically from above and then below. 

Further interesting patterns are apparent in figures shown in the supplementary materials. McCoy Tyner (Figure \ref{fig:rsi_sm_tyner_melody}) is associated with multiple features that outline a perfect fourth interval ($\pm5$), including $(0, -5, -7, 0, -5)$ and $(0, 5, 3, 0)$: Tyner himself has acknowledged that the use of this interval is ``one of the characteristics of [his] style'' \citep[p. 97]{Sidran1992}. Thelonious Monk (Figure \ref{fig:rsi_sm_monk_melody}) is associated with a descending whole-tone pattern $(0, -2, -4, -6, -8)$ that theorist Gunther \citet[p. 231]{Schuller1965} claimed Monk was ``addicted'' to. Finally, Tommy Flanagan (Figure \ref{fig:rsi_sm_flanagan_melody}), perhaps best known for accompanying saxophonist John Coltrane on his album ``Giant Steps'', is strongly associated with a $(0, 2, 4, 7)$ scalar pattern that Coltrane used throughout this record \citep[discussed also in][]{Frieler2018}.

The supplementary materials contains analogous figures for the harmony features (Figures \ref{fig:rsi_sm_ibrahim_harmony}--\ref{fig:rsi_sm_flanagan_harmony}). While we reserve a full analysis for subsequent work, several observations can nonetheless be made here. For instance, Bill Evans (Figure \ref{fig:rsi_sm_evans_harmony}) is strongly associated with the cluster $(0, 1, 5)$, which can be considered a minor voicing containing the ninth, minor third, and fifth; it was in the use of such ``rootless'' voicings that Evans was particularly known for \citep{Levine2011-1}. Many other features can be considered elaborations of an underlying dominant harmony, such as the $(0, 16, 22, 27)$ voicing associated with Cedar Walton (Figure \ref{fig:rsi_sm_walton_melody}) --- commonly known to rock musicians as the ``Jimi Hendrix chord'' \citep{Shapiro1995}. Finally, McCoy Tyner (Figure \ref{fig:rsi_sm_tyner_harmony}) is again associated with voicings built on the fourth interval, including $(0, 5, 10)$ and $(0, 6, 11)$. 

\begin{figure}[]
  \centering
  \includegraphics[width=1\textwidth]{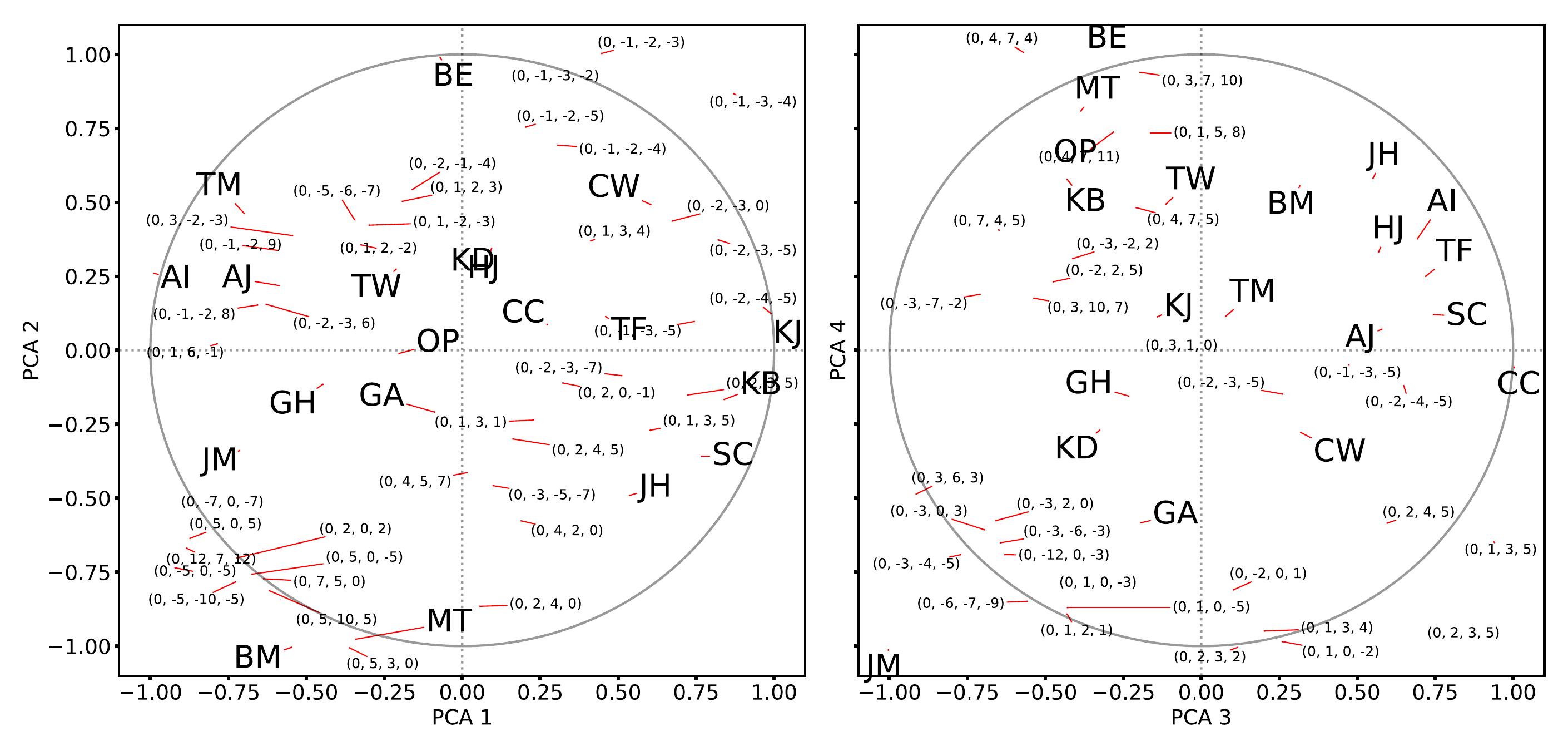}
  \caption{PCA projections. The horizontal and vertical axes respectively show melody feature and performer projections onto (a) principal components 1 and 2 and (b) components 3 and 4. Values are linearly scaled separately for each dimension to within the range $(-1, 1)$. Each melody feature shows the number of semitones relative to the initial note $0$, as in Figure \ref{fig:rsi_feature_counts}. To select the melody features to plot, we divide the two-dimensional representation into eight circular ``slices'' and within each slice plot the 5 features with the highest absolute magnitudes. Performers are shown by their initials, such that \texttt{BE} corresponds to Bill Evans.}
\label{fig:rsi_pca_feature_counts}
\end{figure}

\subsubsection{How are the Styles of Particular Performers Related?}

These preceding analyses focused on individual $n$-grams, but it is also possible to use dimensionality reduction to visualise larger-scale clusters of $n$-gram usage in order to relate the melodic styles of every performer in our dataset. Here we use principal component analysis (PCA); we apply it to melodic $n$-grams of length 4, of which there are 5,838. We process the raw feature counts with TF-IDF as before (see section \ref{sec:rsi_handcrafted_feature_extraction}), then centre them on their means with $z$-transformation and scale the results between the range $(0, 1)$ according to the minimum and maximum values of every feature. We apply PCA to the result to obtain a set of principal components and then project each performer onto the same space by taking the dot product of their averaged feature counts with every component. In Figure \ref{fig:rsi_pca_feature_counts}, we plot all performer classes and a subset of the features that load most strongly onto the first four principal components. 

Features that load positively onto principal component 1 typically consist of small scalar patterns with a span of less than a perfect fifth, such as $(0, -2, -4, -5)$, $(0, 2, 3, 5)$, and $(0, -2, -3, -7)$ \citep[the inverted, minor version of the``Coltrane pattern'': see section \ref{sec:rsi_feature_importance_by_performer} and][]{Frieler2018}. Keith Jarrett, Tommy Flanagan, and Kenny Barron all load positively onto this component: Figure \ref{fig:rsi_sm_flanagan_melody} shows how Flanagan uses several of these melodic patterns. Features that load negatively onto component 1 span a much larger distance, such as $(0, -1, -2, 8)$ and $(0, 1, 6, -1)$, with Abdullah Ibrahim, Ahmad Jamal, and Thelonious Monk also loading negatively here.

Features that load positively onto component 2 involve chromatic patterns, such as $(0, -1, -2, -3)$ and $(0, -1, -3, -2)$, with Bill Evans loading positively onto this component. Features that load negatively consist of the same interval played multiple times in either an ``up-down-up'' or ``down-up-down'' fashion, including $(0, -7, 0, -7)$ and $(0, 5, 0, 5)$, while Brad Mehldau, Junior Mance, and McCoy Tyner all load negatively onto component 2 also. Here, we note that many of these negatively loading features could act as ostinato patterns that would otherwise be played by a bassist: this could explain why Mehldau loads negatively onto component 2, as the majority of his recordings in our dataset are unaccompanied. The majority of these figures also alternate between perfect intervals, including fourths $(0, 5, 0, 5)$ and octaves $(0, 12, 7, 12)$: fourths are associated with Tyner in Figure \ref{fig:rsi_sm_tyner_melody}, and octaves with Mance in Figure \ref{fig:rsi_sm_mance_melody}.

Features that load positively onto component 3 are scalar fragments that approach a perfect fourth, including $(0, 1, 3, 5)$ and $(0, 2, 4, 5)$, while features that load negatively instead outline an augmented fourth, such as $(0, 3, 6, 3)$ and $(0, -3, -6, -3)$. Several features that load positively onto component 4 outline ascending major $(0, 4, 7, 11)$ and minor $(0, 3, 7, 10)$ seventh arpeggios. Perhaps unsurprisingly given the patterns in \ref{fig:rsi_predictive_melody_features}a, Bill Evans loads positively onto this component; interestingly, so also do McCoy Tyner and Oscar Peterson.

Finally, we note that the lower-dimensional space obtained through PCA appears to embody a degree of invariance to melodic inversion, with features that outline the same intervallic patterns such as $(0, -5, -10, -5)$ and $(0, 5, 10, 5)$ typically mapped to similar coordinates regardless of their melodic contour. There may be some invariance to mode, too, with the major scale fragment $(0, -2, -3, -5)$ loading onto component 1 with a similar magnitude as the minor scale equivalent $(0, -2, -4, -5)$ --- repeated, also, for component 4 and the major $(0, 4, 7, 11)$ and minor $(0, 3, 7, 10)$ arpeggios.

\section{Representation Learning Approach}\label{sec:rsi_representation_learning_approach}

One weakness of the architectures explored in section \ref{sec:rsi_handcrafted_features_approach} is that it is difficult to be certain that a handcrafted feature space includes all features that might be predictive of improvisation style. We did not consider, for example, how the arrangement of different chords in sequence might impact voice leading, which is a skill that many jazz pianists devote significant time to mastering \citep{Berliner1994, Levine2011-1, Lyons1983}. Consequently, rather than defining the feature space in advance, it may be attractive to allow the model to learn a representation directly from an input that can be used in classification.

\subsection{Methods}\label{sec:rsi_representation_methods}

We evaluate two convolutional neural networks (CNNs) on our dataset. This type of network possesses a range of inductive biases that makes it effective at modelling musical performances in the symbolic domain. For instance, the convolution operation is equivariant to translation; if an input is shifted (e.g., if the pitches are transposed), so will the corresponding feature map. When applied to ``piano roll'' representations, which put time on the horizontal axis and pitch on the vertical axis, CNNs therefore capture both transposition invariance (a motif's identity is preserved when played at higher or lower pitches) as well as temporal invariance (a motif's identity is preserved when played earlier or later in a recording). CNNs have outperformed both transformer and graph neural network architectures when identifying classical pianists from symbolic representations of their performances \citep{Zhang2023-Symbolic}. 

The first model we test (CRNN) uses eight convolutional layers and a bidirectional gated recurrent unit followed by a fully connected layer and softmax function to generate class probabilities, as in \citet{Kong2020}. The second (ResNet) is the ResNet-50 model, which has previously been used for composer identification in symbolic music \citep{Foscarin2022, Kim2020}, for identifying ``samples'' in music audio recordings \citep{Cheston2025}, and for identifying artists from hand-drawn sketches \citep{Chirosca2024}. Both models are implemented using the \texttt{PyTorch} (version \texttt{2.4.1}) Python library \citep{Paszke2019}. Further description of both models is provided in electronic supplementary materials, section \ref{sec:rsi_sm_cnn_architectures}.

\subsubsection{Input Representation}\label{sec:rsi_representation_input}

Each 30-second MIDI clip extracted from a recording is represented as a one-channel ``image'' of shape $(88, 3000)$, where the channel dimension is the velocity of each note. Following \citet{Kong2020}, we scale this between 0 and 1 (where a value of 1 is equivalent to the note with the highest velocity in the clip) to avoid overfitting to any variation in dynamic range compression between recordings \citep[the ``album effect'': see][]{Edwards2023, Flexer2010, Rodriguez-algarra2019}. 

The height of the input corresponds to the pitch range of the piano, with one key per bin; the width is the duration of the clip multiplied by the number of frames per second used by the transcription model. We do not downsample the width of the input due to the importance of microtiming in prior analyses of expressive jazz performances \citep{Benadon2006, Cheston2024-RSOS, Eppler2014, Ramirez2010}.

\subsubsection{Data Augmentation}\label{sec:rsi:data_augmentation}

We use data augmentation to increase the diversity of our training data and improve model generalisability \citep{Liu2022, Yang2021}. Our augmentation pipeline jointly manipulates the pitch, time (i.e., onset and offset), and velocity parameters of each note within a recording. A brief description of the pipeline is given below, with full details provided in electronic supplementary materials, section \ref{sec:rsi_sm_data_augmentation}.

Pitch augmentation involves shifting all pitch values in a clip by an integer constant sampled from a discrete, symmetric uniform distribution bounded according to the pitch range of the input, with the maximum possible transposition being $\pm 6$ semitones. Time augmentation involves scaling all note onset and offset times by between 0.8 and 1.2 times their original value. Velocity augmentation involves perturbing the velocity of every note by a random variable sampled from a discrete uniform distribution, with upper and lower bounds set to 12 and $-12$, respectively. Finally, during training we randomly adjust the hop between two successive 30-second clips from the same recording to between 15 and 30 seconds (i.e., between 0 and 50\% overlap). Note that a constant hop of 30 seconds is maintained for the validation and test datasets.

As jazz pianists are encouraged to become familiar with playing the same music across multiple keys and tempi \citep{Haerle1994, Levine2011-1}, one possibility would be to apply data augmentation to every clip in the training dataset. However, it is equally likely that performers may display an innate preference for playing certain material in particular keys or at certain tempi \citep{Berliner1994}, and that it could prove beneficial for the model to learn this preference. As a compromise, for every clip seen in a single training epoch, we only assign a 50\% probability that data augmentation will be applied.

\subsubsection{Training}\label{sec:rsi_representation_training}

We train two versions of both models, with and without data augmentation. Each model is trained for 100 epochs on a single NVIDIA A100 SXM4 GPU with categorical cross-entropy loss, which we noted was sufficient to minimise the validation loss. With the exception of batch size (which we set to 20), we tune hyperparameters for both models separately. For training the CRNN, we set the learning rate to 0.001 and use the Adam optimiser; for the ResNet, we use the hyperparameter configuration described in \citet{Kim2020}. After training, we use the saved model weights with the best accuracy on the validation split to predict the held-out test data, for compatibility with the process described earlier in section \ref{sec:rsi_handcrafted_training}. Training takes approximately five hours for the ResNet and six and a half hours for the CRNN.

\subsection{Results}\label{sec:rsi_representation_results}

\subsubsection{Who is the Performer of an Unknown Musical Piece?}\label{sec:rsi_representation_evaluation}

We show the accuracy of each model when predicting the performer of an unseen test recording in Table \ref{tab:rsi_sm_representation_learning_results}. The best performing model from these experiments is the ResNet trained using our data augmentation pipeline ($0.944$ accuracy). This model considerably outperforms both the previous-best LR model ($0.767$) and the CRNN with augmentation ($0.825$) and can be considered to set a new state-of-the-art in the task of automatic performer identification. 

The fact that the ResNet outperformed the CRNN (both with and without augmentation) could be explained by the deeper architecture of this model or the larger dimensionality of the final feature vector. Our data augmentation pipeline substantially improves the predictive capacity for both the CRNN ($0.769$ without augmentation vs. $0.825$ with) and ResNet ($0.875$ vs. $0.944$).

\subsubsection{Which Features are Associated With Particular Performers?}\label{sec:rsi_representation_lime}

One possible way to understand the local musical features that are associated with particular performers is to explore the areas of a single input that most influence the decisions made by the model. The Locally-Interpretable Model Explanations (LIME) technique is well-established here and was initially used to create human-interpretable explanations of black-box computer vision models \citep{Ribeiro2016}. In the audio domain, LIME has been applied to explain music recommender \citep{Melchiorre2021} and singing voice detection systems \citep{Mishra2017}. In the symbolic domain, it has been used to identify the regions of a MIDI piano roll that pushed a convolutional neural network to classify the clip as being from a particular genre of popular music \citep{Dervakos2022}. LIME works by creating multiple versions of the original input with slight perturbations, generating predictions for these versions using the original model, and then approximating these predictions using a simple (interpretable) model.

\begin{figure}[]
  \centering
  \includegraphics[width=1\textwidth]{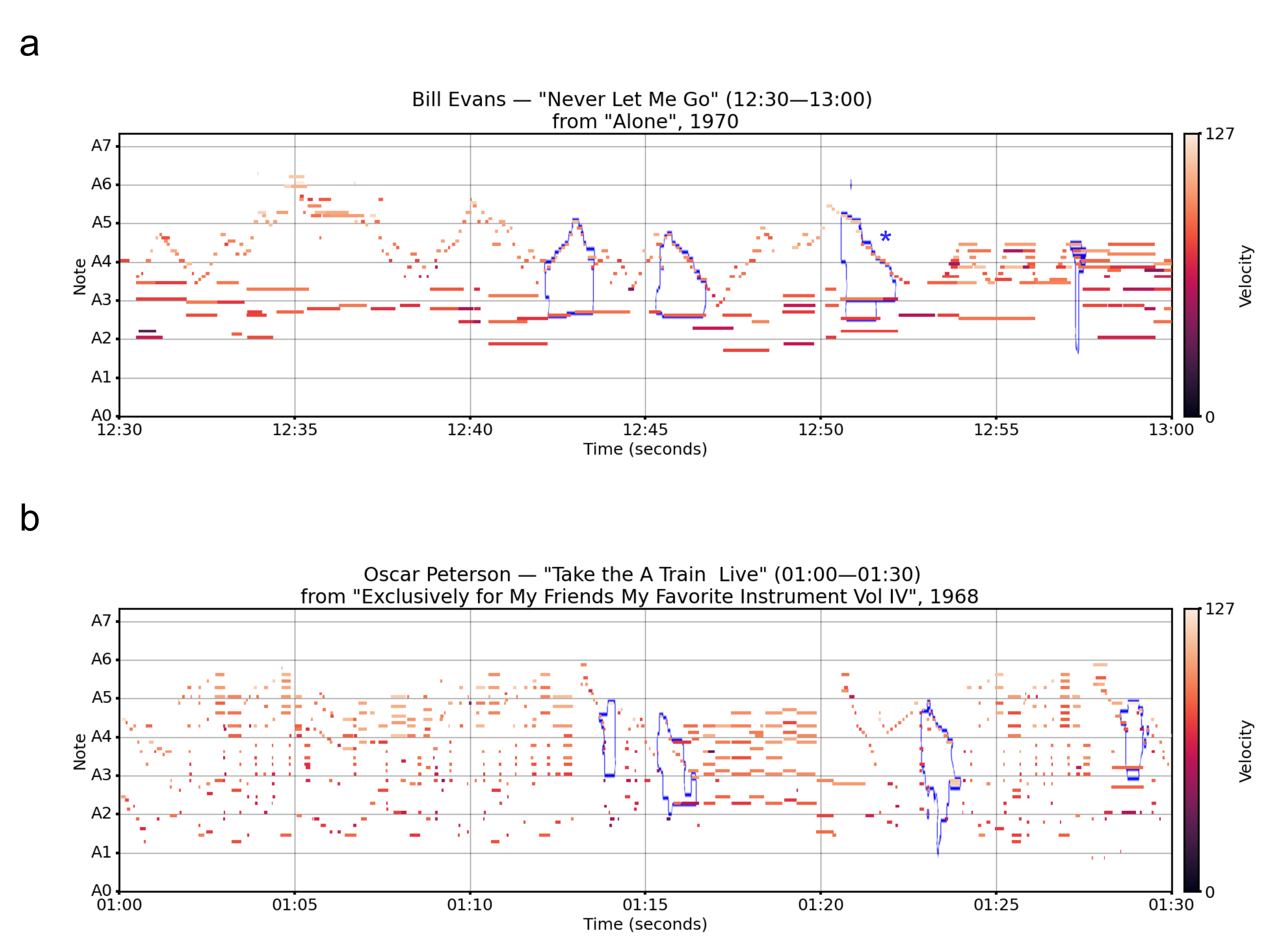}
  \caption{LIME piano rolls. Both panels show a single clip from a performance by (a) Bill Evans and (b) Oscar Peterson. Highlighted in blue are the five areas of the piano roll that contribute the most towards predicting the target label. The asterisk in (a) indicates the descending arpeggio $(0, -4, -7, -11)$ pattern identified in Figure \ref{fig:rsi_predictive_melody_features}}
\label{fig:rsi_lime_plots_main}
\end{figure}

We use LIME to interpret test-split predictions made by the ResNet trained with augmentation. We use the default settings in the Python library (version \texttt{0.2.0.1}) provided by the authors of the original paper \citep{Ribeiro2016}. Figure \ref{fig:rsi_lime_plots_main} highlights the four regions with the strongest LIME attributions (i.e., that push the model to positively identify the target class) for clips by Bill Evans and Oscar Peterson. Similar figures for clips by Abdullah Ibrahim, Chick Corea, and Keith Jarrett (the third, fourth, and fifth pianists with the greatest number of recordings: see Figure \ref{fig:rsi_dataset}a) are provided in Figure \ref{fig:rsi_sm_lime_plots}. 

The LIME technique does seem to emphasise musical gestures that could reasonably be considered distinctive for each performer. These include particular ascending and descending melodic patterns for Evans and several chord voicings for Peterson: interestingly, in the case of Evans, one of the attributed regions even contains the descending $(0, -4, -7, -11)$ arpeggio observed previously in Figure \ref{fig:rsi_predictive_melody_features}a. Nonetheless, it is difficult to understand exactly what aspects of those gestures the model is paying attention to. Considering several of the scalar patterns highlighted for Evans, their explanatory power could reasonably derive from the intervallic structure of the scale, the harmony that it outlines, or the rhythmic and dynamic trends that shape it. Exactly which of these possibilities is the case is difficult to grasp, as all of these musical dimensions are entangled within the input \citep[see][]{Dervakos2022}.

\section{Multiple Input Representations Approach}\label{sec:rsi_factorised_inputs_approach}

One possible solution to this problem would be to extract a single high-level musical dimension from the original transcription and train a model with only this information. Melody, for instance, could be isolated by applying the skyline algorithm. However, isolating only a single musical dimension would likely reduce predictive accuracy, due to the model seeing less information than the original transcription. In addition, unless all other high-level musical dimensions are fixed, the model may exploit contextual ``shortcuts'' that do not encode the desired feature but instead those correlating with it \citep{Wei2024} --- such as increases in velocity appearing alongside an ascending melodic contour. 

Instead, in this section we outline a novel architecture that learns separate representations for four fundamental musical domains --- melody, harmony, rhythm, and dynamics \citep[see][]{Zhang2023-Horowitz} --- but then combines information from these four domains to make its predictions. Each domain is represented as a piano roll with the same shape as the original transcription. We then train small convolutional sub-networks to learn from each representation and aggregate the outputs together to generate a single feature vector that can be used to make predictions \citep[as in][]{Ramoneda2024}. 

This multi-\textit{input} architecture is different to the multi-\textit{task} learning approach described earlier by \citet{Velenis2023}, where a single input representation from a jazz ``lead sheet'' was used to generate multiple downstream predictions (for e.g., composer, tonality, form identification). Instead, our approach can be thought of as conceptually similar to a mixture-of-experts model without a gating network, allowing us to explicitly control the contribution of each input and sub-network (``expert'') towards a single downstream task (performer identification).

\subsection{Methods}\label{sec:rsi_factorised_methods}

\subsubsection{Input Representations}\label{sec:rsi_factorised_description}

Given a one-channel piano roll with the dimensionality $(88, 3000)$, we want to split this into four separate rolls, all with the same height and width but which contain the melodic, harmonic, rhythmic, or dynamic content from the input. While it would be possible to use different dimensions for each roll (for instance, representing each melody note using a single pixel), maintaining the same dimensionality ensures that each roll remains directly comparable to the original input and to one another (receiving, for instance, the same number of learnable parameters). We provide a brief description below of how each roll is created; further information is given in electronic supplementary materials, section \ref{sec:rsi_sm_generating_factorised_representations}.

Melody rolls are generated by quantising and applying the skyline algorithm to the transcription (see section \ref{sec:rsi_handcrafted_feature_extraction}), setting the velocity of each note to binary, and adjusting the inter-onset interval for each note to a consistent value dependent on the total number of melody notes and the width of the input piano roll. Harmony rolls are generated by quantising the transcription, keeping quantised frames with at least three notes, adjusting inter-onset intervals for notes within one bin, and binarising the velocity of every note. Rhythm rolls maintain the onset and offset time for each note, but randomise the pitch (i.e., height) and binarise the velocity. Dynamics rolls are generated by binning notes by onset time, setting their inter-onset intervals to a consistent value, and randomising their pitch --- i.e., maintaining only the initial velocity of each note.

\subsubsection{Model Architecture}\label{sec:rsi_factorised_architecture}

Our proposed architecture uses four convolutional sub-networks to generate an $x$-dimensional embedding from every input roll. Each sub-network shares the same architecture, but the weights are updated separately. We test a range of architectures in our experiments, including 50-, 34-, and 18-layer ResNets --- the latter being the smallest implemented in \texttt{PyTorch}. We also test the CRNN architecture described in electronic supplementary materials, section \ref{sec:rsi_sm_cnn_architectures}. As a form of regularisation, during training we experiment with masking the outputs of particular sub-networks with zeroes. There is a 10\% chance that one, two, or three sub-networks will be masked when processing a clip (i.e., a 30\% total probability of masking any number of sub-networks), with 70\% probability that all four sub-networks will be used during prediction.

\begin{figure}[]
  \centering
  \includegraphics[width=1\textwidth]{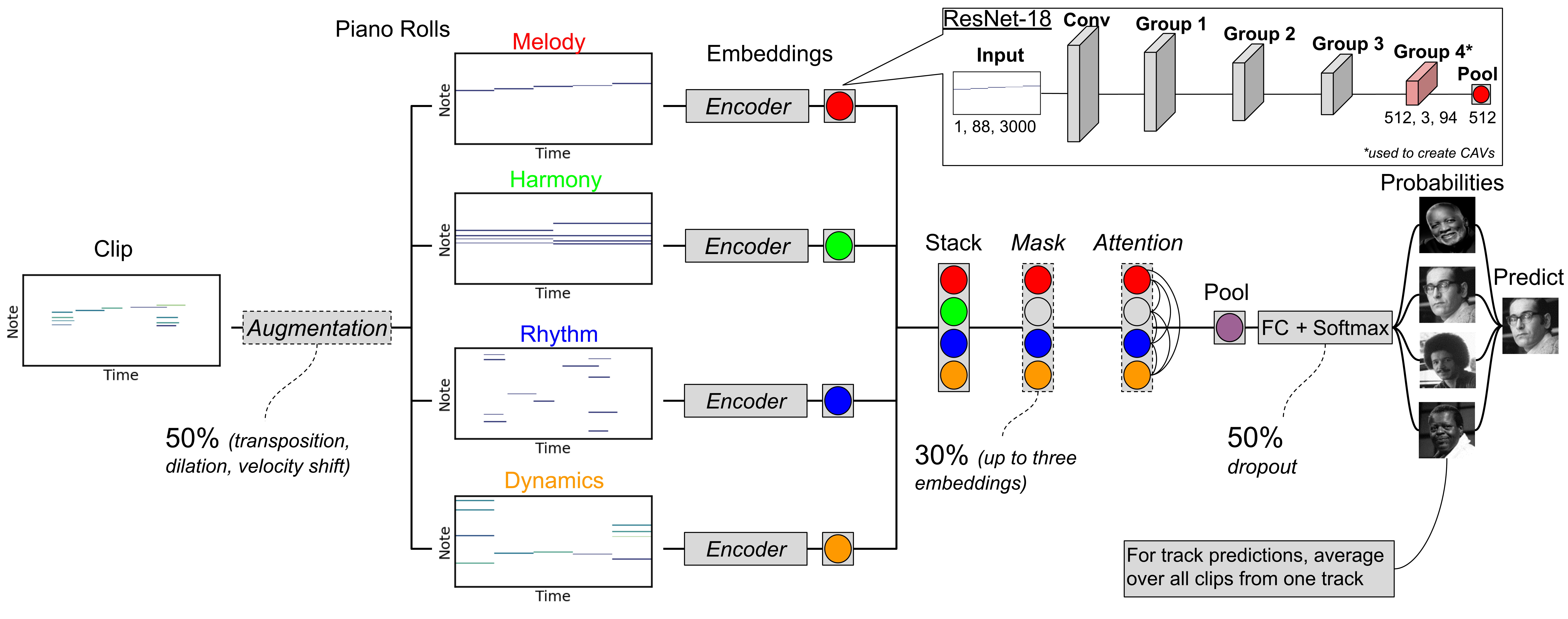}
  \caption{Proposed architecture. An input transcription is represented using four piano rolls, each relating to a musical domain. Each roll is processed with a separate sub-network, the embeddings are then pooled, and class probabilities are generated. The architecture of an 18-layer ResNet is shown here; different architectures are tested in our experiments (see Table \ref{tab:rsi_factorised_input_experiment_results}). Optional components are represented with dashed lines.}
\label{fig:rsi_proposed_architecture}
\end{figure}

The outputs from each network are then stacked vertically to create an array with shape $(4, x)$. It would be straightforward to generate the final $x$-dimensional feature vector by either taking the average or maximum of the neurons in every ``column'' of this array. However, this might not be sufficient to allow the model to capture interactions between different musical domains. Consequently, we experiment with using a self-attention layer (with between four and sixteen heads) across embeddings prior to pooling. After aggregation, the resulting $x$-dimensional feature vector is fed into a fully connected layer (with dropout at a rate of 50\%) and softmax function to generate class probabilities. Figure \ref{fig:rsi_proposed_architecture} shows an outline of the proposed architecture. 

\subsubsection{Training}\label{sec:rsi_factorised_training}

We train the model for 100 epochs with a batch size of 20, using thirty-second clips and aggregating class probabilities to obtain track-level predictions. We use the Adam optimiser with categorical cross-entropy loss and a learning rate of 0.0001. Training takes between twelve and twenty-one hours on a single NVIDIA A100 SXM4 GPU, depending on the size of the model. We make a model card \citep{Mitchell2019} available on the same open source repository that contains the model code (see note \ref{note:rsi_webapp}).

\subsection{Results}\label{sec:rsi_factorised_results}

\subsubsection{Who is the Performer of an Unknown Musical Piece?}\label{sec:rsi_factorised_evaluation}

\begin{sidewaystable}[]
    \caption{Experiment results for models trained with multiple inputs. Note that, when the number of attention heads is 0, self-attention is not used.}
    \label{tab:rsi_factorised_input_experiment_results}
    \centering
        \begin{tabular}{l c c c c c c}
        \toprule
        \textbf{Input Encoder} & \textbf{Attention heads} & \textbf{Pooling} & \textbf{Augmentation} & \textbf{Masking} & \multicolumn{2}{c}{\textbf{Accuracy}} \\
        & & & & & \textbf{Clip} & \textbf{Track} \\
        \midrule
        \multicolumn{7}{c}{\textit{Network Architecture}} \\
        \midrule
        ResNet-18 & 0 & Average & Y & Y & \textbf{0.767} & \textbf{0.913} \\
        ResNet-34 & 0 & Average & Y & Y & 0.744 & 0.906 \\
        ResNet-50 & 0 & Average & Y & Y & 0.707 & 0.863 \\
        CRNN & 0 & Average & Y & Y & 0.622 & 0.812 \\
        \midrule
        \multicolumn{7}{c}{\textit{Self-Attention}} \\
        \midrule
        ResNet-18 & 4 & Average & Y & Y & 0.691 & 0.825 \\
        ResNet-18 & 8 & Average & Y & Y & 0.721 & 0.856 \\
        ResNet-18 & 16 & Average & Y & Y & 0.707 & 0.844 \\
        \midrule
        \multicolumn{7}{c}{\textit{Pooling}} \\
        \midrule
        ResNet-18 & 0 & Max & Y & Y & 0.753 & 0.906 \\
        ResNet-18 & 4 & Max & Y & Y & 0.683 & 0.831 \\
        \midrule
        \multicolumn{7}{c}{\textit{Data Augmentation \& Masking}} \\
        \midrule
        ResNet-18 & 0 & Average & N & Y & 0.644 & 0.789 \\
        ResNet-18 & 0 & Average & Y & N & 0.743 & 0.863 \\
        ResNet-18 & 0 & Average & N & N & 0.619 & 0.806 \\
        \bottomrule
        \end{tabular}
\end{sidewaystable}

The results for all experiments are shown in Table \ref{tab:rsi_factorised_input_experiment_results}. We obtain the most accurate predictions of the held-out test recordings using the smallest ResNet with 18 layers; increasing the size of each sub-network beyond this decreases performance, which would suggest that overparameterisation occurs with larger architectures \citep{Zhang2023-Symbolic}. We find that accuracy does not improve with the addition of self-attention. This could imply that there are no meaningful interactions between the different musical domains as they are represented here. As the network has only a single fully connected layer, this would instead imply that the contributions of each sub-network are additive. Finally, we observe slight performance improvements from using masking.

Compared with the other models described here, our proposed architecture using multiple inputs achieves impressive accuracy in identifying twenty jazz pianists. For instance, it almost matches the performance of the ResNet trained with augmentation on a unified piano roll representation (0.913 vs. 0.944 accuracy). There is therefore some trade-off between interpretability and predictive performance, at least in our case; however, it is impressive that the multi-input model can nearly match the state-of-the-art in this task.

\subsubsection{Which Domains are Important for Musical Style?}\label{sec:rsi_factorised_domain_importance}

\begin{figure}[]
  \centering
  \includegraphics[width=1\textwidth]{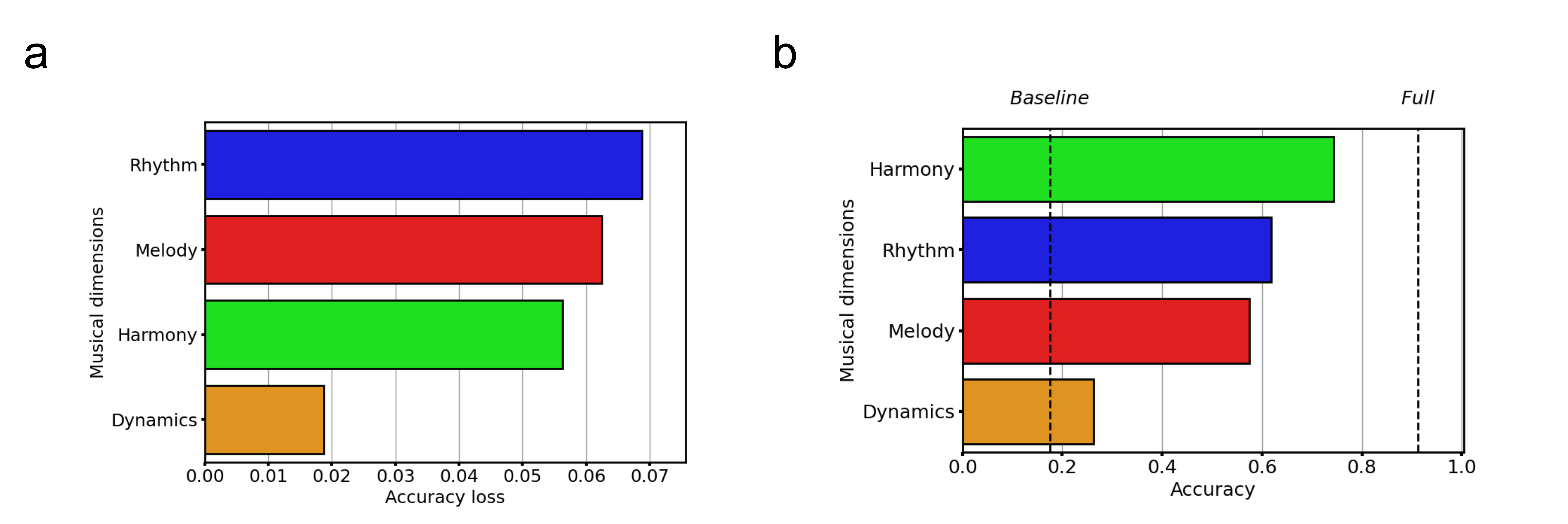}
  \caption{Multi-input model evaluation. (a) shows the loss in accuracy compared with the full model when masking a single sub-network. (b) shows the accuracy of predictions made using only a single sub-network. The dotted lines in (b) show the accuracy from predicting the majority class and the accuracy of the ``full'' model (using all four sub-networks).}
\label{fig:rsi_factorised_evaluation}
\end{figure}

There are at least two ways to evaluate these high-level musical domains. The first considers how important each representation is to the full model by computing the loss in accuracy when the output of a single sub-network is masked with zeroes (Figure \ref{fig:rsi_factorised_evaluation}a). Masking either the melody, rhythm, or harmony sub-network leads to a small, relatively consistent drop in accuracy, with the greatest loss observed for rhythm (rhythm accuracy loss: $-0.069$, melody: $-0.063$, harmony: $-0.056$). Masking the output of the ``dynamics'' sub-network leads to a considerably smaller loss in accuracy, however ($-0.019$), which suggests that dynamics are relatively unimportant when predicting the improvisation style of a jazz pianist. 

A second way to consider the importance of each domain is to compute how well a single representation predicts by itself, when the output of the other three sub-networks is masked (Figure \ref{fig:rsi_factorised_evaluation}b). As before, predictive accuracy is lowest when using only the dynamics sub-network ($0.263$ accuracy) --- and is only slightly better than a model which simply predicts the majority class ($0.175$). Both melody and rhythm sub-networks yield similar predictive accuracy ($0.575$ and $0.619$, respectively). The most accurate predictions are obtained using only the harmony sub-network, with nearly three-quarters of unseen recordings being classified correctly ($0.744$). We show the predictive accuracy obtained for all combinations of sub-networks in Figure \ref{fig:rsi_sm_factorised_lollipop}.

The importance of melodic content for distinguishing between different improvising jazz musicians has been demonstrated previously in the computational modelling literature \citep{Frieler2016, Weis2018}. The same is true, also, for rhythm \citep{Cheston2024-RSOS}, with \citet[pp. 86-88]{Benadon2006} noting how ``expressive features of `time feel' serve to define the stylistic profile of [particular] jazz musicians''. Finally, that harmony should be particularly indicative of performance style is perhaps unsurprising: in his monograph on jazz style, Mark \citet[p. 95]{Gridley1994} describes how ``each pianist's particular approach to ... chording [is] a signature for [their] style''.

An interesting discrepancy here is that harmony is seemingly the most predictive domain when used on its own, while rhythm is the most important when all other domains are included. One possible interpretation could be that some melodic information bled into the harmony representation, and vice-versa. As we did not remove the output of the skyline algorithm before extracting chords (see section \ref{sec:rsi_handcrafted_feature_extraction}), there could be a degree of overlap between the features learned by the harmony and melody sub-networks. This is difficult to avoid, however, as the top note of any chord can theoretically have both a harmonic and melodic function in jazz \citep{Levine2011}.

\subsubsection{Which Domains Best Distinguish Particular Performers?}\label{sec:rsi_performer_domain_distinctiveness}

\begin{figure}[]
  \centering
  \includegraphics[width=1\textwidth]{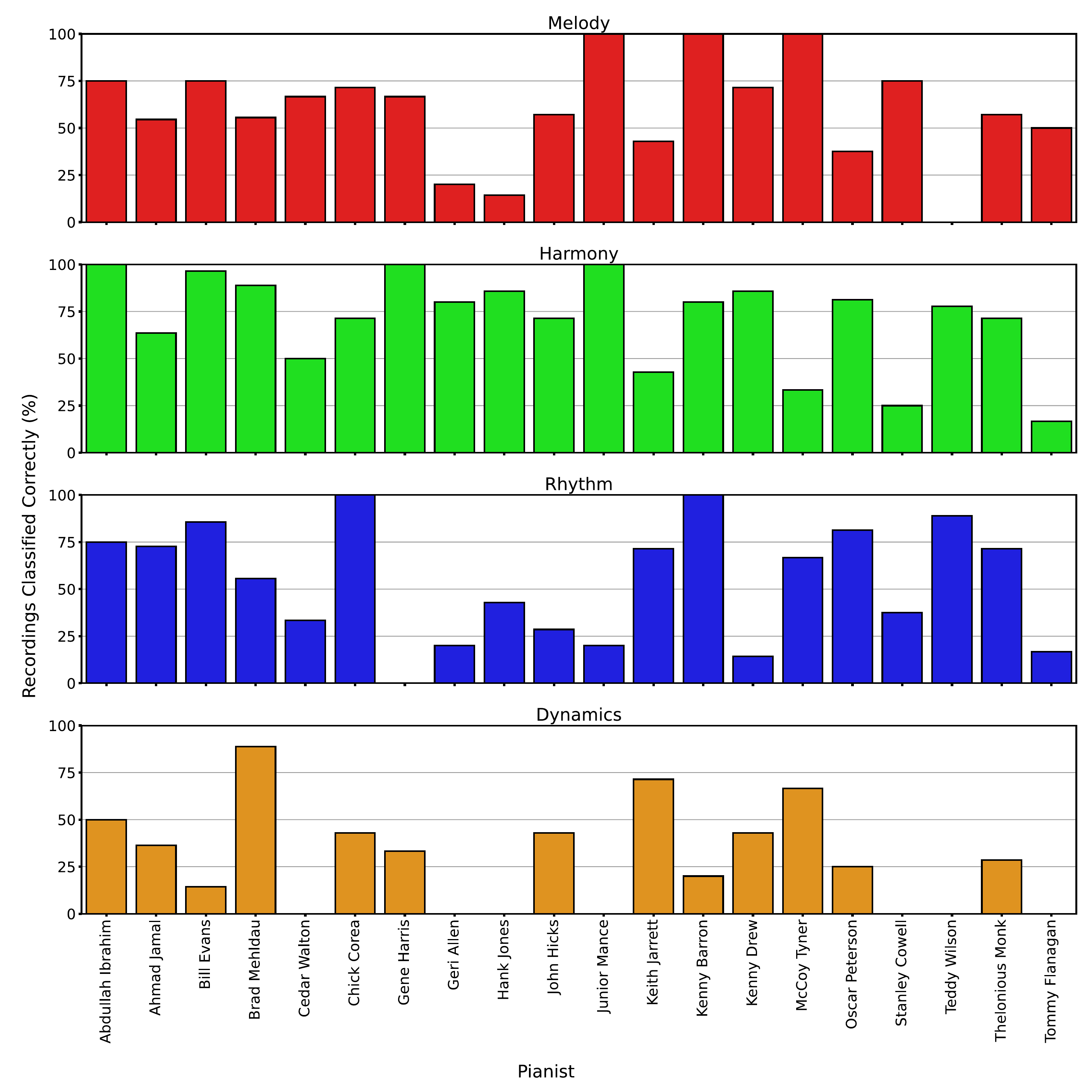}
  \caption{Class accuracy per musical domain. Each facet shows the percentage of recordings in the held-out test split classified correctly for each pianist when predicting using only a single sub-network.}
\label{fig:rsi_class_accuracy_per_domain}
\end{figure}

In Figure \ref{fig:rsi_class_accuracy_per_domain}, we show the per-class accuracy for predictions made using a single sub-network, which allows us to consider the degree to which particular domains best distinguish individual performers. The 22 held-out recordings by Bill Evans, for instance, can be predicted with 0.964 accuracy using only the harmony sub-network. In his description of Evans' improvisation style, Brian Priestley \citep[p. 159]{Carr1988} notes that ``his left-hand [chord] voicings were utterly distinctive although universally imitated''. Related to this is how numerous pianists --- all of whom were born after Evans --- are misclassified by our model as Evans when predicting with only the harmony sub-network (Figure \ref{fig:rsi_sm_factorised_heatmap}). For Ted \citet[p. 275]{Gioia2011}, Evans' voicings ``almost serv[ed] as a default standard among later pianists''. Several of the pianists frequently misidentified as Evans have directly cited him as an influence, including Keith Jarrett \citep{Jarrett2009}.

Our results also suggest that several musicians are particularly distinctive for their use of rhythm. This includes Kenny Barron and Chick Corea, as all of their recordings in the held-out test set can be correctly classified solely using the rhythm sub-network, with considerably lower accuracy obtained when using only the harmony sub-network (0.800 for Kenny Barron, 0.714 for Chick Corea). A possible explanation could be that Corea is one of the only pianists in our dataset to have extensively recorded Latin American music \citep{Gioia2011}. The distinctiveness of every domain for each performer can be explored as part of our interactive web application (see note \ref{note:rsi_webapp}, subheading ``style'').

However, we also note that the performers distinguished most accurately using the rhythm sub-network of our multi-input model differ from those identified by the model described in \citet{Cheston2024-RSOS}. This model was trained on a handcrafted set of rhythmic features derived from prior empirical writings on jazz, such as swing ratio and onset density. There are nine performers in common between the datasets used to train both models. Ranking the accuracy of classifications for these performers and then correlating the ranks demonstrates little connection between the performers that can be distinguished solely using rhythm ($r = -0.296$). One possibility is that the earlier, handcrafted model averages features across an entire recording to produce a global value, while our multi-input model may instead have learned to pick up on more local rhythmic features. 

\subsubsection{What is the Most ``Characteristic'' Example of a Performer's Style?}

Our web application also allows the user to listen to a single example deemed most ``distinctive'' for a given combination of performer and musical domain (e.g., which recording demonstrates the most ``Bill Evans-like'' melodic lines?). This is established by finding the clip from the held-out test set that maximises the logits for the target class when predicting using each of the four sub-networks individually. 

Listening to some of these examples, we can speculate that our multi-input model may indeed have learned to pick up on some of the same features discovered by our handcrafted model (see section \ref{sec:rsi_feature_importance_by_performer}). For instance, the most ``distinctive'' melody clip for Bill Evans includes many of the arpeggio figurations contained in Figure \ref{fig:rsi_predictive_melody_features}a; likewise, the most distinctive harmony clip contains some of the ``rootless'' voicings shown in Figure \ref{fig:rsi_sm_evans_harmony}.

\subsubsection{Which Features are Associated With Particular Performers?}\label{sec:rsi_factorised_cavs}

Finally, we can relate the features learned by our model to a set of music-theoretic concepts extracted from a textbook used widely in jazz education \citep{Haerle1994} in order to understand how these might manifest in each performer's playing. This book comprises twenty chapters, with each chapter containing multiple exercises that demonstrate a single harmonic concept or progression --- beginning with simple ideas (``Block Chords'', ``Diatonic 7th Chords'') and progressing to more complex ones by the end (``Tritone II-V-Is'', ``Dominant Polychord Groups''). For a complete description of each chapter, see Table \ref{tab:rsi_sm_haerle_concepts}. We reproduce a single exercise from this book in Figure \ref{fig:rsi_sm_haerle_example}.

To relate these exercises to the model's predictions, we use concept analysis techniques initially developed for computer vision \citep{Kim2018} and more recently applied to classical music composer identification \citep{Foscarin2022}. To summarise, this method works by using ``activation vectors'' that represent the contributions of human interpretable concepts \citep[e.g., difficult-to-play music, contrapuntal textures in][]{Foscarin2022} to explain a model's decision after it has been trained. Unless stated otherwise, our methods follow those given in \citet{Foscarin2022}.

We derive our concepts from the exercises in each chapter, transposed to C minor/major. As data augmentation, we include versions of each exercise with (1) all possible transpositions up to $\pm 6$ semitones, (2) all possible inversions, and (3) both root and ``rootless'' forms. These exercises are represented in the same two-dimensional piano roll format as used to train our harmony sub-network (see electronic supplementary materials, section \ref{sec:rsi_sm_generating_factorised_representations}). We compute the activations for an exercise by flattening the output of the final layer of this sub-network (``Group 4'', Figure \ref{fig:rsi_proposed_architecture}) from the best-performing multi-input model into a one-dimensional representation with shape $(C \times H \times W)$.

Next, we train binary logistic regressions to separate the activations for one concept with those from a random dataset, consisting of an equivalent number of exercises sampled randomly across every other chapter in Haerle's book. We then take a clip from either the validation or test split of the dataset and score its sensitivity to a concept by taking the dot product of the layer activations with the corresponding concept activation vector. Finally, we define the sign-count ratio for a given performer and concept as the proportion of their clips for which the score is positive. If the sign-count ratio is greater than 0.5, this suggests that this concept encourages the classification of the performer; vice-versa, a sign-count ratio below 0.5 suggests that it discourages classification.

\begin{figure}[]
  \centering
  \includegraphics[width=1\textwidth]{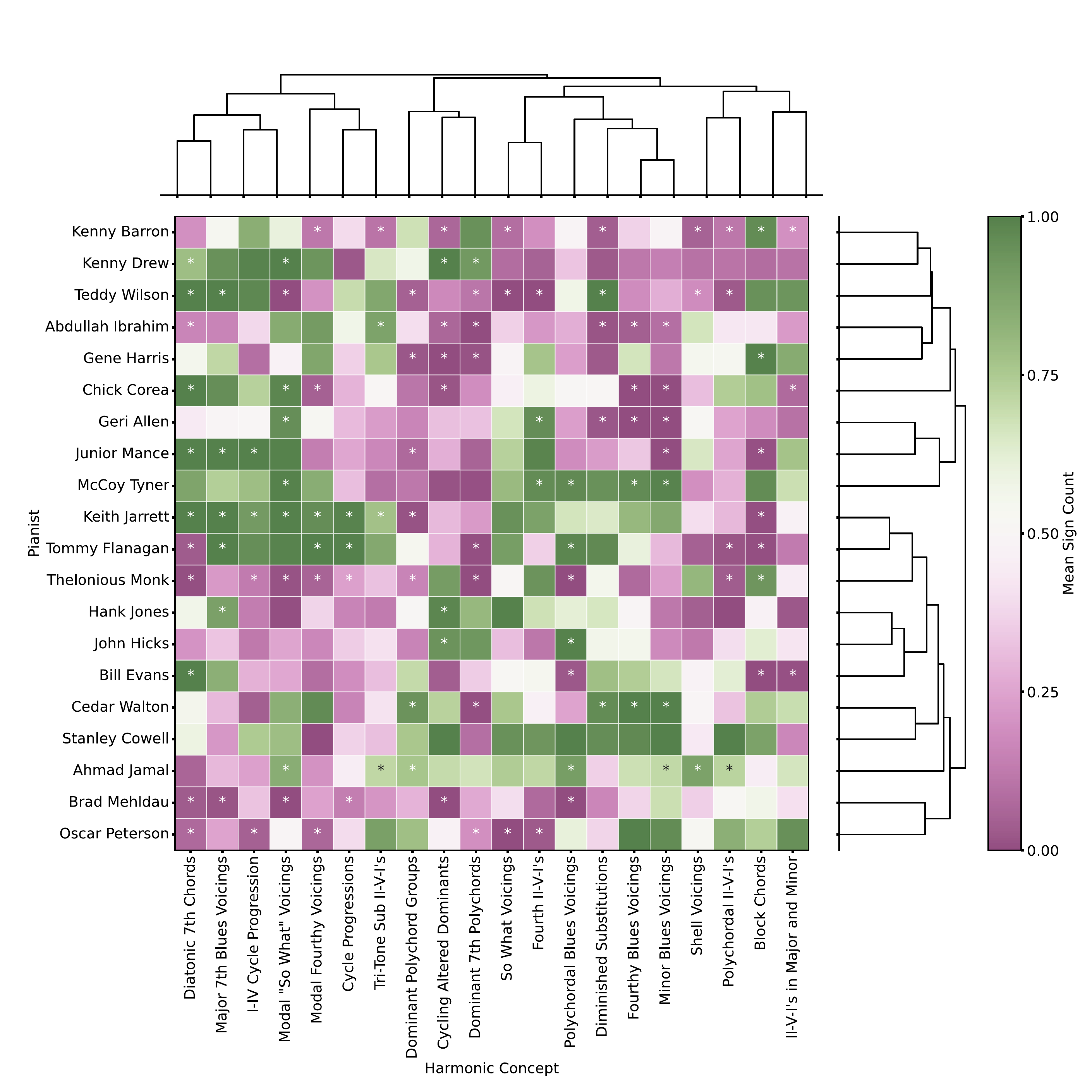}
  \caption{Concept scores. The heatmap shows the mean sign-count scores obtained for each performer ($y$-axis) to every harmonic concept ($x$-axis). The dendrograms show the results of applying agglomerative hierarchical clustering on the sign-count scores within each ``column'' or ``row''; pianists or concepts joined at lower positions are estimated to be more similar to those joined at higher values. Darker green colours indicate a positive influence of a concept on the classification of a performer; darker purples show negative influence. Asterisks show the significance of the sign-count scores ($^* \ p < .05$).}
\label{fig:rsi_cav_sign_counts}
\end{figure}

For every concept, we compute $N$ sign-count ratios for each performer by fixing the concept dataset and drawing a new random dataset for every iteration. We then obtain a null distribution of $N$ sign-count ratios for each performer by performing an equivalent analysis using two non-overlapping random datasets. We set $N = 10$, as in \citet{Foscarin2022}. We then perform a two-sided hypothesis test (Wilcoxon sign-rank) for every performer and concept and apply Bonferroni correction to control the false discovery rate on the performer level, with the number of hypotheses set to the number of concepts (i.e., 20). We show the mean sign-count ratio for each pianist and concept in Figure \ref{fig:rsi_cav_sign_counts}. 

\begin{figure}[]
  \centering
  \includegraphics[width=1\textwidth]{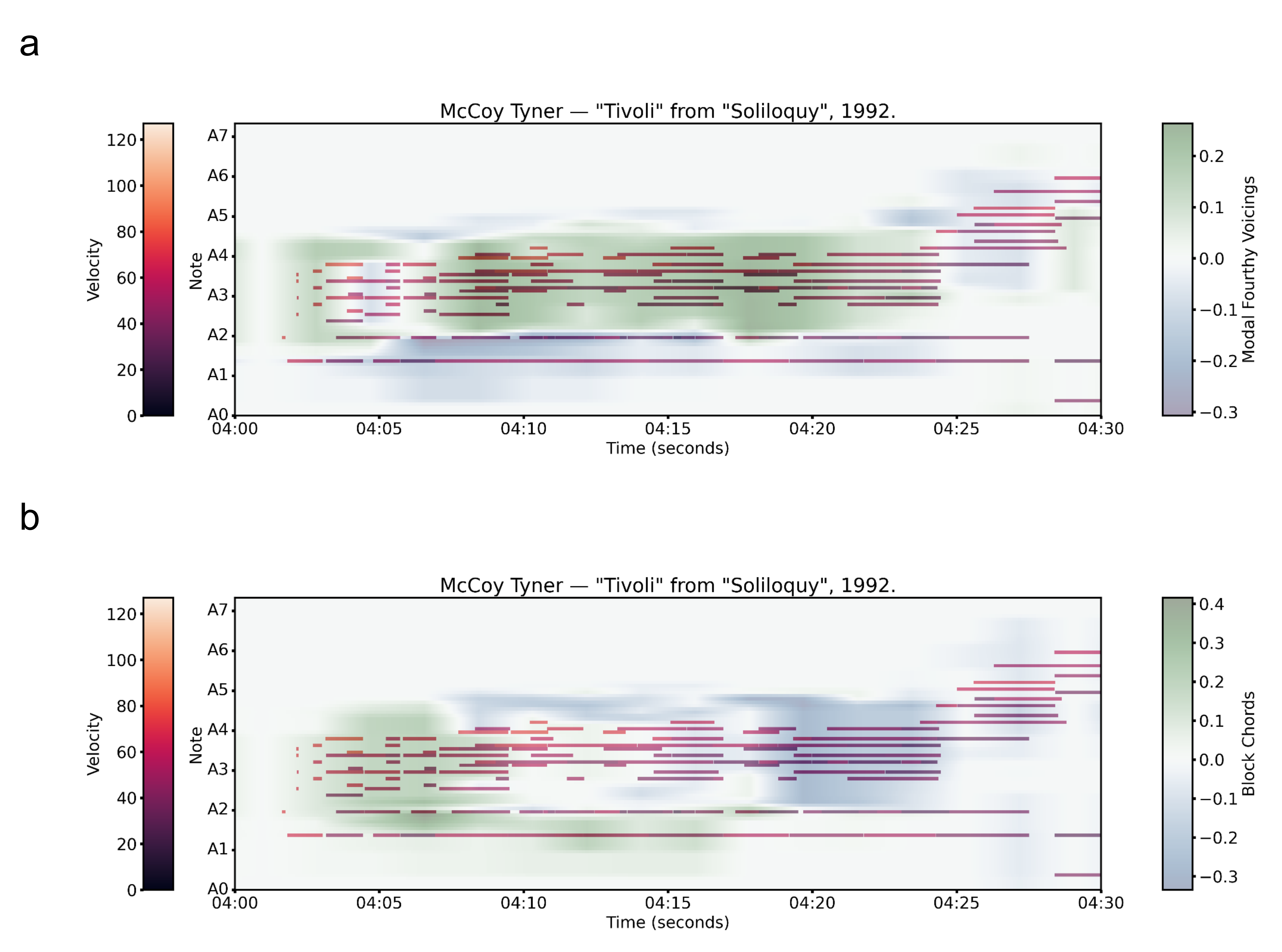}
  \caption{Visualising changes in concept score with masking. Each panel contains a MIDI piano roll with a heatmap overlaid to show the change in score to the (a) ``Modal Fourthy Voicings'' and (b) ``Block Chords'' concepts. Intuitively, green coloured sections contribute positively to the score calculated between the clip and the concept (i.e., they ``embody'' the target concept) and blue coloured sections contribute negative to the score. Values are scaled such that 0 is equivalent to the original score, calculated without masking. Musical notation is provided in Figure \ref{fig:rsi_sm_tivoli_transcription}}
\label{fig:rsi_tivoli_cav}
\end{figure}

To validate our pipeline, we use a masking technique (described in electronic supplementary materials, section \ref{sec:rsi_sm_masked_concept_sensitivity}) to visualise which parts of a performance contribute most to a concept. Figure \ref{fig:rsi_tivoli_cav} shows how the score calculated between a clip and the ``Modal Fourthy Voicings'' concept is based on the appearance of stacked, parallel fourth intervals. The same sections of this clip are not sensitive to the ``Block Chords'' concept, which involves chords based on stacked third intervals. Musical notation for this excerpt is given in Figure \ref{fig:rsi_sm_tivoli_transcription}. We make interactive versions of these figures available for many other performances and concepts as part of our web application (see note \ref{note:rsi_webapp}, subheading ``harmony'').

A number of observations can be made from this analysis. For instance, Brad Mehldau --- the youngest pianist in the dataset --- cannot positively be associated with any concept, which could suggest that contemporary jazz harmony draws from progressions not covered in Haerle's book. Indeed, several of the voicings most strongly associated with Mehldau by the model described in section \ref{sec:rsi_handcrafted_features_approach} consist of inversions of simple triads without any extensions, such as $(0, 8, 15)$: here, see Figure \ref{fig:rsi_sm_mehldau_harmony}. These are not included in any of Haerle's chapters. John Hicks is associated with both the ``Cycling Altered Dominants'' and ``Dominant 7th Polychords'' concept. For Hicks, conceptualising ``two chords a tritone apart'' as a means to alter a dominant seventh by adding a sharpened eleventh and flattened ninth --- the same type of voicing contained within these concepts --- reportedly gave him ``a new freedom'' in his playing \citep[p. 160]{Berliner1994}.

Yet there are also some surprising results. Thelonious Monk is negatively associated with the ``Dominant 7th Polychords'' concept; yet, for Gunther \citet[p. 231]{Schuller1965}, it is in the use of such chords that Monk was ``one of the most imaginative innovators'' in jazz. Equally unexpected is that Bill Evans is not associated with concepts based on the ``So What'' chord voicing, despite himself being the originator of this voicing in work with trumpeter Miles Davis \citep{Gioia2011}. Similarly, Junior Mance was ``known principally for [his] highly ... bluesy approach'' \citep[p. 319]{Carr1988}, but is negatively associated with the ``Minor Blues Voicings'' concept.

These cases should be interpreted with caution, however. Just because the classifier does not consider a particular concept relevant for a musician does not mean that this is not present in their work \citep{Foscarin2022}. While Haerle's book is a comprehensive overview of jazz harmony it is not exhaustive, and there are ways (for instance) to create ``Blues-y'' sounding voicings that are not covered in this book \citep[see, for instance, the method adopted in][]{Adegbija2023}. Another possibility is that our dataset only considers trio and solo improvisations: Evans may have used the ``So What'' voicing primarily during his quintet recordings with Davis, for example.

\subsubsection{How are the Styles of Particular Performers Related?}

Finally, we relate the harmonic style of each performer using agglomerative hierarchical clustering. We compute the correlation coefficient $r$ between the sign-count ratios obtained for two performers across all concepts (i.e., $N = 200$: see Figure \ref{fig:rsi_sm_pianist_correlations}), transform this into a distance matrix by taking $1 - r$, and perform hierarchical clustering using the mean inter-cluster dissimilarity to link performers together \citep[as in][]{Cheston2024}. A dendrogram created from this analysis is shown above the heatmap in Figure \ref{fig:rsi_cav_sign_counts}. 

The final split in the dendrogram partitions the pianists into two groups containing nine and eleven musicians, respectively. The pianists in group 1 are associated more frequently with chords based on fourths (e.g., ``Modal `So What' Voicings'', ``Modal Fourthy Voicings''). McCoy Tyner and Chick Corea are included in this group (see Figures \ref{fig:rsi_sm_tyner_harmony}, \ref{fig:rsi_sm_corea_harmony}). Vice-versa, the pianists in group 2 are instead associated with altered and extended chords (e.g., ``Cycling Altered Dominants'', ``Polychordal Blues Voicings''). Cedar Walton and John Hicks are included in this group (see Figures \ref{fig:rsi_sm_walton_harmony} and \ref{fig:rsi_sm_hicks_harmony}).

To further validate our pipeline, we can apply the clustering technique to all of the sign-count ratios obtained for each individual concept (see also Figure \ref{fig:rsi_sm_cav_correlations}). Concepts with inherent musical similarities --- e.g., ``Fourth II-V-Is'' and ``So What Voicings'' (both based on the perfect fourth interval), and``Fourthy Blues Voicings'' and ``Minor Blues Voicings'' (both based on the twelve-bar blues progression) --- are clustered together earlier than those with contrasting harmonic constructions --- e.g., ``Block Chords'' and ``Modal `So What' Voicings'' (built on stacked third and fourth intervals, respectively). A dendrogram from this analysis is shown on the right of the heatmap in Figure \ref{fig:rsi_cav_sign_counts}.

\section{Discussion}\label{sec:rsi_discussion}

The purpose of this research was to consider how machine learning can help deconstruct the ``style'' of individual artists --- the signature elements that make their work recognisable and distinct. We focussed on musical style (and jazz in particular), owing to the richness of critical literature on the subject and the fact that musicians commonly refer to many of these same theoretical concepts. We trained a series of supervised learning models to identify the performer on 84 hours of recordings made by twenty iconic jazz pianists and interpreted their decision-making processes \textit{post hoc}. Models we trained using handcrafted input features proved to be both accurate and interpretable, capable of answering questions relating both to the distinctiveness of high-level musical domains (harmony, melody) and local musical features (particular melodic patterns and chord voicings). Models we trained to learn representations directly from an input proved highly accurate (achieving 94\% accuracy and improving upon the current state-of-the-art), but considerably harder to interpret meaningfully when compared with the handcrafted models. Finally, a novel multi-input architecture we developed that analyses a musical transcription in terms of four high-level dimensions (melody, harmony, rhythm, and dynamics) proved to be both accurate (achieving 91\% accuracy, close to the state-of-the-art) while also remaining interpretable in terms of these four domains. Ultimately, however, it seems that there is no single computational model capable of answering every question we might conceivably have about artistic style: rather, we found that different models were effective at answering different questions. 

This computational approach to the analysis of art has several intrinsic benefits. While human experts can assess style within small numbers of individual artworks \citep[see, for instance,][]{Monson1996}, our computational modelling allows much larger datasets to be processed in ways that are nevertheless still interpretable to humans. In some cases, our analyses also directly reflect stylistic characteristics that artists themselves refer to, such as the association between McCoy Tyner and quartal harmony (see Figures \ref{fig:rsi_tivoli_cav}, \ref{fig:rsi_sm_tyner_harmony}). A dismissive view of this work might, therefore, simply claim that our models show artists to ``do what they say they do''. However, in many cases --- including interviews with several pianists studied here, such as Abdullah Ibrahim, and Keith Jarrett \citep{Sidran1992} --- artists are not always willing or able to divulge the characteristics of their own style to others \citep{Berliner1994}. With this in mind, our work takes an important step to registering objectively the ways in which the unique style of different artists forms.

Our work also has important pedagogical implications. Much of the critical writing on artistic style depends on language (including, for instance, ``swing'', ``groove'', and ``feel'' in jazz) that can be used to describe the style of particular artists, but finding clear examples of how this manifests in their work can be challenging. We have demonstrated that computational models are able to automatically perform this task of associating well-known music performers with particularly distinctive features and concepts, with interesting implications for demystifying their style. A few examples include the use of arpeggios spanning a seventh for Bill Evans (Figures \ref{fig:rsi_predictive_melody_features}a, \ref{fig:rsi_lime_plots_main}a), the use of melodic enclosures by Oscar Peterson (Figure \ref{fig:rsi_predictive_melody_features}b), and the use of quartal harmony and the perfect fourth interval by McCoy Tyner (Figures \ref{fig:rsi_cav_sign_counts}, \ref{fig:rsi_tivoli_cav}, \ref{fig:rsi_sm_tyner_harmony}).

As a further pedagogical contribution, we have created a web application that enables users to explore the stylistic signatures of the twenty jazz pianists we consider in this work in more detail (see note \ref{note:rsi_webapp}). This application allows the user to explore recordings where the melodic patterns they are associated with in section \ref{sec:rsi_feature_importance_by_performer} appear \citep[with a similar presentation to][]{Frieler2018}, as well as their sensitivity to the harmonic progressions outlined in section \ref{sec:rsi_factorised_cavs}. A final section of the application allows the user to explore the overall ``distinctiveness'' of each pianist's style using the individual representations described in section \ref{sec:rsi_performer_domain_distinctiveness}, and to explore which individual recordings best distinguish each performer using the four musical domains we consider here. 

We can foresee at least three limitations of this work. For one, musical dimensions overlap somewhat, making it difficult to treat them as independently as we did with our multi-input model. For example, some melodic information may bleed into the harmony piano roll as a result of voice leading: for Keith Jarrett, ``voice-leading is melody-writing in the center of the harmony'' \citep{Jarrett2009}. However, explicitly isolating individual musical domains can be useful, insofar as it allows us to analyse and manipulate each component separately --- as is the case for our concept-based analysis. This can yield insights that are harder to obtain from fully entangled representations, as demonstrated with the LIME analysis in Figure \ref{fig:rsi_lime_plots_main}.

A second limitation concerns the inputs themselves. The ``skyline'' algorithm we used to extract melody is simplistic; unfortunately more sophisticated methods have yet to be optimised for the types of expressive piano performances in our dataset \citep[e.g.,][, see Figure \ref{fig:rsi_sm_melody_extraction_results}]{Chou2024}. Similarly, our harmony input only includes chords where all of the notes have similar onset times. This has certain advantages --- it enables chords to be identified without requiring assumptions about the underlying harmony they outline, for instance. However, it also prevents some musical features from being studied, such as spread chords or pedal points. More complex methods for segmenting harmony from MIDI do exist, such as the work of \citet{Masada2018} and \citet{Pardo2002}, but these particular methods have not been extensively evaluated on jazz piano.

A third limitation concerns our concept-based evaluation method. Unlike \citet{Foscarin2022}, we did not systematically validate our pipeline by generating labelled examples that demonstrate particular concepts; instead, we used the labels already provided in a textbook. Additionally, it is sometimes unclear exactly what is being encoded within a concept: considering the sensitivity heatmaps in Figure \ref{fig:rsi_tivoli_cav}, this could reasonably include the intervallic structure of particular chords, or their overall harmonic trajectory. A development of this method could involve applying our concept-based technique to activations from different layers of the harmony sub-network --- perhaps starting from the assumption that shallower layers may better capture individual chords (equivalent to the ``edges'' of an image) due to their smaller receptive field.

Nonetheless, this work unlocks several exciting possibilities for further research. We considered ``style'' to relate to an individual artist. However, style can also be considered with respect to geographical, historical, and cultural trends: one example in jazz is the ``Detroit school'' of pianists \citep[see][]{Gioia2011}. Our model could also be used to investigate these possibilities, for instance by studying the evolution of artistic style over time \citep{Hamilton2024, Broze2013}. Training equivalent models on (e.g.) classical performances would also enable interesting comparisons of style across musical genres to be made \citep{Harrison2018-harmony}. Equally interesting would be to consider whether human judgments of artistic style reflect the trends of our models. 

Taken together, our work synthesises previous research on the explainable modelling of artistic style and introduces a music performer identification architecture that achieves near state-of-the-art classification accuracy while maintaining interpretability. These models provide an opportunity for computer science and machine learning researchers to engage with the humanities through the development of ``human friendly'' explanations of artistic style. We hope that the release of our codebase and pre-trained models will encourage future researchers to continue embracing explainability as a core design principle motivating this research.

\section{Inclusion \& Ethics}

This work did not require ethical approval from a human subject or animal welfare committee.

\section{Data Availability}

Data for this work comes from two published and openly accessible datasets of symbolic musical transcriptions, listed in the manuscript references as \citet{Cheston2024} and \citet{Edwards2023}. We have archived the preprocessed versions of this data and the concept dataset (with no restrictions on data availability) within the Zenodo repository, listed in the manuscript references as \citet{Cheston2025-dataset}. 

\section{Code Availability}

Relevant code (including model cards), alongside the web application developed as part of this work, is accessible in the repository linked at \url{https://cms.mus.cam.ac.uk/jazz-piano-style-ml}. Checkpoints for the trained models are also accessible within the Zenodo repository, listed in the manuscript references as \citet{Cheston2025-dataset}. 

\section{Declaration of AI Use}

We have not used AI-assisted technologies in creating this article.

\section{Authors' Contributions}

H.C.: conceptualization, data curation, formal analysis, investigation, methodology, software, supervision, visualization, writing --- original draft and writing --- review \& editing. 
R.B.: conceptualization, investigation and methodology. 
P.M.C.H.: conceptualization, supervision and writing --- review and editing.

\section{Conflict of Interest Declaration}

We declare that we have no competing interests.

\section{Funding}

H.C. was supported by a postgraduate research scholarship from the Cambridge Trust.

\printbibliography

\clearpage


\setcounter{equation}{0}
\setcounter{figure}{0}
\setcounter{table}{0}
\setcounter{page}{1}
\setcounter{table}{0}
\setcounter{section}{0}
\makeatletter
\renewcommand{\thesection}{S.\arabic{section}}
\renewcommand{\theequation}{S.\arabic{equation}}
\renewcommand{\thefigure}{S.\arabic{figure}}
\renewcommand{\thetable}{S.\arabic{table}}

\section{Supplementary Material: Deconstructing Jazz Piano Style Using Machine Learning}

\subsection{Supplementary Methods and Analyses}

\subsubsection{Extracting Maximally Predictive Features}\label{sec:rsi_sm_maximally_predictive_features}

Given a matrix of feature weights $W \in \mathbb{R}^{I \times J}$ extracted from the LR model, where $I$ is the number of performers (with $I = 20$) and $J$ is the number of features (with $J = 17,918$ for melody, $J = 3,752$ for harmony), we transform this into the matrix $W' \in \mathbb{R}^{I \times K}$ by first taking the maximum absolute value of each feature $j \in \{1, \dots, J\}$ across all performers $i \in \{1, \dots, I\}$ and then taking the top-$K$ of sorted values of $j$, such that
\begin{align}
W'_{i, k} = \text{sort} \left( \max_{i \in \{1, \dots, I\}} \left( |W_{i, j}| \right), \; \forall j \in \{1, \dots, J\} \right)_{i, k}, \ k \in \{1, \dots, K\}.
\end{align}
We then use $W'$ to fit separate models to each dataset (see section \ref{sec:rsi_feature_importance_by_dataset}).

\subsubsection{Convolutional Neural Network Architectures}\label{sec:rsi_sm_cnn_architectures}

Here, we describe the architectures of the two convolutional neural networks evaluated in section \ref{sec:rsi_handcrafted_features_approach} of the full paper. The first model is inspired by the success of convolutional recurrent neural networks (CRNNs) in prior performer and composer identification work \citep{Edwards2023, Kong2021, Mahmudrafee2023}. The input is processed using eight convolutional layers (with average pooling after layers two, four, and six) and a bidirectional gated recurrent unit with two layers and a hidden dimension of 256. This is then followed by a global max pooling layer to yield a 512-dimensional feature vector that passes through a fully connected layer and softmax function to generate class probabilities. Dropout is used with 50\% probability on the classification head, as in \citet{Kong2020}. This model has a total of 14.9 million learnable parameters. 

The second model consists of the classical ``ResNet-50'' architecture, initially developed for computer vision and subsequently used for classical composer identification by \citet{Kim2020} and \citet{Foscarin2022}. The input is processed using an initial convolutional and max pooling layer, before passing into four residual blocks comprising three, four, six, and three layers of convolutional and batch normalization modules, each with skip connections. Average pooling is used after the final block to generate a 2048 dimensional feature vector, which again passes through a fully connected layer (without dropout) and softmax to generate class probabilities. This model has a total of 23.6 million learnable parameters. Both models are implemented in the \texttt{PyTorch} (version \texttt{2.4.1}) Python library \citep{Paszke2019}. 

\subsubsection{Data Augmentation Pipeline}\label{sec:rsi_sm_data_augmentation}

Here, we describe the data augmentation pipeline used to train all deep-learning models described within the full paper. This pipeline consists of pitch, time, and velocity augmentation applied jointly to a single input clip and is broadly inspired by the procedures in \citet{Liu2022}.

Pitch augmentation involves shifting all note pitch values $p_0$ in the clip by the value $s$, such that $p_{shift} = p_0 + s$, where $s \sim \mathcal{U}(-S, S)$ semitones and $\mathcal{U}$ is a discrete uniform distribution of integers. The upper limit $S$ that values of $s$ can take is defined separately for each individual clip (with a maximum value of $\pm 6$ semitones) so as to always satisfy the inequality $20 < p_{shift}^k < 109, \ k = 1, 2, \dots, K$, where $K$ is the total number of MIDI notes. This is to ensure that augmented pitch values lie within the range of the piano keyboard and have the same intervallic structure as the original input.

Time augmentation applies dilation by scaling all note onset $x_0$ and offset $y_0$ times in the clip by the constant $t$, where $t \sim \mathcal{U}(0.8, 1.2)$ and $\mathcal{U}$ is a continuous uniform distribution. The new onset and offset times are defined as $x_{shift} =x_0 \times t$ and $y_{shift}= y_0 \times t$, respectively. When $t > 1$, any notes where $x_{shift} > 30$ or $y_{shift} > 30$ (i.e., that now lie outside of the clip boundaries: see section \ref{sec:rsi_dataset}) are removed. Vice-versa, when $t < 1$, the right ``edge'' of the clip will potentially now include notes just outside the original clip, with their onset and offset times also scaled by $t$. 

Velocity augmentation involves perturbing the velocity of every note $v_0$ by the random variable $d$ such that the new values $v_{shift}=v_0+d$, where $d \sim \mathcal{U}(-12, 12)$ and $\mathcal{U}$ is a discrete uniform distribution of integers. Note that, unlike both pitch and time augmentation, we sample a new value of $d$ from $\mathcal{U}$ for every value of $v_0$. Additionally, values of $v_{shift}$ are clipped to satisfy the inequality $0 < v_{shift} < 128$ to conform with MIDI specifications.

The final part of our pipeline involves randomly adjusting the boundaries that are used to segment a single recording into the 30-second clips used during training (see section \ref{sec:rsi_dataset} in the full paper). We randomly adjust the hop between two successive clips from the same recording to between 15 and 30 seconds (i.e., between 0 and 50\% overlap). Note that a constant hop of 30 seconds is maintained for the validation and test datasets.

\subsubsection{Generating Multiple Input Representations}\label{sec:rsi_sm_generating_factorised_representations}

Here, we describe the process by which the individual inputs used to train the multi-input model described in section \ref{sec:rsi_factorised_inputs_approach} of the full paper are created.

Similar to how $n$-grams were extracted for the handcrafted models (see section \ref{sec:rsi_handcrafted_feature_extraction} of the full paper), to generate the melody representation we apply the skyline algorithm to the transcription to generate a vector of note events $M$. We remove velocity information from each note $m \in M$ by converting the channel dimension to binary, setting a value of 1 for when a note is played and 0 when it is not --- such that $m_{velocity}\in \{0, 1\}$. We remove rhythmic information by setting the inter-onset interval for each note to a constant value dependent on the length of M. Thus, 
\begin{align}
m_{on}^{i+1}-m_{on}^i = \dfrac{T}{|M|}, \ i \in \{1, 2, \dots, |M|\} \ ,
\end{align}
where $T$ is the width of the original input (i.e., 3,000). Note that, when $i = 1$, $m_{on} = 0$ and, when $i=|M|$, $m_{off}^{|M|} = T$. We then adjust the onset and offset times between successive notes to remove any overlap and maintain the original dimensionality of the input, such that $m_{on}^{i+1} = m_{off}^i$.

To generate the harmony representation, we quantise the transcription by grouping near-simultaneous notes into non-overlapping bins according to their onset time, with the size of each bin again set to 100 milliseconds as in section \ref{sec:rsi_handcrafted_feature_extraction} of the full paper. This yields a vector of quantised bins $H$, where the size of every bin $|h| > 3$ for $h \in H$. Unlike the models trained on the handcrafted feature set (see section \ref{sec:rsi_handcrafted_feature_extraction}), however, we do not need to set an upper limit to the number of notes in the chord. We set the inter-chord interval for all notes in each bin to a consistent value, adjust the onset and offset times to remove any overlap between successive chords, and convert the channel dimension to binary.

To generate the rhythm representation, we want to maintain only the onset and offset times for every note $r$ in the original transcription. The influence of dynamics can be removed by converting the channel dimension to binary, as before. While it would be possible to remove the pitch dimension entirely as well, this could lead to situations where two notes with overlapping onset and offset times in the transcription are indistinguishable from each other in the derived representation. Instead, we randomise the pitch of each note $r$, such that $r_{pitch} \sim \mathcal{U}(21, 108)$, where $\mathcal{U}$ is a discrete uniform distribution of integers.

To generate the dynamics representation, we want to maintain only the velocity of each note in a transcription. We follow the process described to generate the harmony representation by binning near-simultaneous notes and adjusting their onset and offset times, but we do not set a lower boundary for the number of notes contained within a single bin (i.e., conceptually similar to the ``snap to grid'' function implemented in many commercial MIDI editors and digital audio workstations). Next, we randomise the pitch of each note in every bin. The representation derived from this process is thus the only one of the four used to train our model where the channel dimension of a note is a floating-point value, rather than an integer.

Compared with the models described in section \ref{sec:rsi_sm_cnn_architectures}, the best-performing multi-input model in Table \ref{tab:rsi_factorised_input_experiment_results} has a total of 44.7 million learnable parameters.

\subsubsection{Measuring Concept Sensitivity with Masking}\label{sec:rsi_sm_masked_concept_sensitivity}

We now describe the technique used to visualise the parts of a performance that contribute the most towards a given concept annotation (see Figure \ref{fig:rsi_tivoli_cav} in the full paper). We compute the original concept score $S_{x, y}$ as 
\begin{align}\label{eq:rsi_sm_cav_calculation}
S_{x, y} = f(x) \circ y \ .
\end{align}
Here, $x$ is a MIDI piano roll with input size $(88, 3000)$, $f$ transforms $x$ into layer activations with shape $(C \times W \times H)$, and $y$ is the vector of coefficients (with the same shape) extracted from a binary classifier trained to separate concept activations from random activations. 

We then slide a two-dimensional rectangular kernel with shape $(24, 250)$ (i.e., 2 octaves, 2.5 seconds) and stride $(2, 200)$ over $x$ and remove all MIDI notes with onset times contained within the span of the kernel. This produces the masked piano roll $x'$. We can then compute the masked concept score $S'_{x', y}$ and express this with relation to the original score:
\begin{align}
S_{x', y} = \frac{S_{x, y} - \biggl(f(x') \circ y \biggr)}{S_{x, y}}
\end{align}

Finally, after computing $S_{x', y}$ at every kernel position, we transform the result back into a two-dimensional image with the same shape as $x$ using linear interpolation in order to create the heatmaps shown in Figure \ref{fig:rsi_tivoli_cav}.\newpage

\subsection{Supplementary Figures}

\begin{figure}[h!]
  \centering
  \includegraphics[width=1\textwidth]{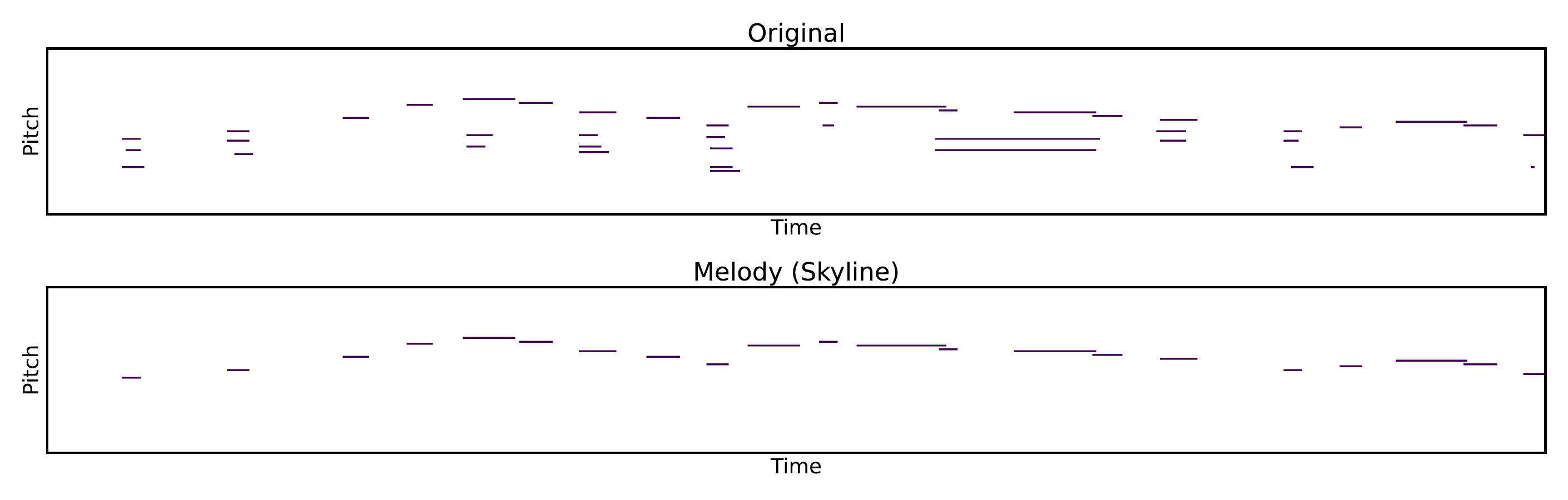}
  \caption{Melody extraction results. The top panel shows ten seconds of a MIDI transcription and the bottom panel shows the same transcription after applying the implementation of the ``Skyline'' algorithm \citep{Uitdenbogerd1998} used in this work. When the same transcription was processed using the model described by \citet{Chou2024}, no notes were detected as being part of the melody.}
\label{fig:rsi_sm_melody_extraction_results}
\end{figure}

\begin{figure}[h!]
  \centering
  \includegraphics[width=1\textwidth]{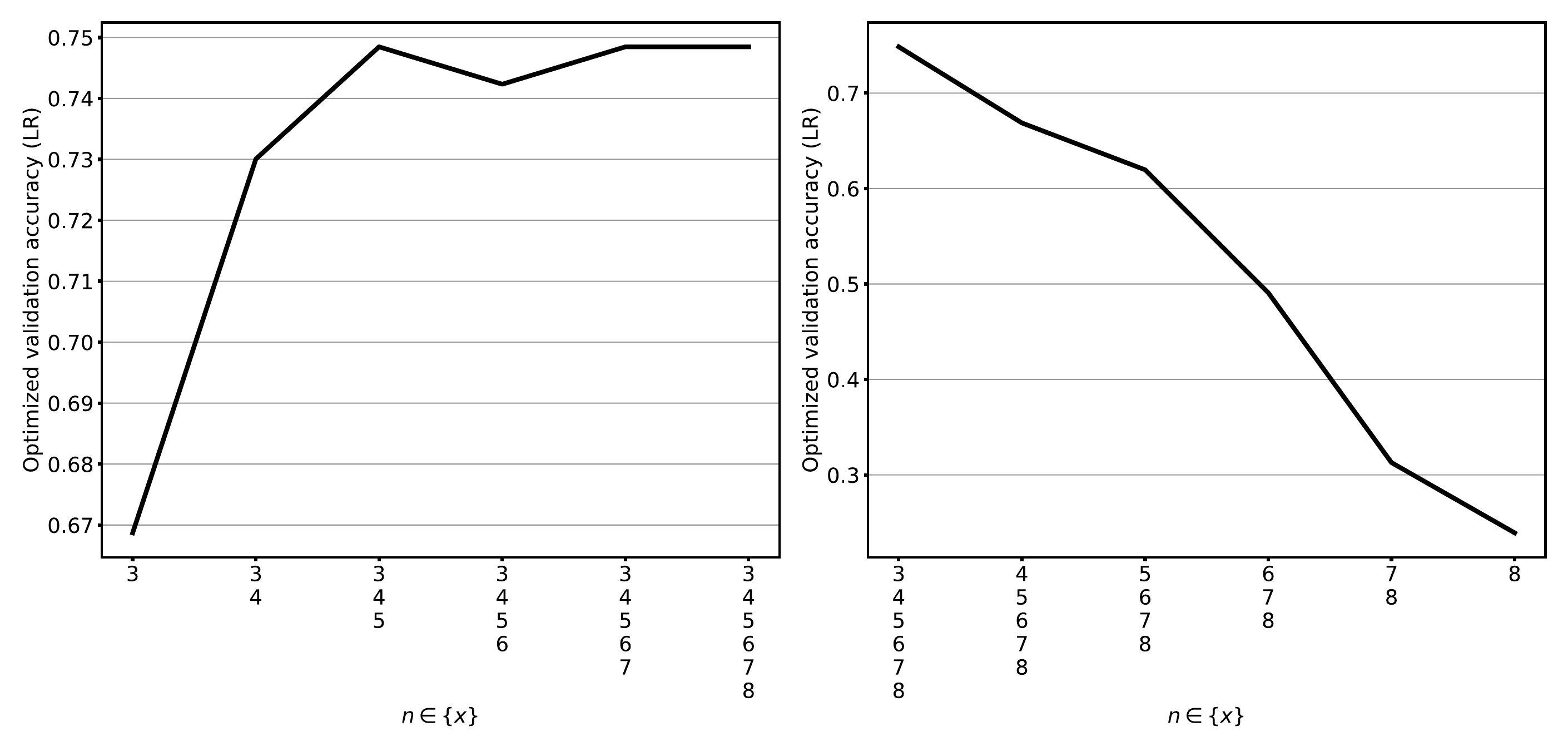}
  \caption{LR accuracy with different values of $n$. The left panel shows changes in accuracy when the maximum value of $n$ used to extract melody and harmony features increases, from 3 to 8 inclusive. The right panel shows equivalent changes when the minimum value of $n$ increases. All accuracy scores are the results of optimising hyperparameters separately using random sampling for each value of $n$ (see section \ref{sec:rsi_handcrafted_training} in the full text) and are obtained for the validation split of the dataset.}
\label{fig:rsi_sm_lr_extraction_at_n}
\end{figure}

\begin{figure}[h!]
  \centering
  \includegraphics[width=1\textwidth]{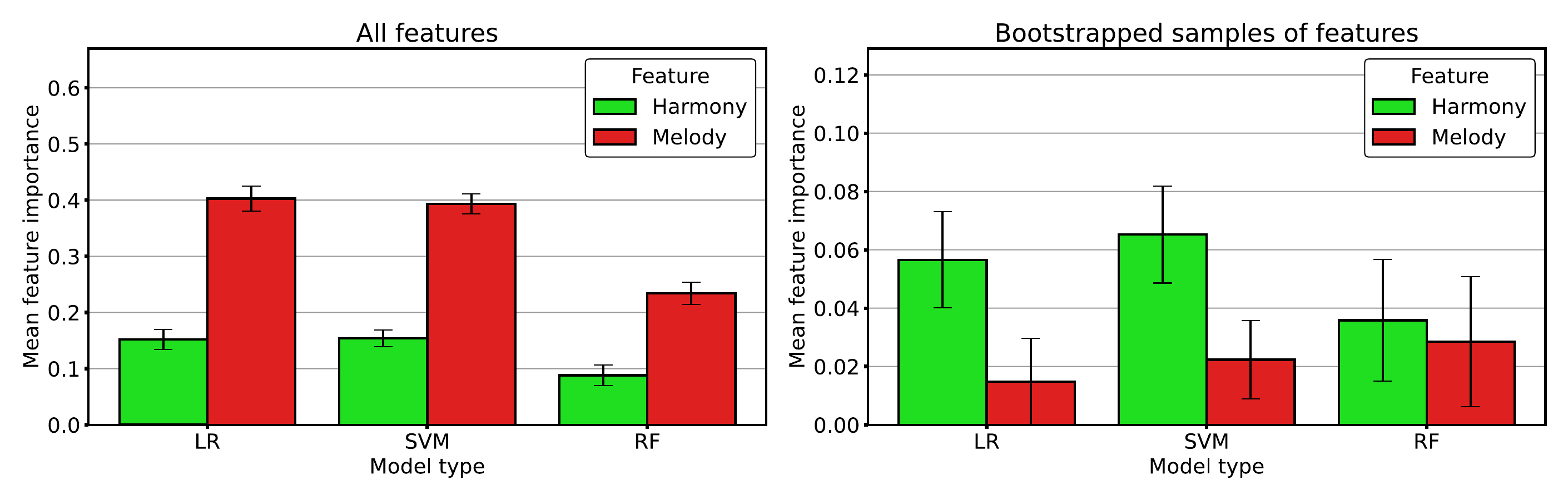}
  \caption{Feature importance by domain and model type. Each bar shows the loss in test accuracy from permuting all melody or harmony features (left panel) and bootstrapped subsamples of 2,000 features (right panel), stratified by model type. Error bars show standard deviations ($N = 1,000$ iterations).}
\label{fig:rsi_sm_feature_importance_by_handcrafted_model_type}
\end{figure}

\begin{figure}[h!]
  \centering
  \includegraphics[width=1\textwidth]{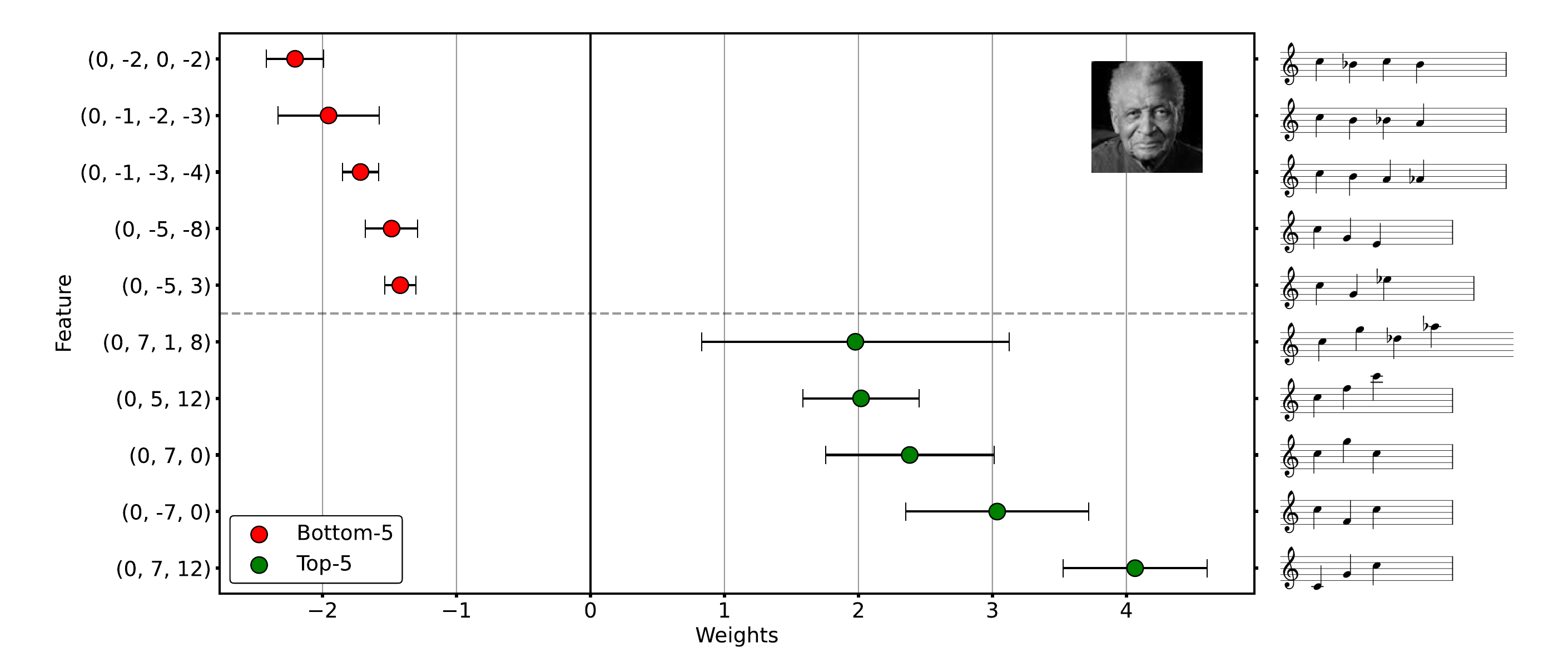}
  \caption{Predictive melody features, Abdullah Ibrahim. This figure (and those following) shows standardised odds ratios obtained for every performer, following Figure \ref{fig:rsi_predictive_melody_features} in the full paper.}
\label{fig:rsi_sm_ibrahim_melody}
\end{figure}

\begin{figure}[h!]
  \centering
  \includegraphics[width=1\textwidth]{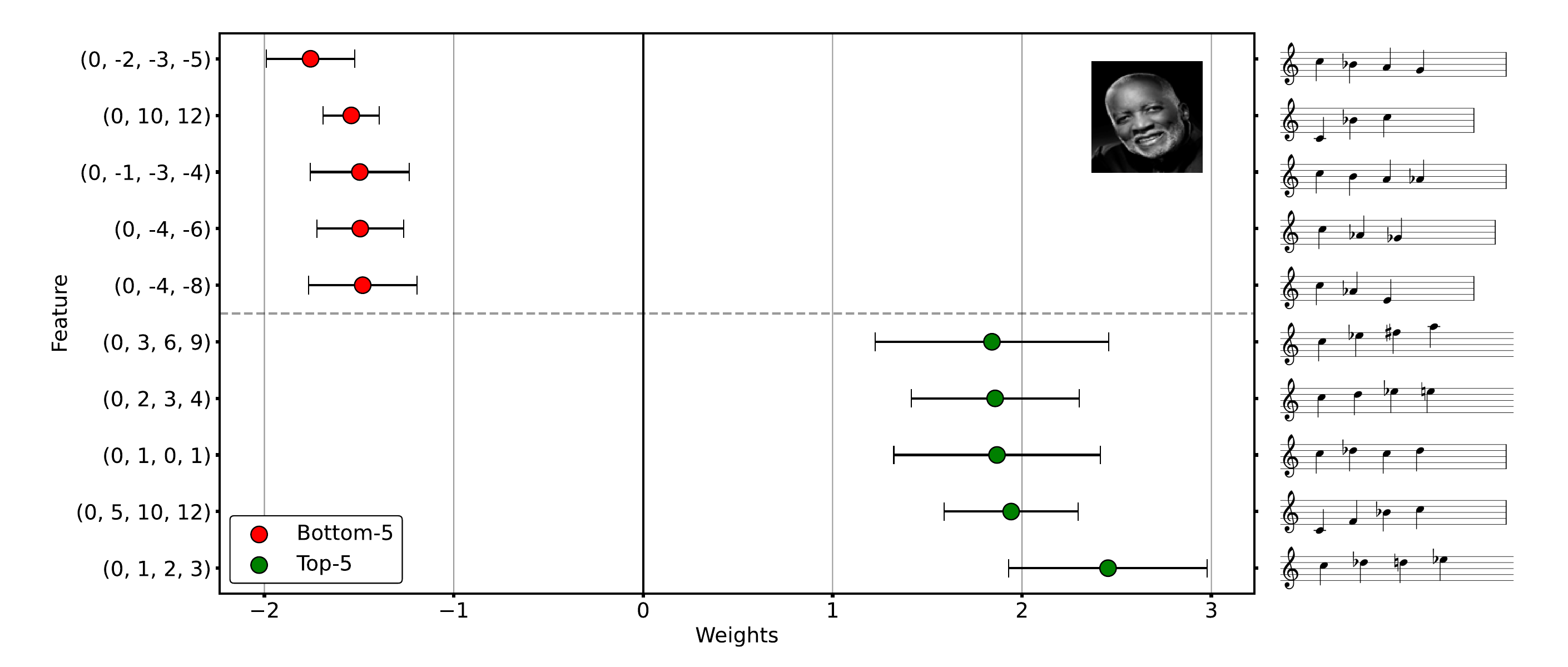}
  \caption{Predictive melody features, Ahmad Jamal.}
\label{fig:rsi_sm_jamal_melody}
\end{figure}

\begin{figure}[h!]
  \centering
  \includegraphics[width=1\textwidth]{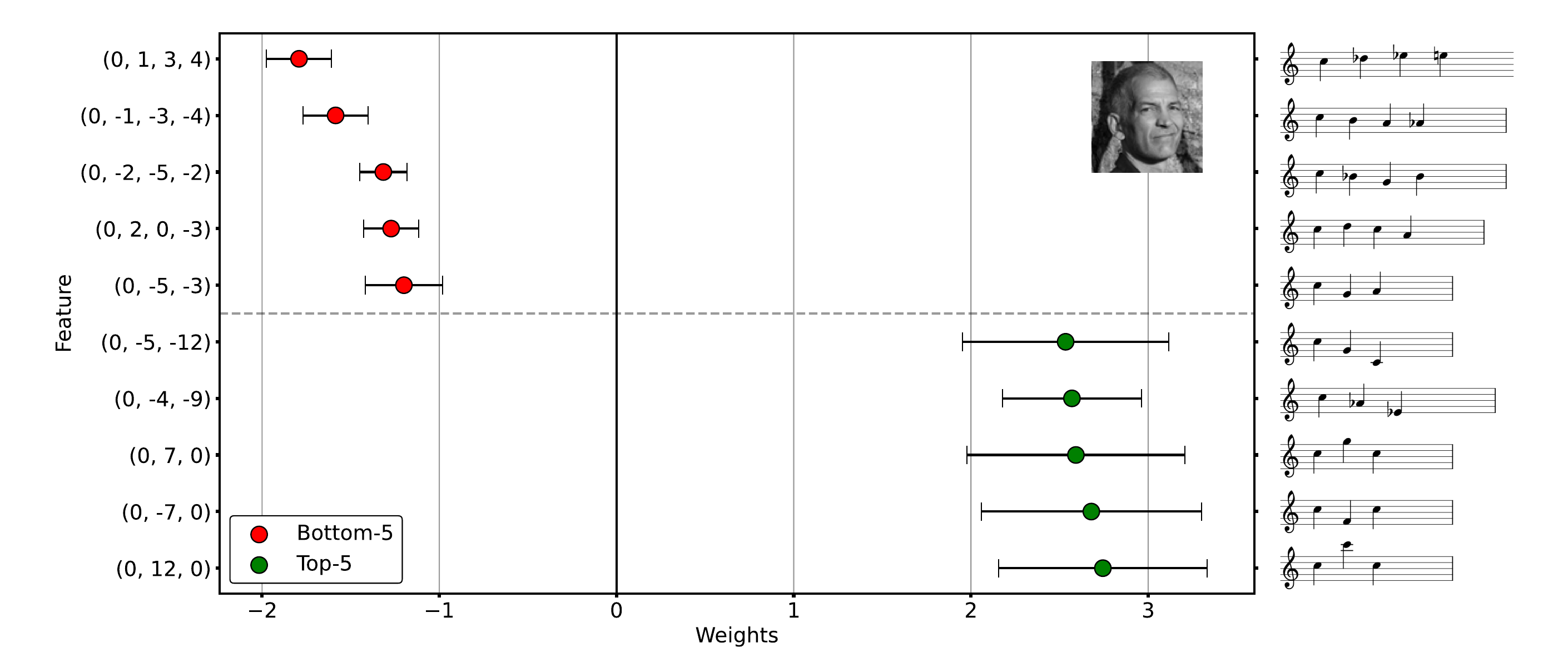}
  \caption{Predictive melody features, Brad Mehldau.}
\label{fig:rsi_sm_mehldau_melody}
\end{figure}

\begin{figure}[h!]
  \centering
  \includegraphics[width=1\textwidth]{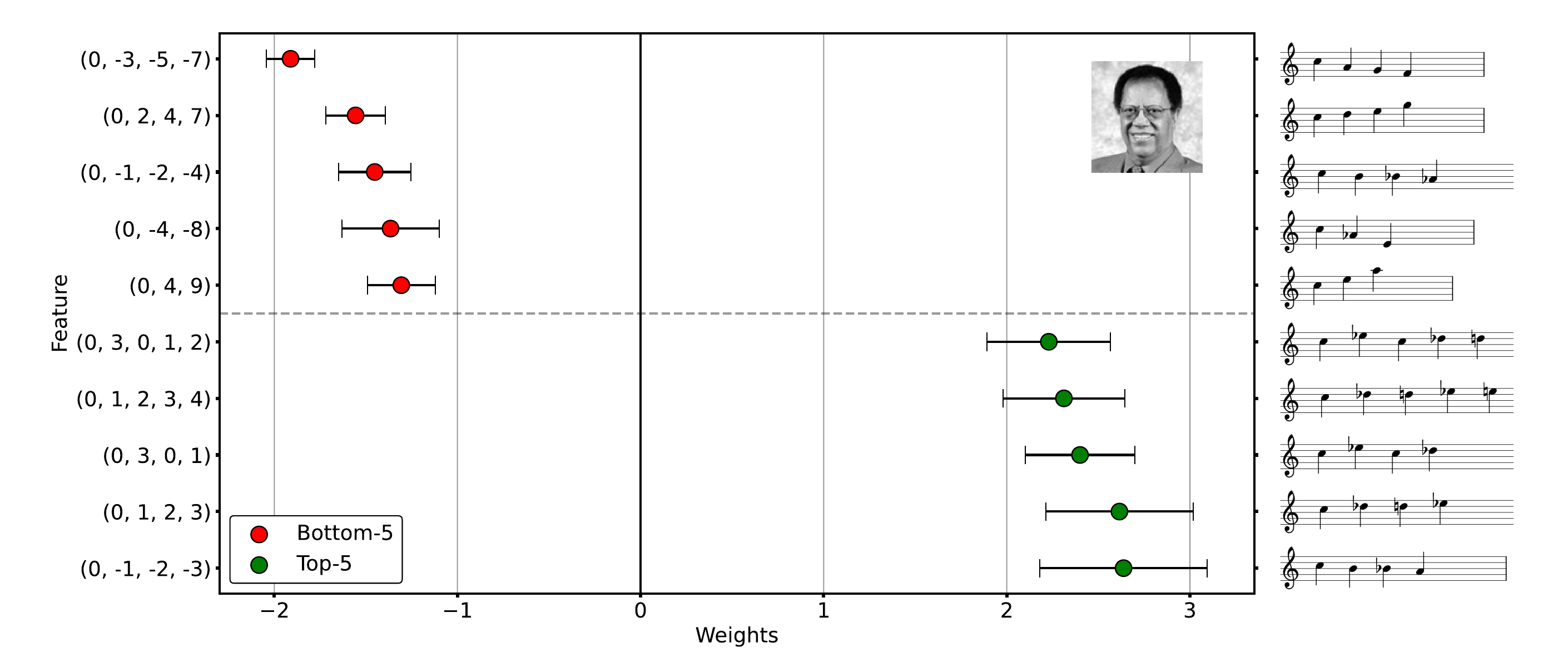}
  \caption{Predictive melody features, Cedar Walton.}
\label{fig:rsi_sm_walton_melody}
\end{figure}

\begin{figure}[h!]
  \centering
  \includegraphics[width=1\textwidth]{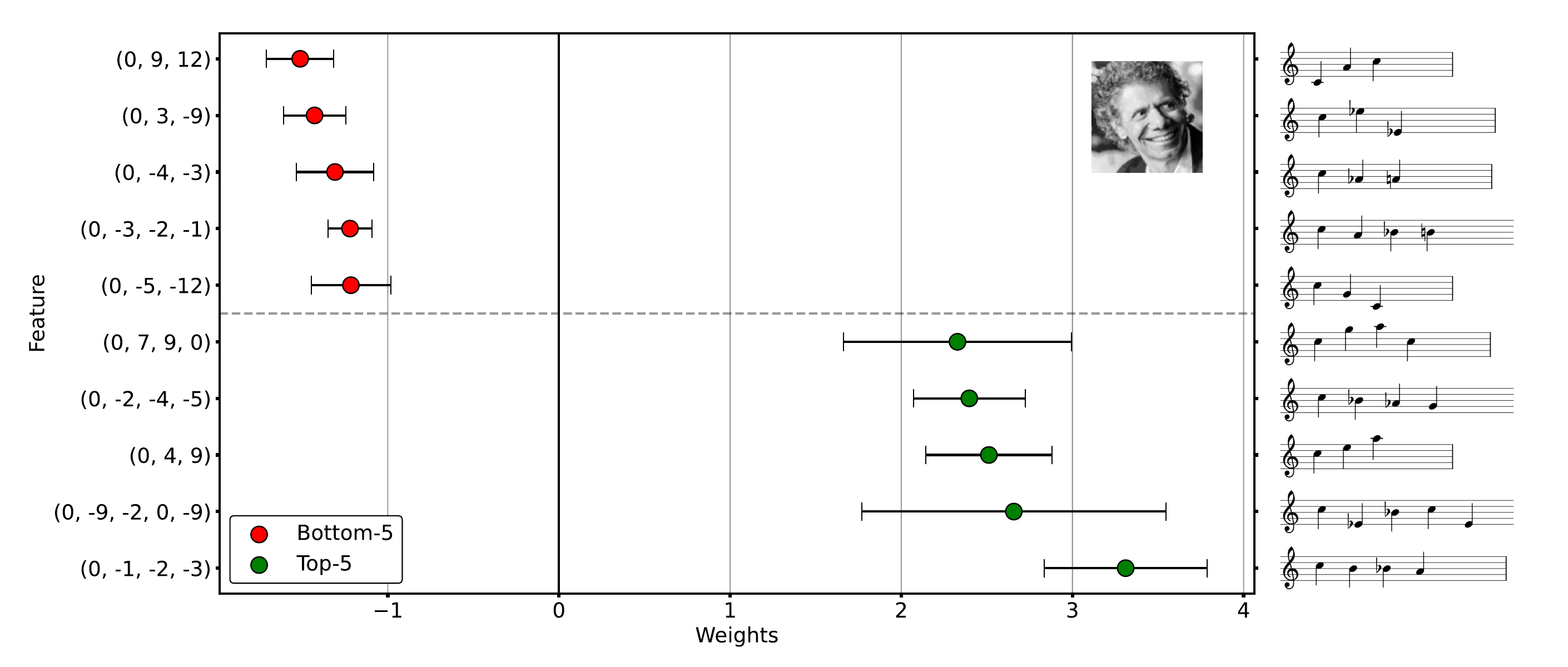}
  \caption{Predictive melody features, Chick Corea.}
\label{fig:rsi_sm_corea_melody}
\end{figure}

\begin{figure}[h!]
  \centering
  \includegraphics[width=1\textwidth]{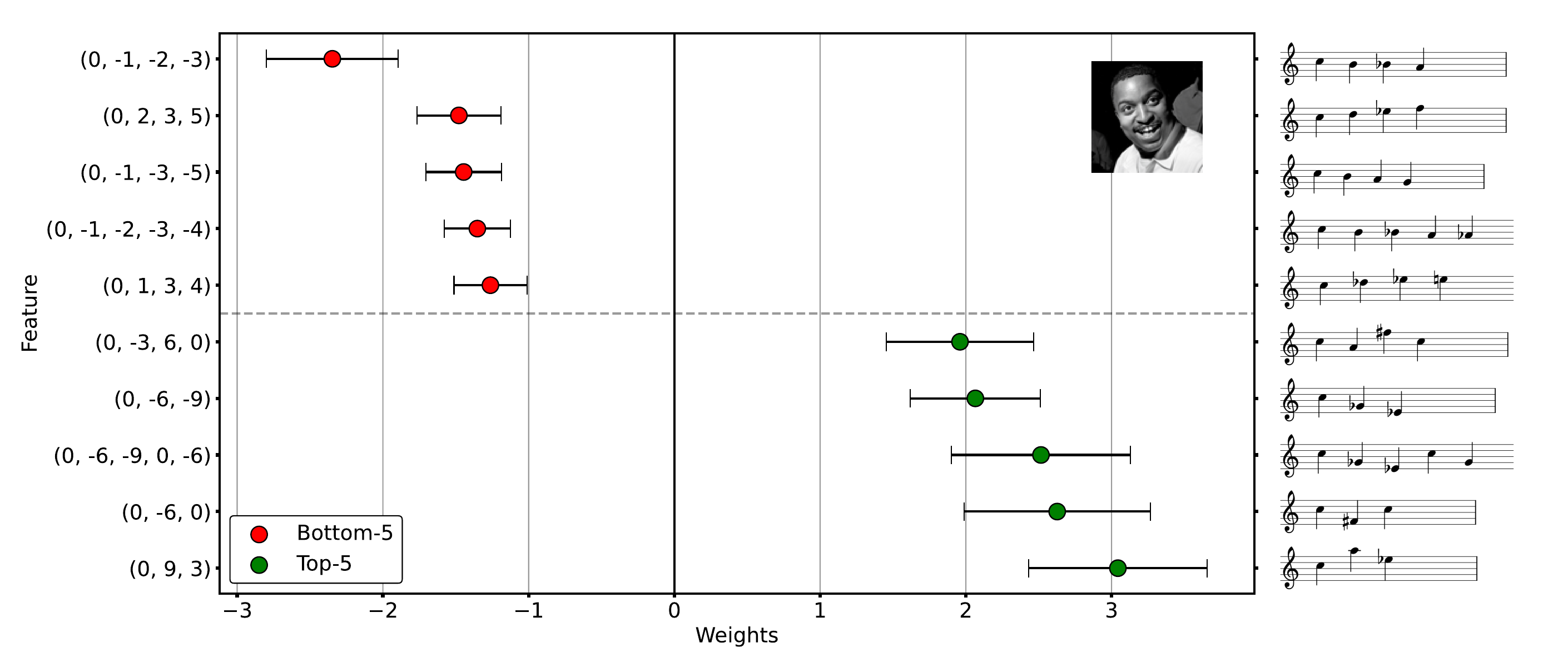}
  \caption{Predictive melody features, Gene Harris.}
\label{fig:rsi_sm_harris_melody}
\end{figure}

\begin{figure}[h!]
  \centering
  \includegraphics[width=1\textwidth]{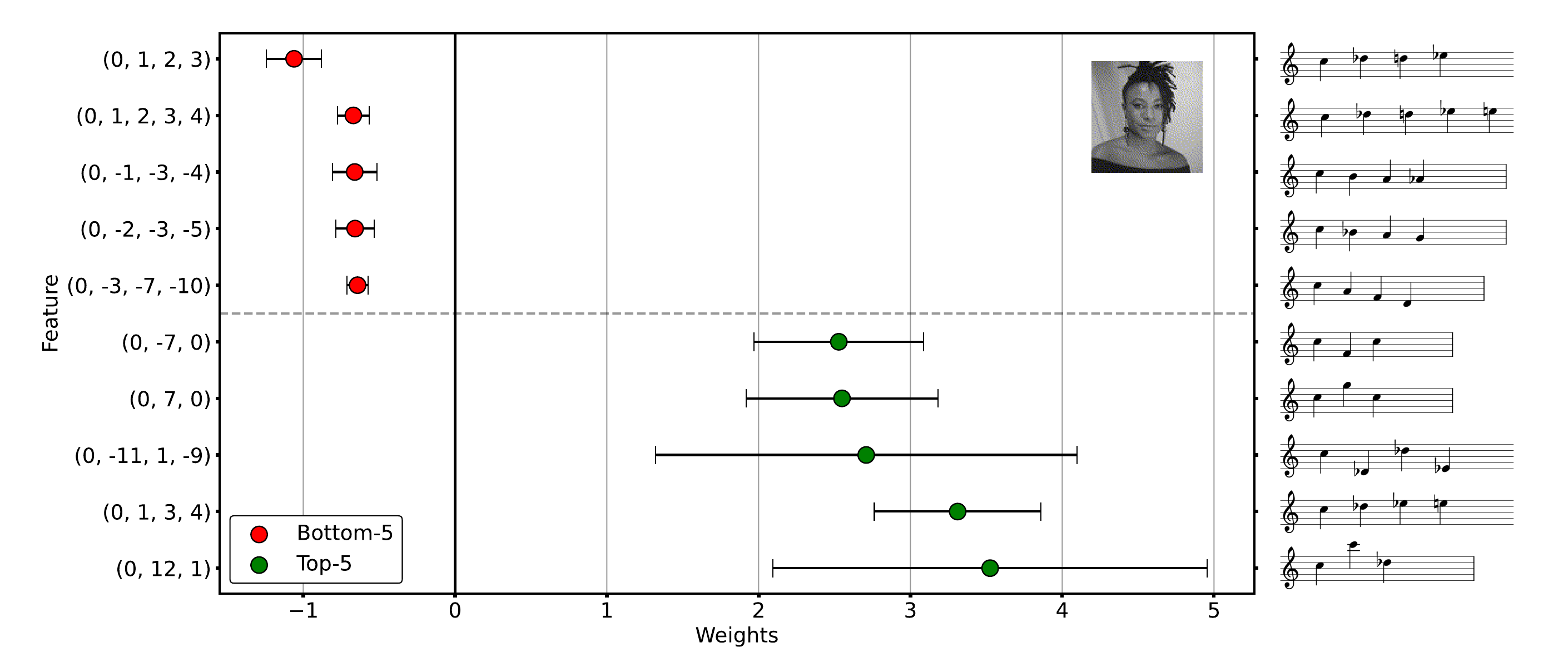}
  \caption{Predictive melody features, Geri Allen.}
\label{fig:rsi_sm_allen_melody}
\end{figure}

\begin{figure}[h!]
  \centering
  \includegraphics[width=1\textwidth]{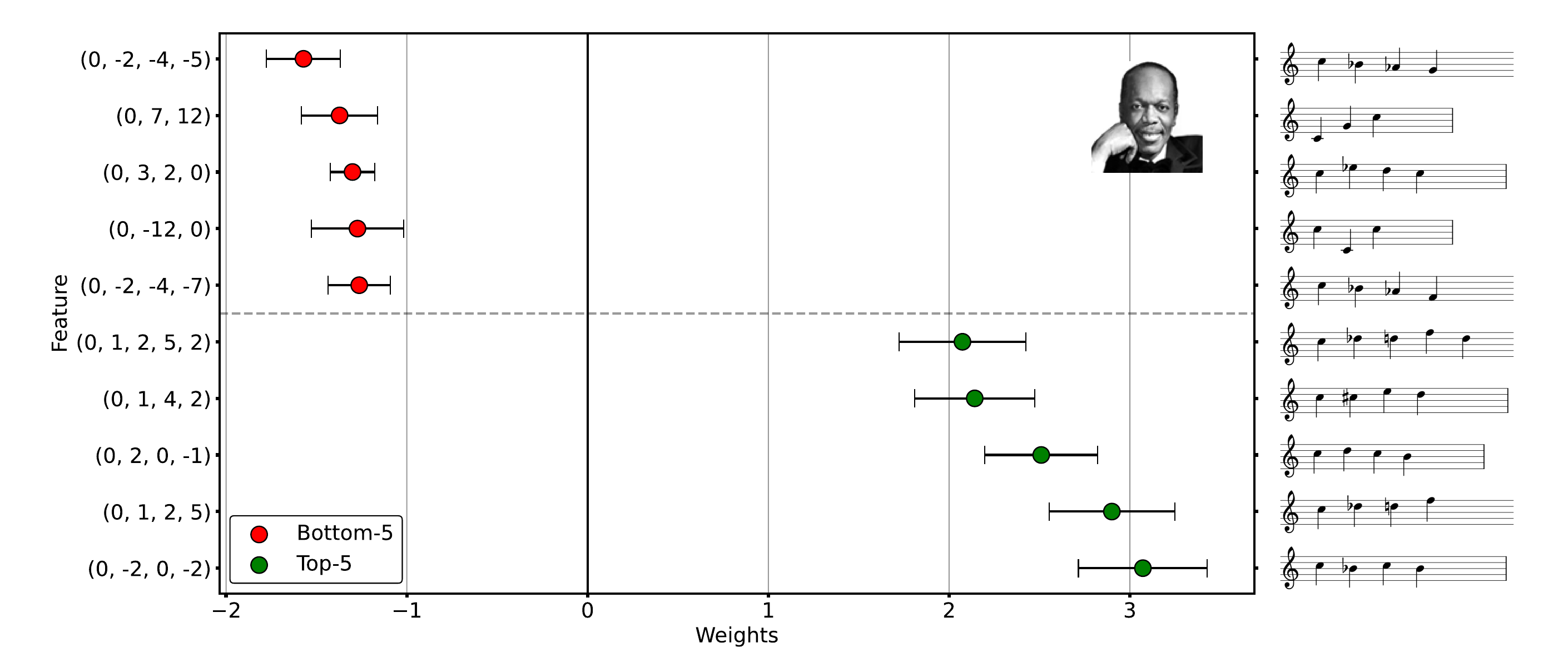}
  \caption{Predictive melody features, Hank Jones.}
\label{fig:rsi_sm_jones_melody}
\end{figure}

\begin{figure}[h!]
  \centering
  \includegraphics[width=1\textwidth]{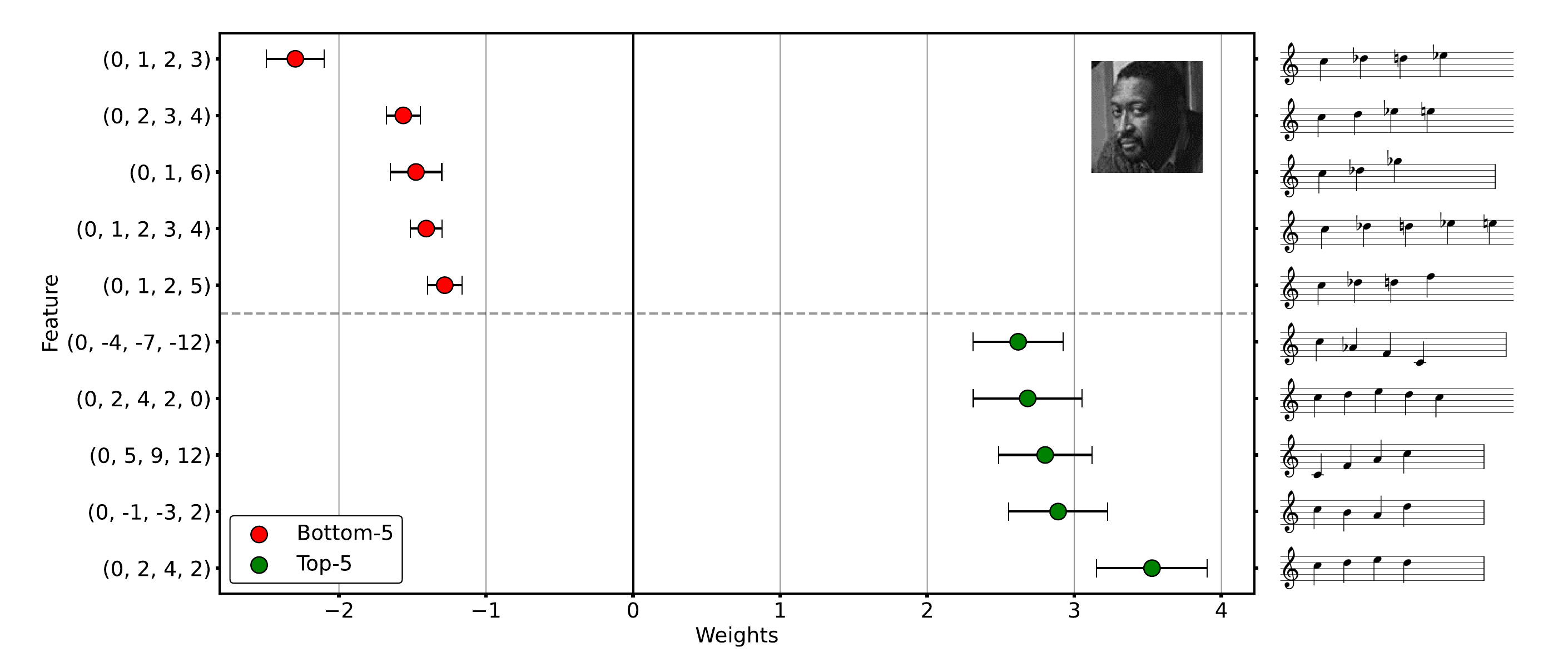}
  \caption{Predictive melody features, John Hicks.}
\label{fig:rsi_sm_hicks_melody}
\end{figure}

\begin{figure}[h!]
  \centering
  \includegraphics[width=1\textwidth]{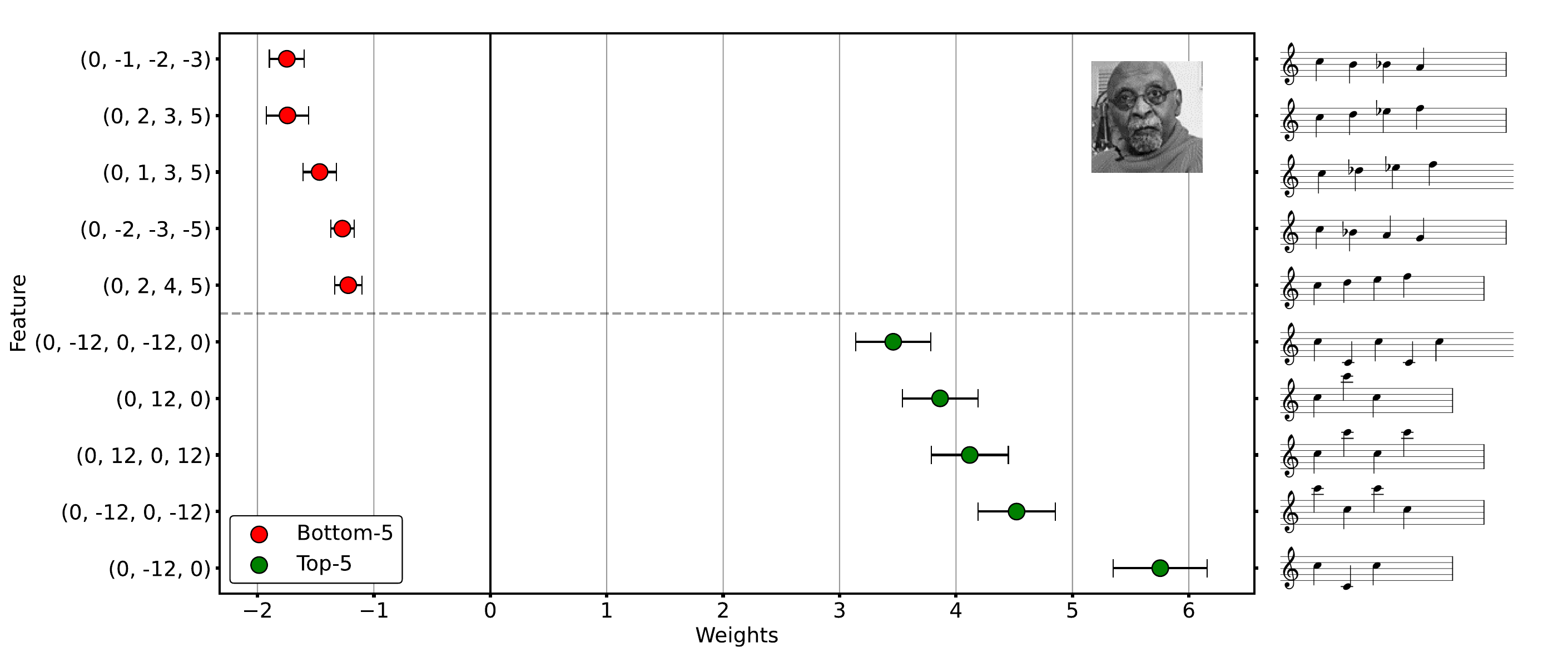}
  \caption{Predictive melody features, Junior Mance.}
\label{fig:rsi_sm_mance_melody}
\end{figure}

\begin{figure}[h!]
  \centering
  \includegraphics[width=1\textwidth]{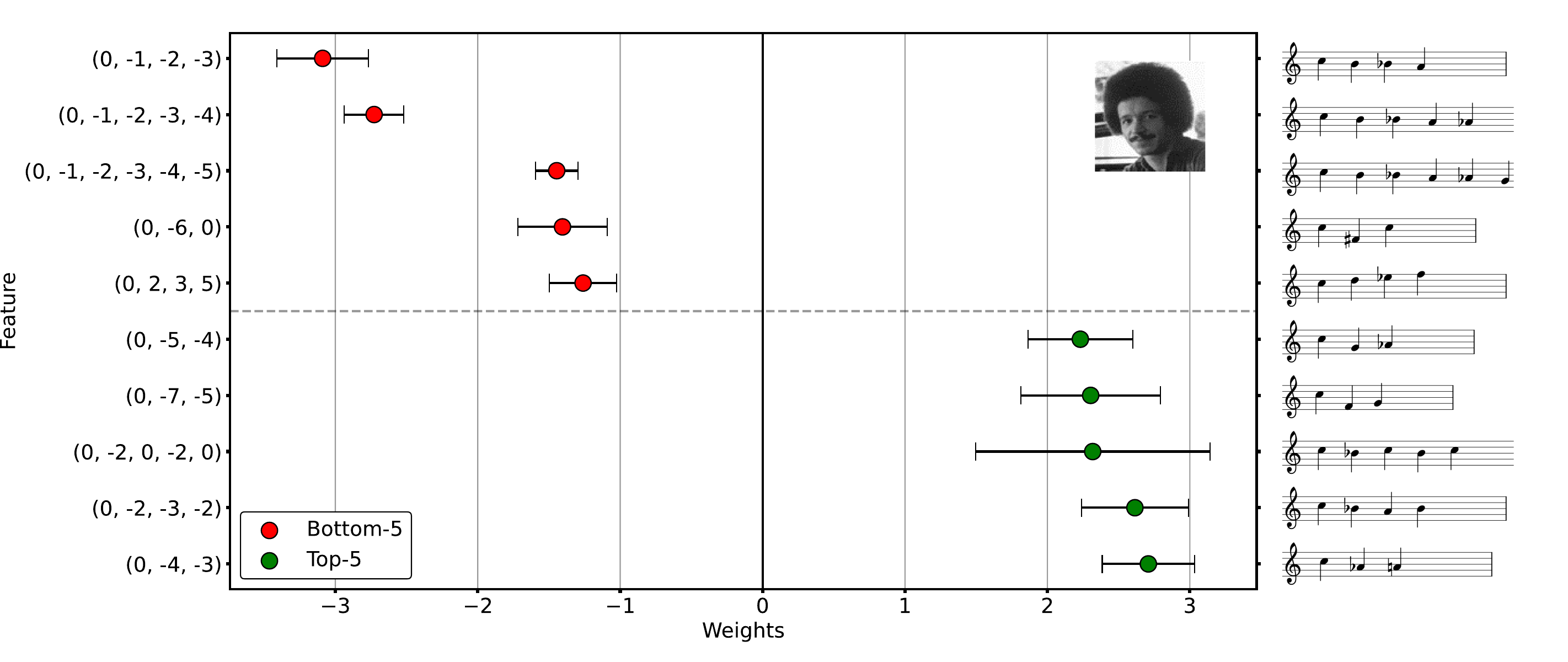}
  \caption{Predictive melody features, Keith Jarrett.}
\label{fig:rsi_sm_jarrett_melody}
\end{figure}

\begin{figure}[h!]
  \centering
  \includegraphics[width=1\textwidth]{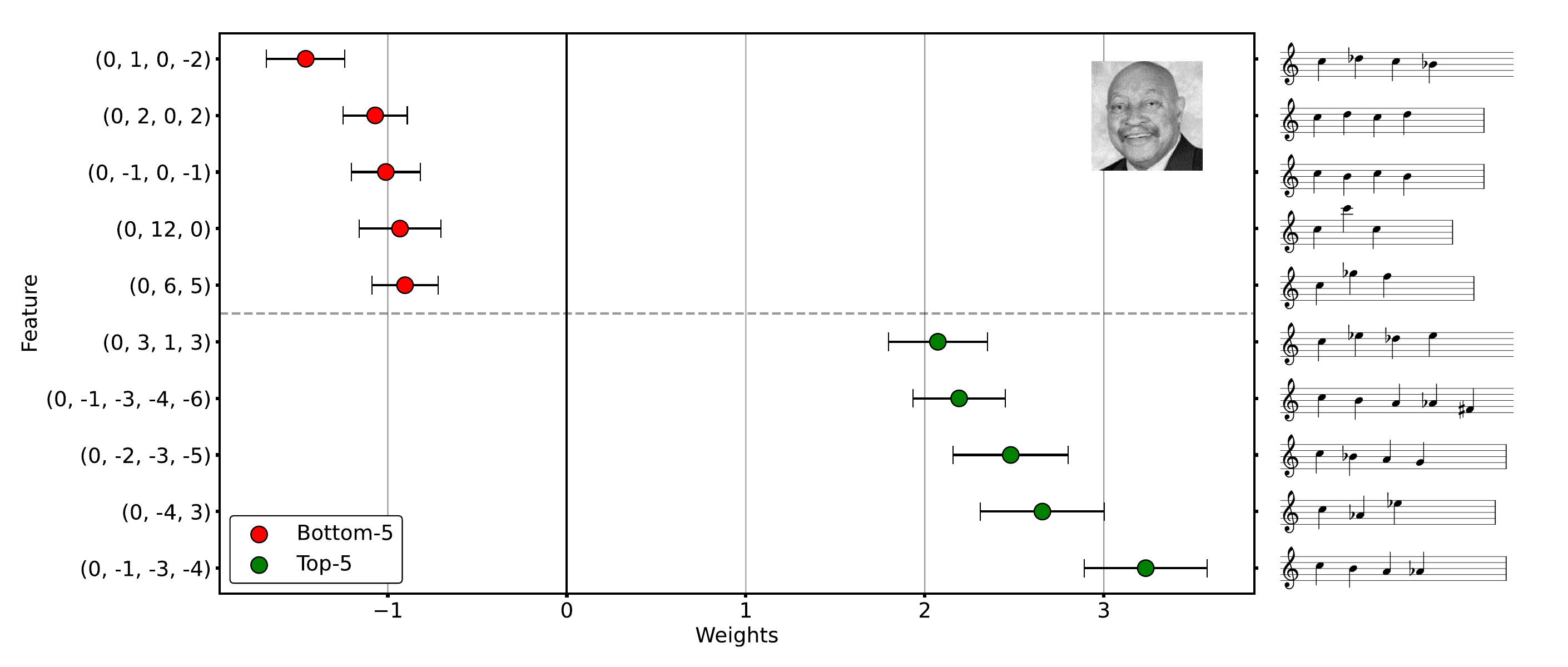}
  \caption{Predictive melody features, Kenny Barron.}
\label{fig:rsi_sm_barron_melody}
\end{figure}

\begin{figure}[h!]
  \centering
  \includegraphics[width=1\textwidth]{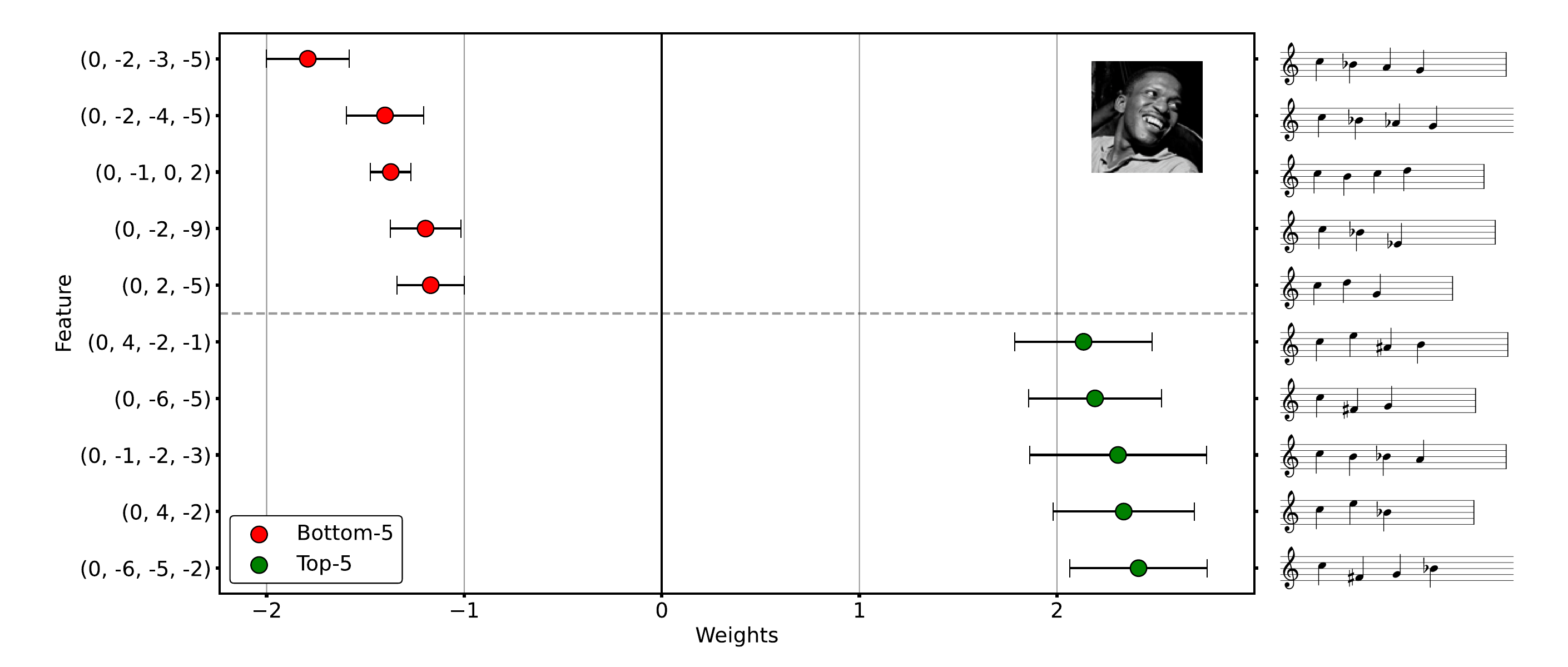}
  \caption{Predictive melody features, Kenny Drew.}
\label{fig:rsi_sm_drew_melody}
\end{figure}

\begin{figure}[h!]
  \centering
  \includegraphics[width=1\textwidth]{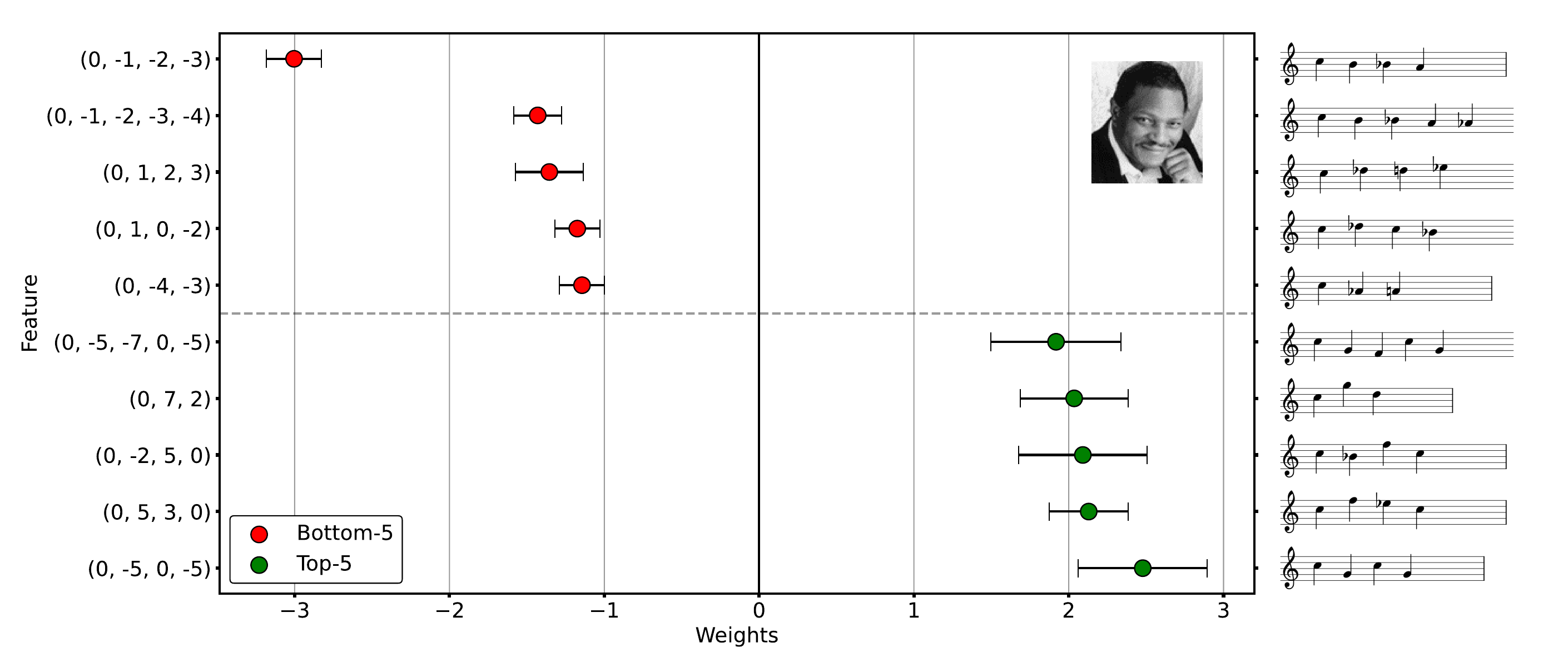}
  \caption{Predictive melody features, McCoy Tyner.}
\label{fig:rsi_sm_tyner_melody}
\end{figure}

\begin{figure}[h!]
  \centering
  \includegraphics[width=1\textwidth]{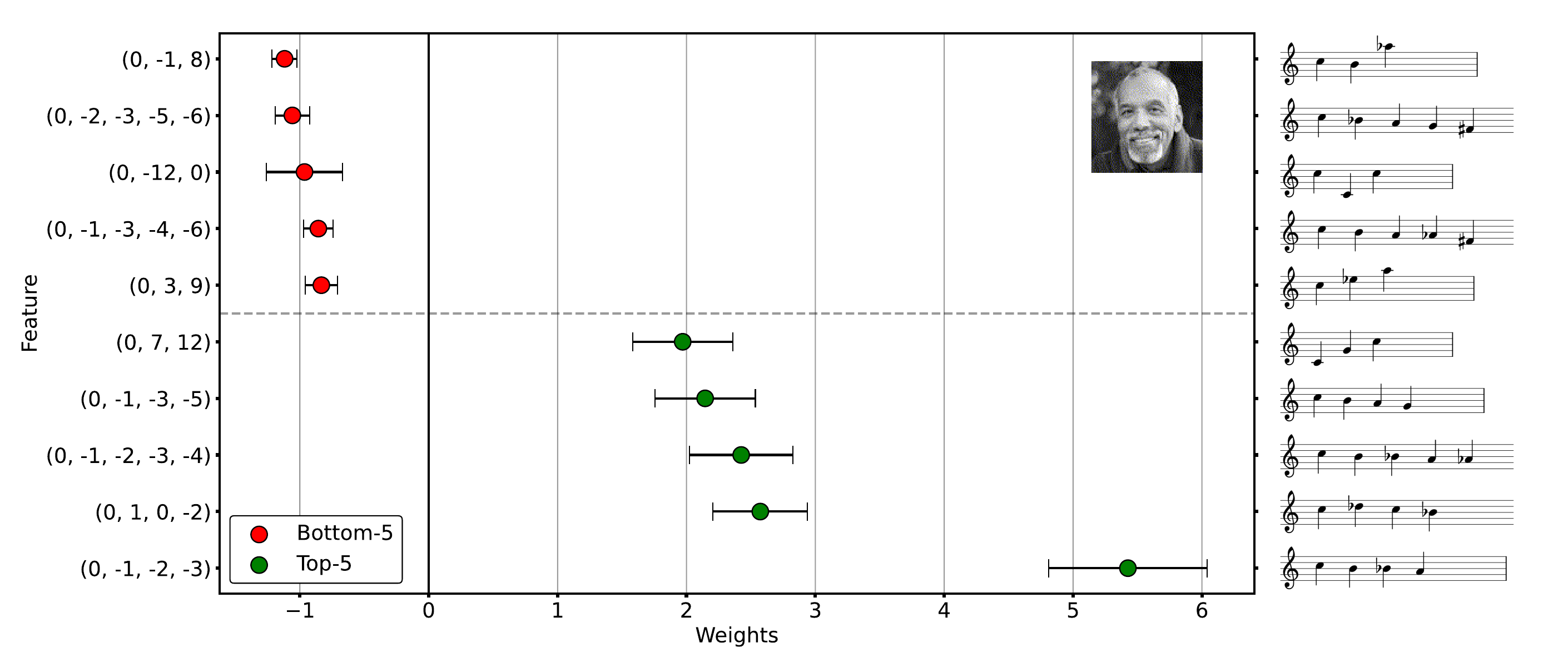}
  \caption{Predictive melody features, Stanley Cowell.}
\label{fig:rsi_sm_cowell_melody}
\end{figure}

\begin{figure}[h!]
  \centering
  \includegraphics[width=1\textwidth]{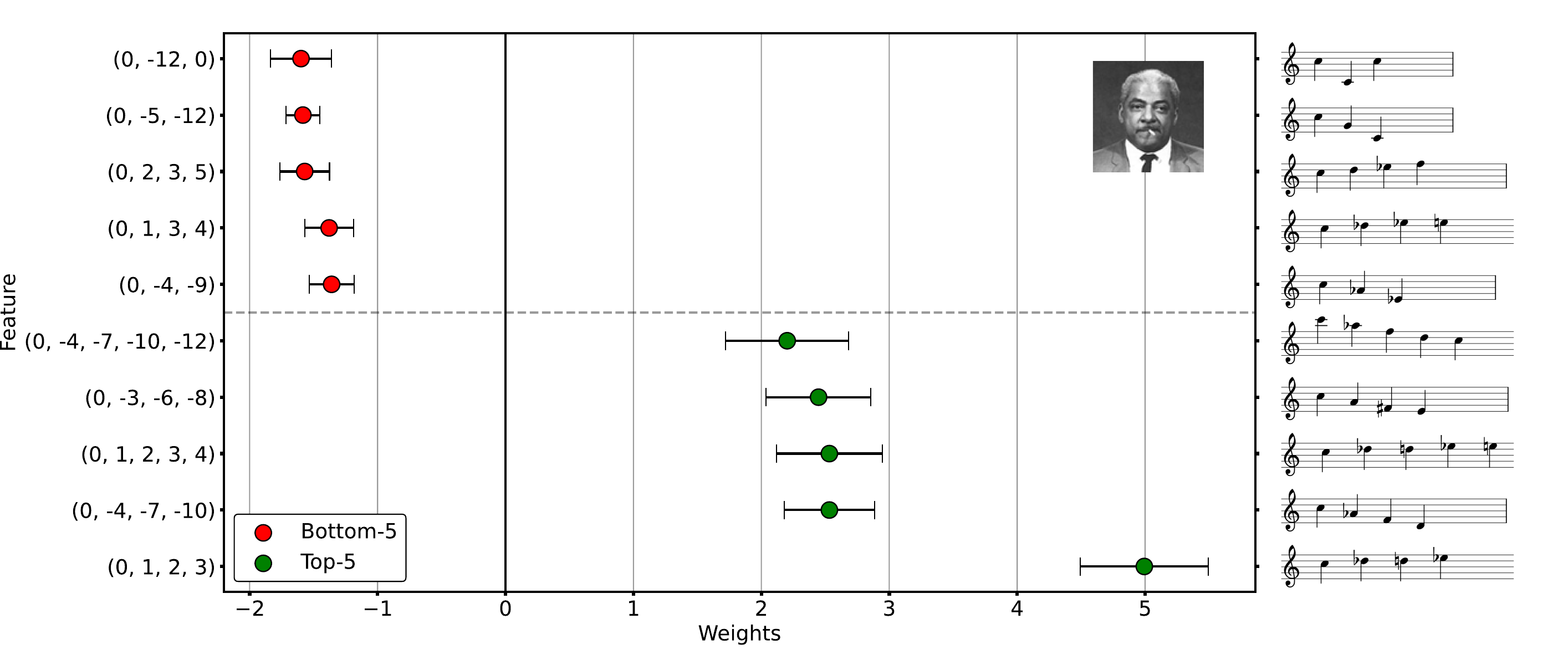}
  \caption{Predictive melody features, Teddy Wilson.}
\label{fig:rsi_sm_wilson_melody}
\end{figure}

\begin{figure}[h!]
  \centering
  \includegraphics[width=1\textwidth]{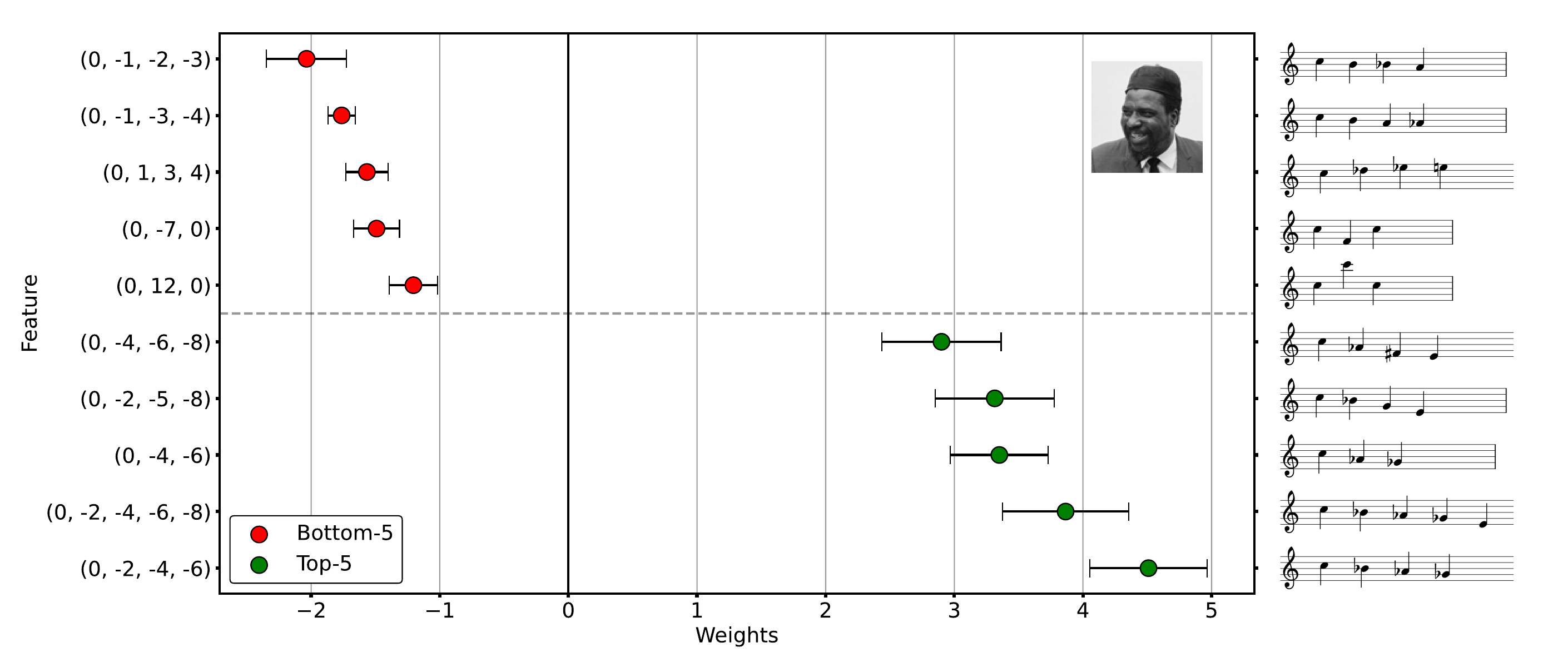}
  \caption{Predictive melody features, Thelonious Monk.}
\label{fig:rsi_sm_monk_melody}
\end{figure}

\begin{figure}[h!]
  \centering
  \includegraphics[width=1\textwidth]{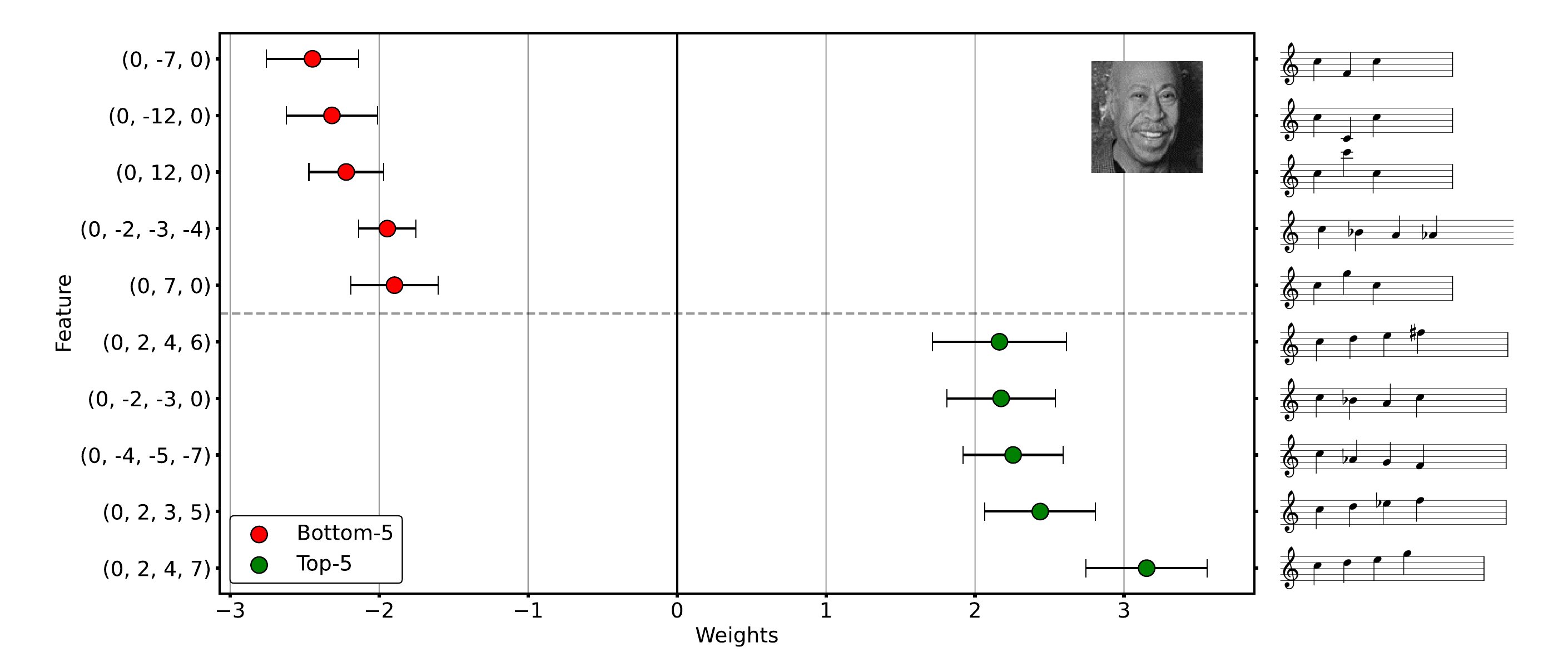}
  \caption{Predictive melody features, Tommy Flanagan.}
\label{fig:rsi_sm_flanagan_melody}
\end{figure}

\begin{figure}[h!]
  \centering
  \includegraphics[width=1\textwidth]{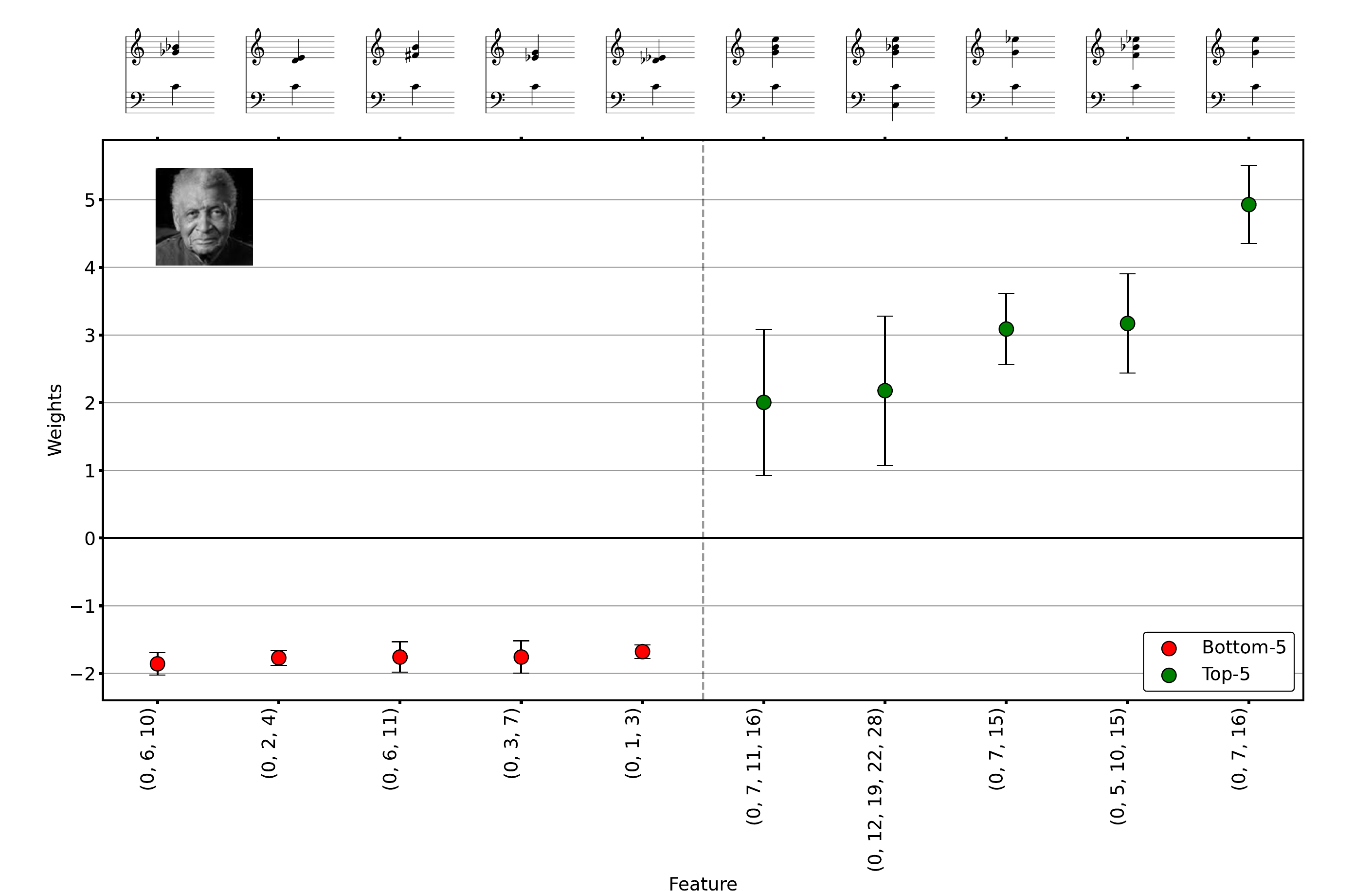}
  \caption{Predictive harmony features, Abdullah Ibrahim. Note that the direction of the axis in these figures is reversed compared to the previous figures, in order to accommodate the ``grand staff'' notation.}
\label{fig:rsi_sm_ibrahim_harmony}
\end{figure}

\begin{figure}[h!]
  \centering
  \includegraphics[width=1\textwidth]{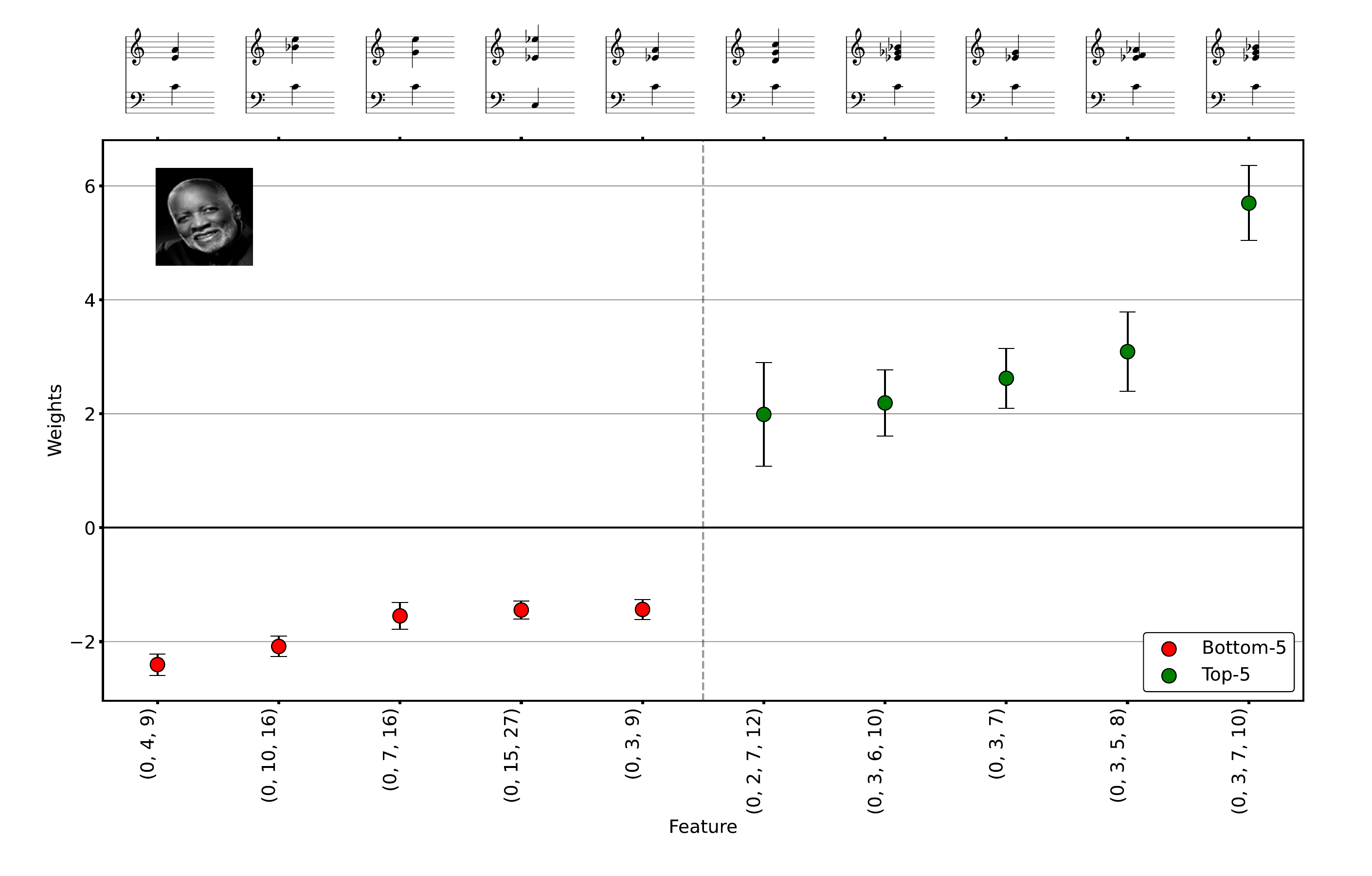}
  \caption{Predictive harmony features, Ahmad Jamal.}
\label{fig:rsi_sm_jamal_harmony}
\end{figure}

\begin{figure}[h!]
  \centering
  \includegraphics[width=1\textwidth]{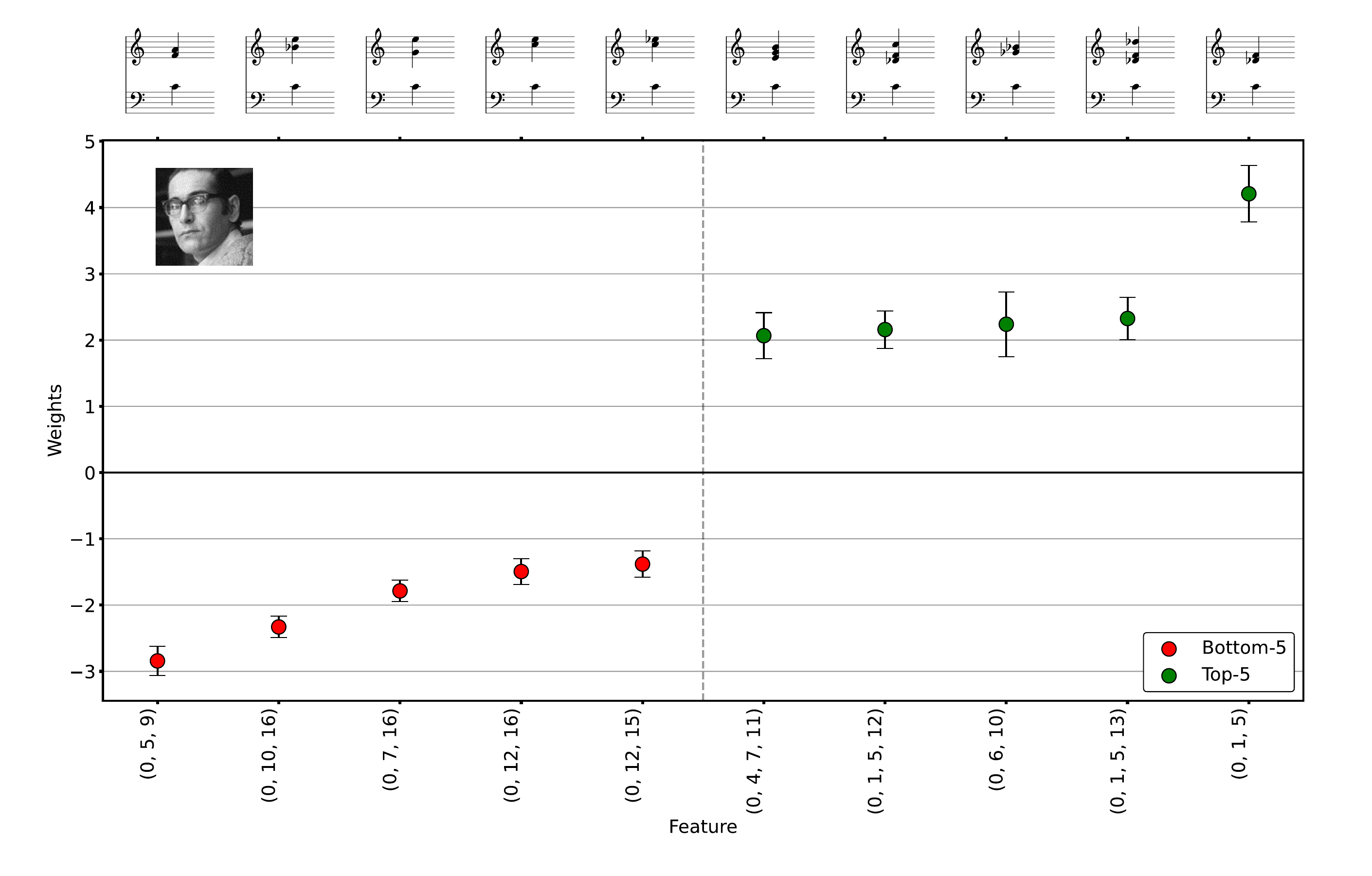}
  \caption{Predictive harmony features, Bill Evans.}
\label{fig:rsi_sm_evans_harmony}
\end{figure}

\begin{figure}[h!]
  \centering
  \includegraphics[width=1\textwidth]{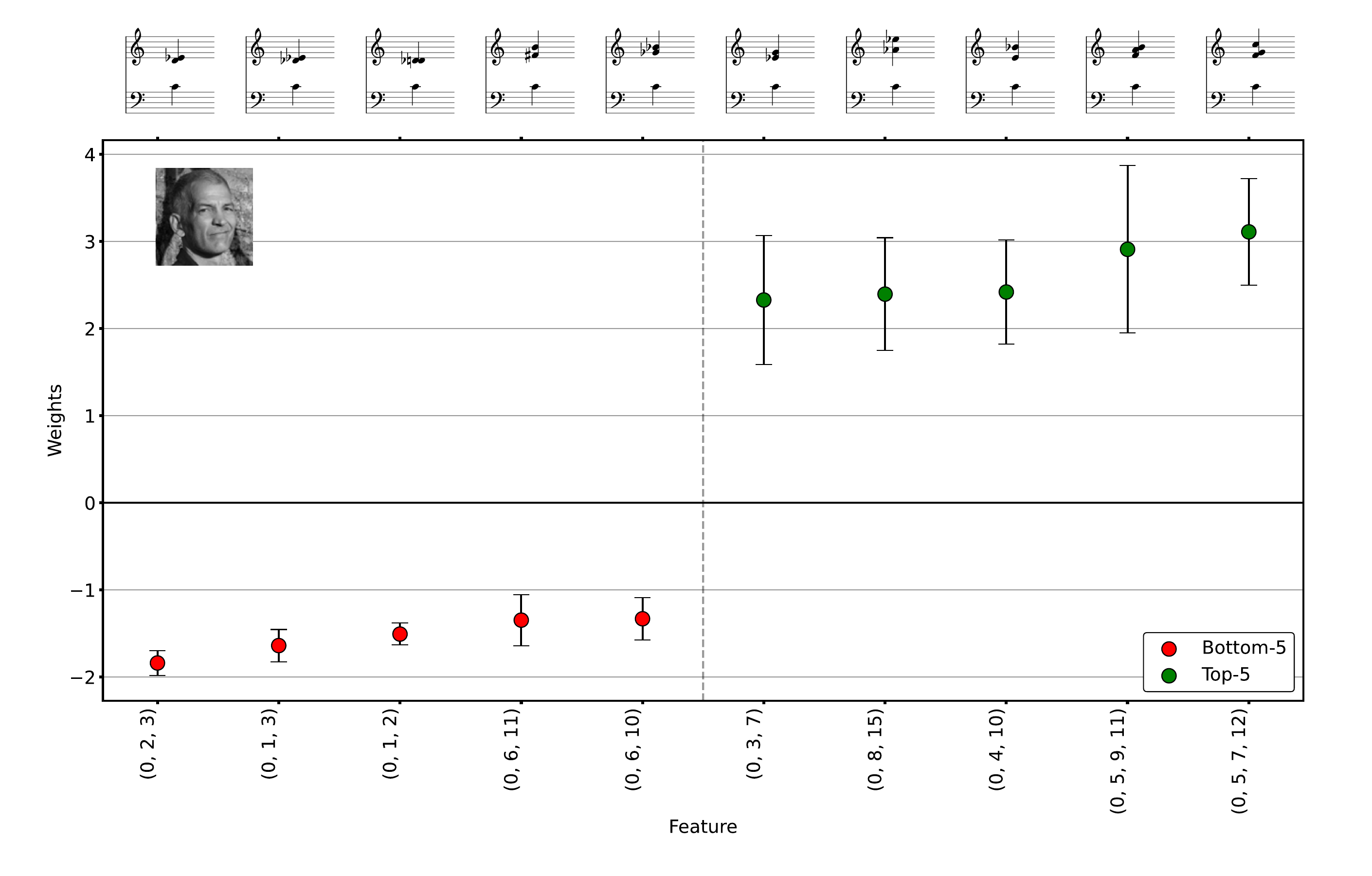}
  \caption{Predictive harmony features, Brad Mehldau.}
\label{fig:rsi_sm_mehldau_harmony}
\end{figure}

\begin{figure}[h!]
  \centering
  \includegraphics[width=1\textwidth]{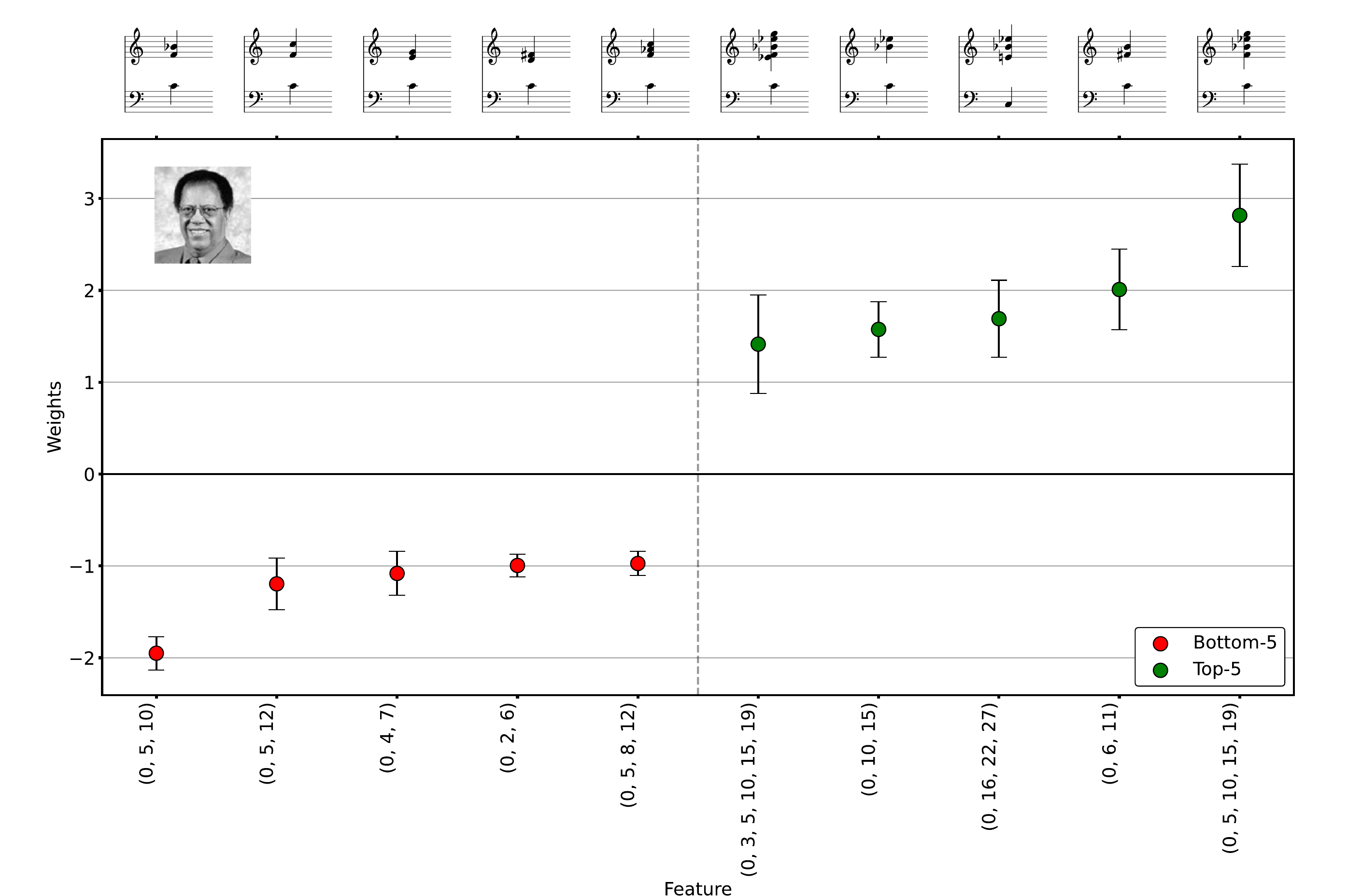}
  \caption{Predictive harmony features, Cedar Walton.}
\label{fig:rsi_sm_walton_harmony}
\end{figure}

\begin{figure}[h!]
  \centering
  \includegraphics[width=1\textwidth]{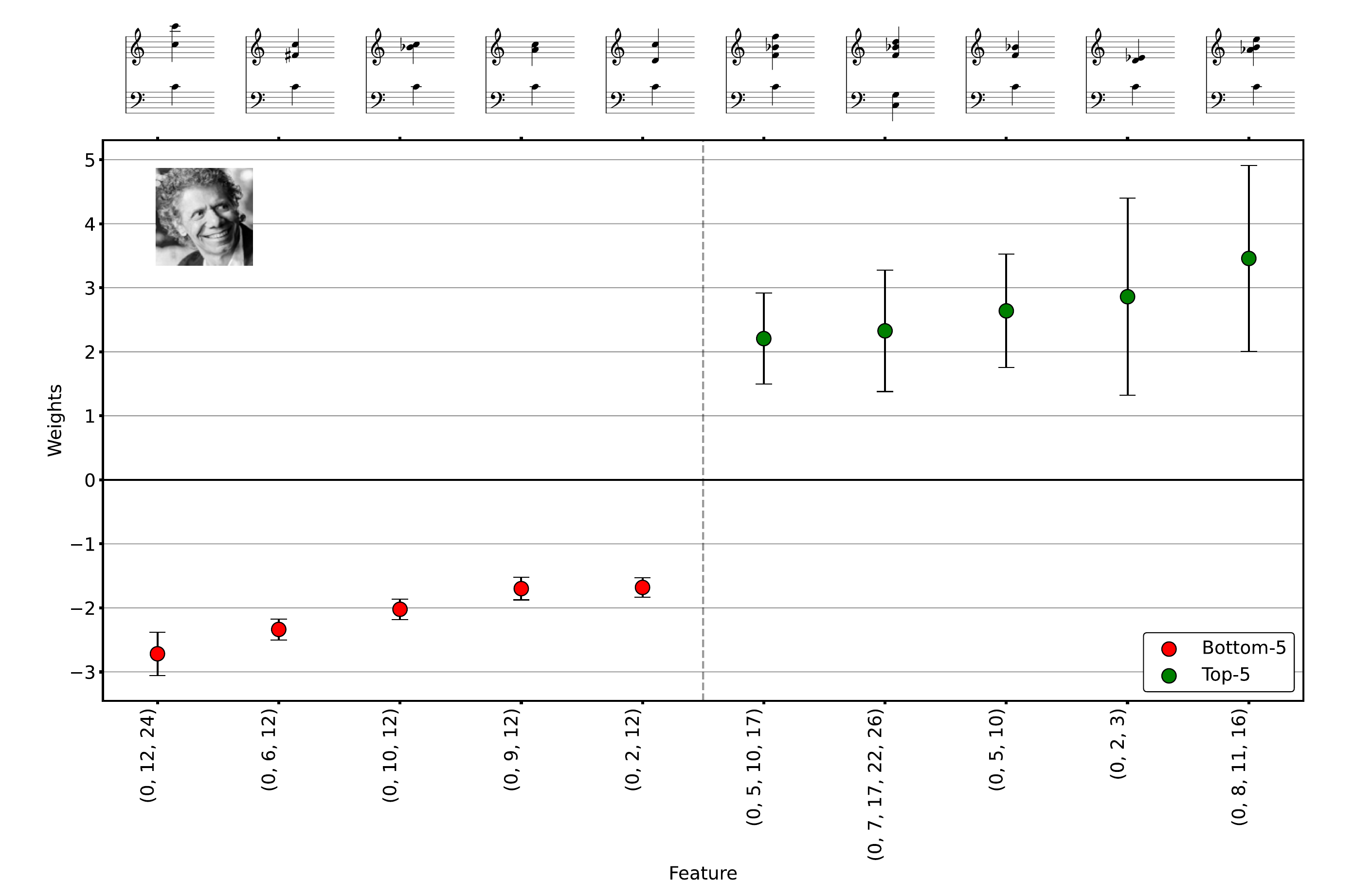}
  \caption{Predictive harmony features, Chick Corea.}
\label{fig:rsi_sm_corea_harmony}
\end{figure}

\begin{figure}[h!]
  \centering
  \includegraphics[width=1\textwidth]{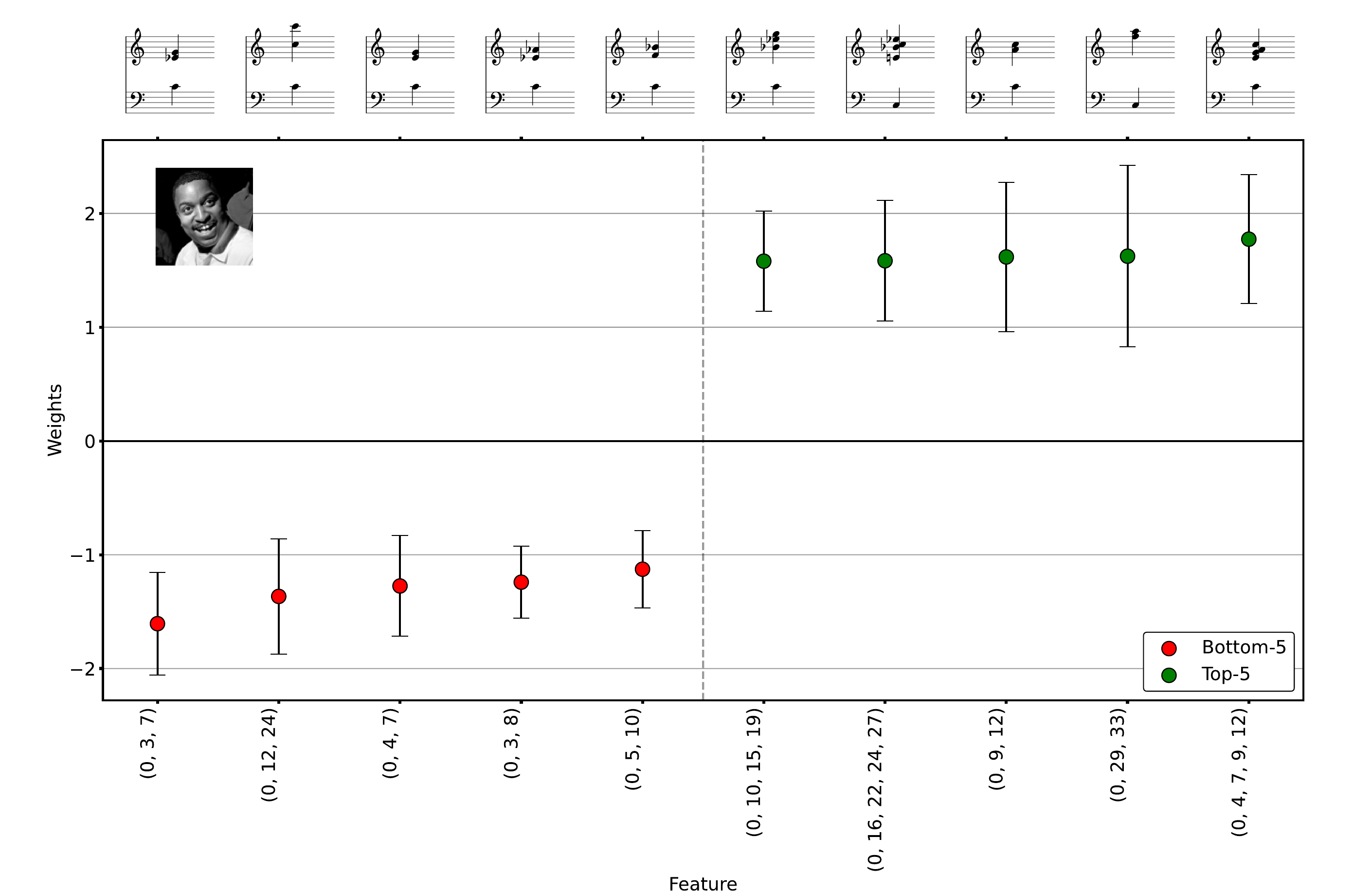}
  \caption{Predictive harmony features, Gene Harris.}
\label{fig:rsi_sm_harris_harmony}
\end{figure}

\begin{figure}[h!]
  \centering
  \includegraphics[width=1\textwidth]{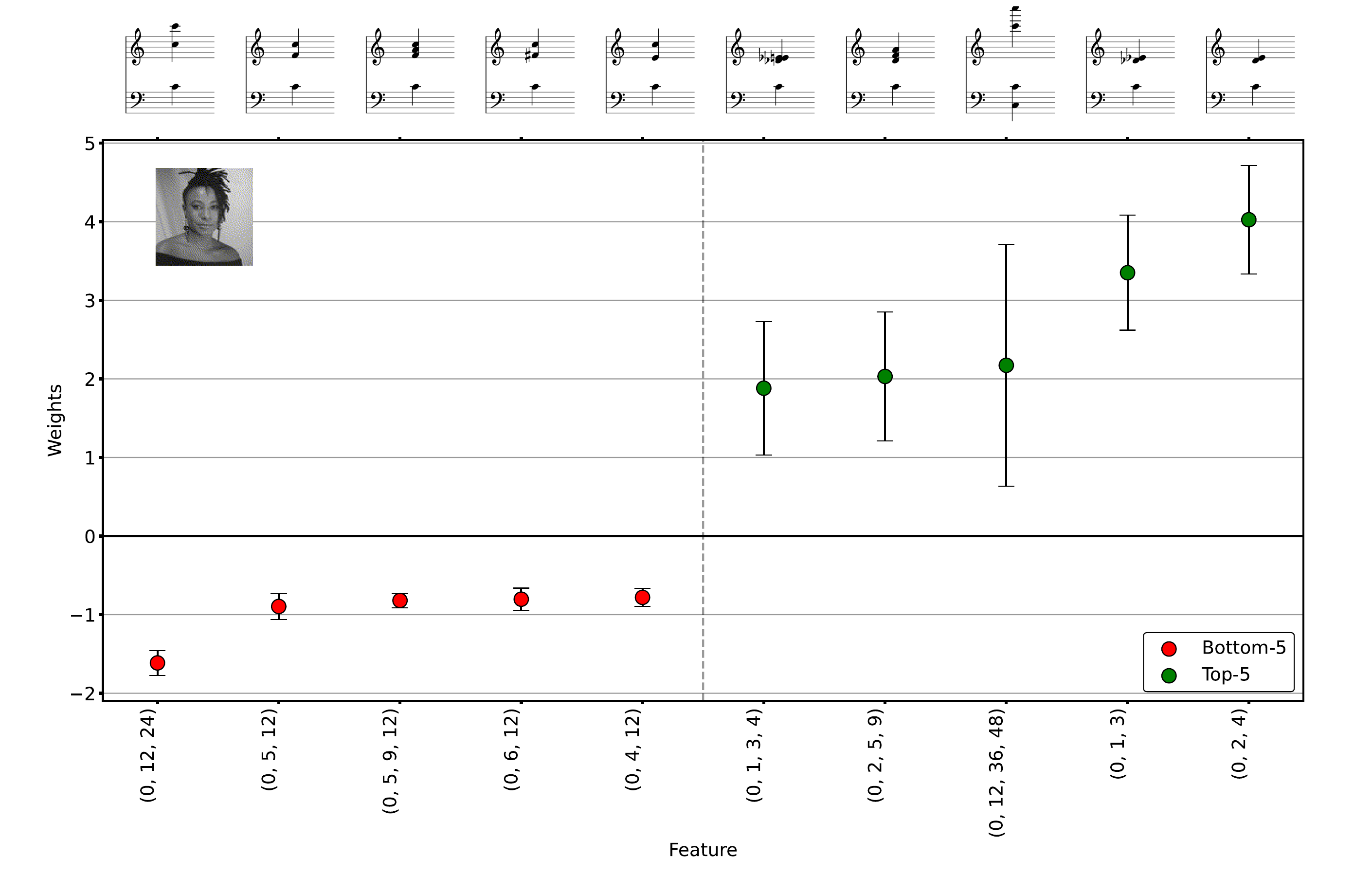}
  \caption{Predictive harmony features, Geri Allen.}
\label{fig:rsi_sm_allen_harmony}
\end{figure}

\begin{figure}[h!]
  \centering
  \includegraphics[width=1\textwidth]{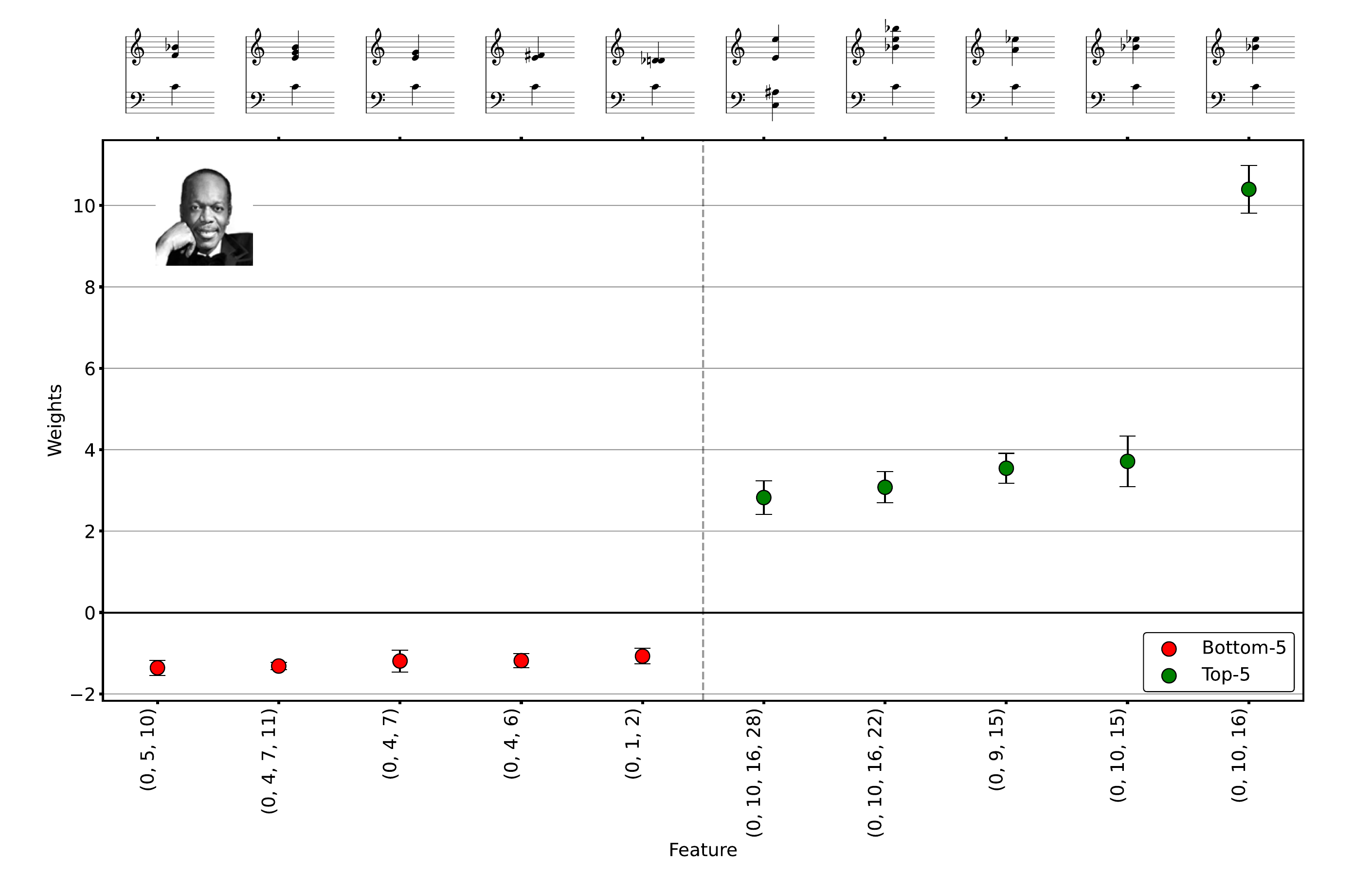}
  \caption{Predictive harmony features, Hank Jones.}
\label{fig:rsi_sm_jones_harmony}
\end{figure}

\begin{figure}[h!]
  \centering
  \includegraphics[width=1\textwidth]{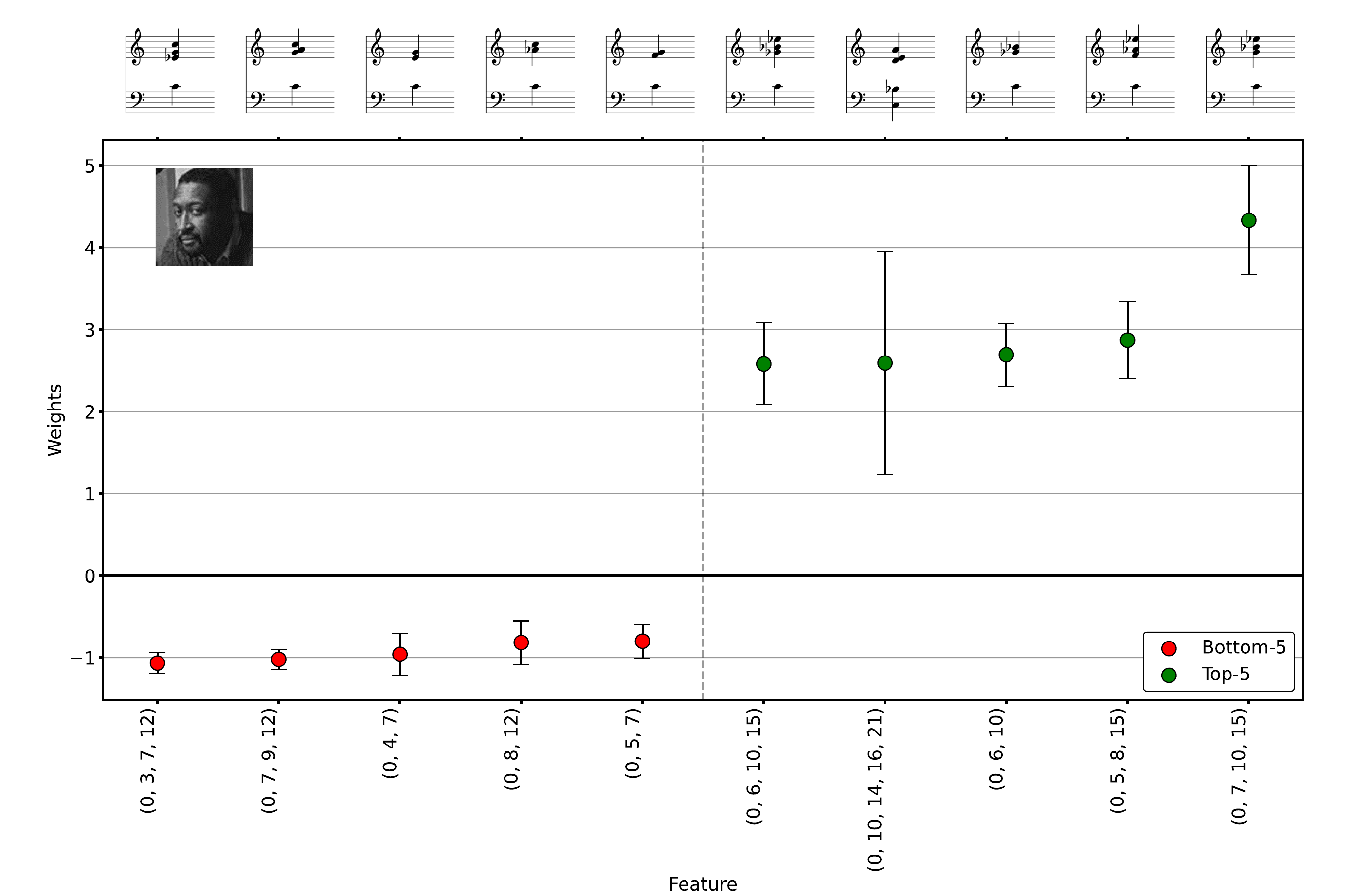}
  \caption{Predictive harmony features, John Hicks.}
\label{fig:rsi_sm_hicks_harmony}
\end{figure}

\begin{figure}[h!]
  \centering
  \includegraphics[width=1\textwidth]{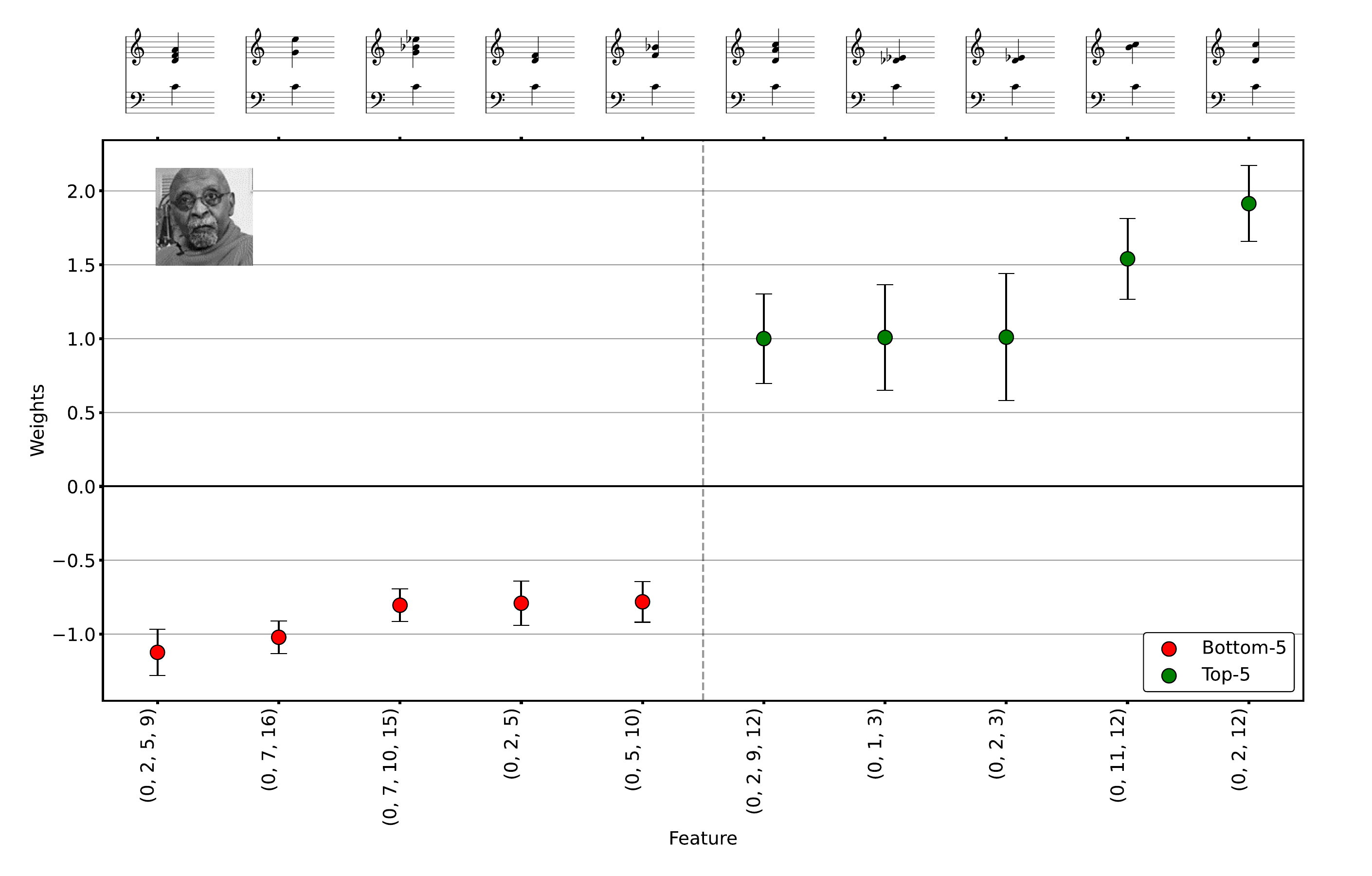}
  \caption{Predictive harmony features, Junior Mance.}
\label{fig:rsi_sm_mance_harmony}
\end{figure}

\begin{figure}[h!]
  \centering
  \includegraphics[width=1\textwidth]{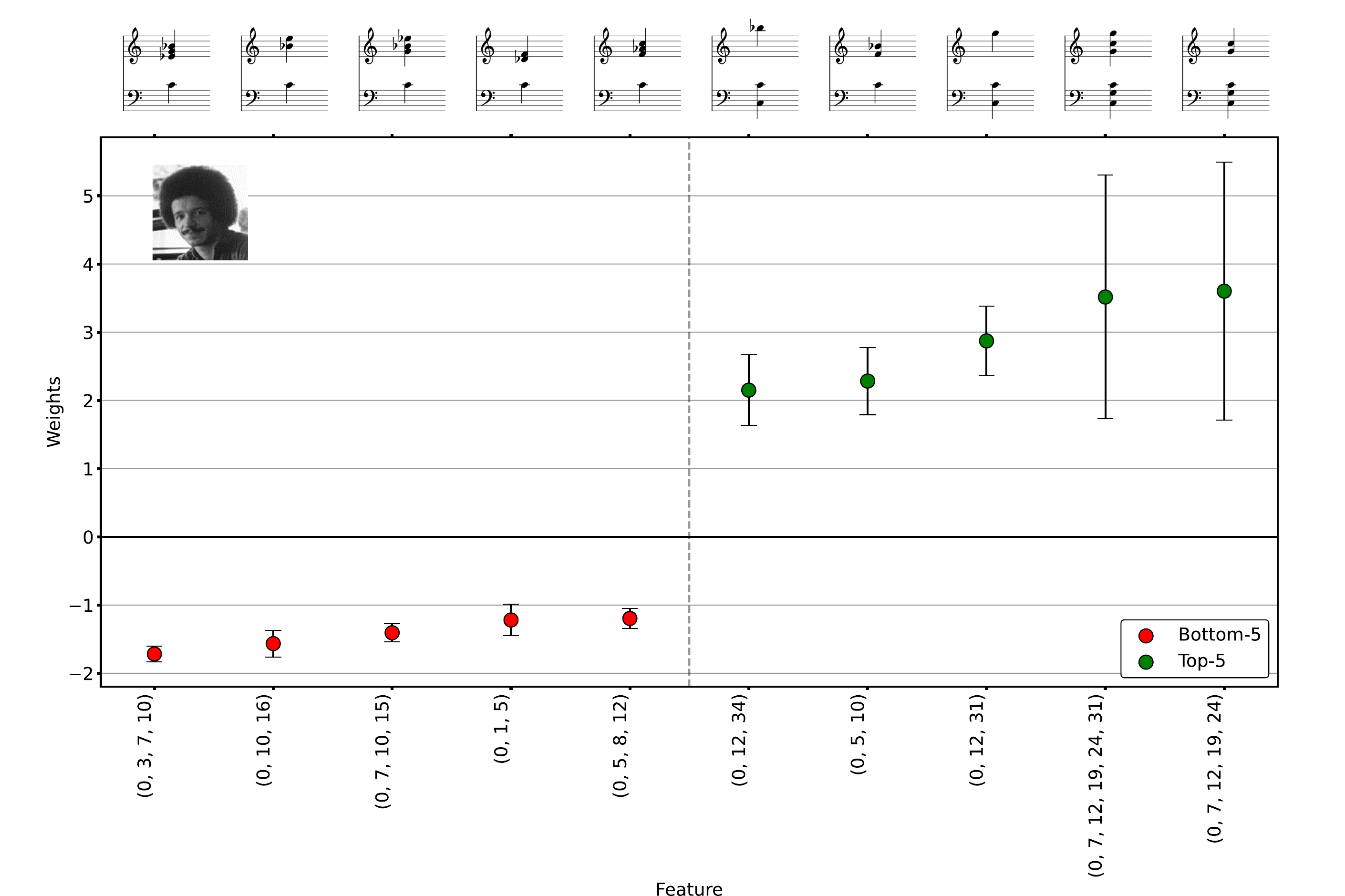}
  \caption{Predictive harmony features, Keith Jarrett.}
\label{fig:rsi_sm_jarrett_harmony}
\end{figure}

\begin{figure}[h!]
  \centering
  \includegraphics[width=1\textwidth]{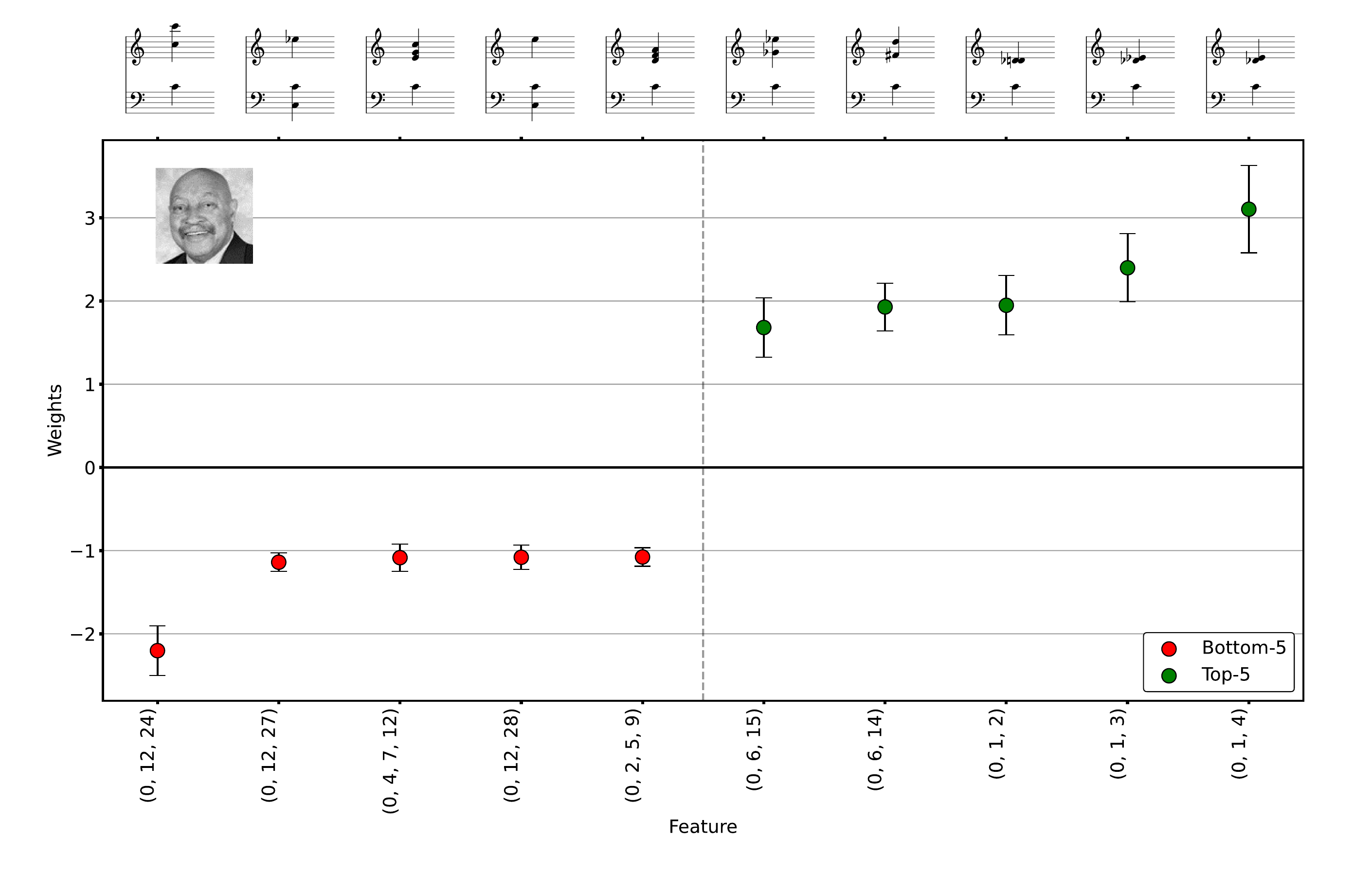}
  \caption{Predictive harmony features, Kenny Barron.}
\label{fig:rsi_sm_barron_harmony}
\end{figure}

\begin{figure}[h!]
  \centering
  \includegraphics[width=1\textwidth]{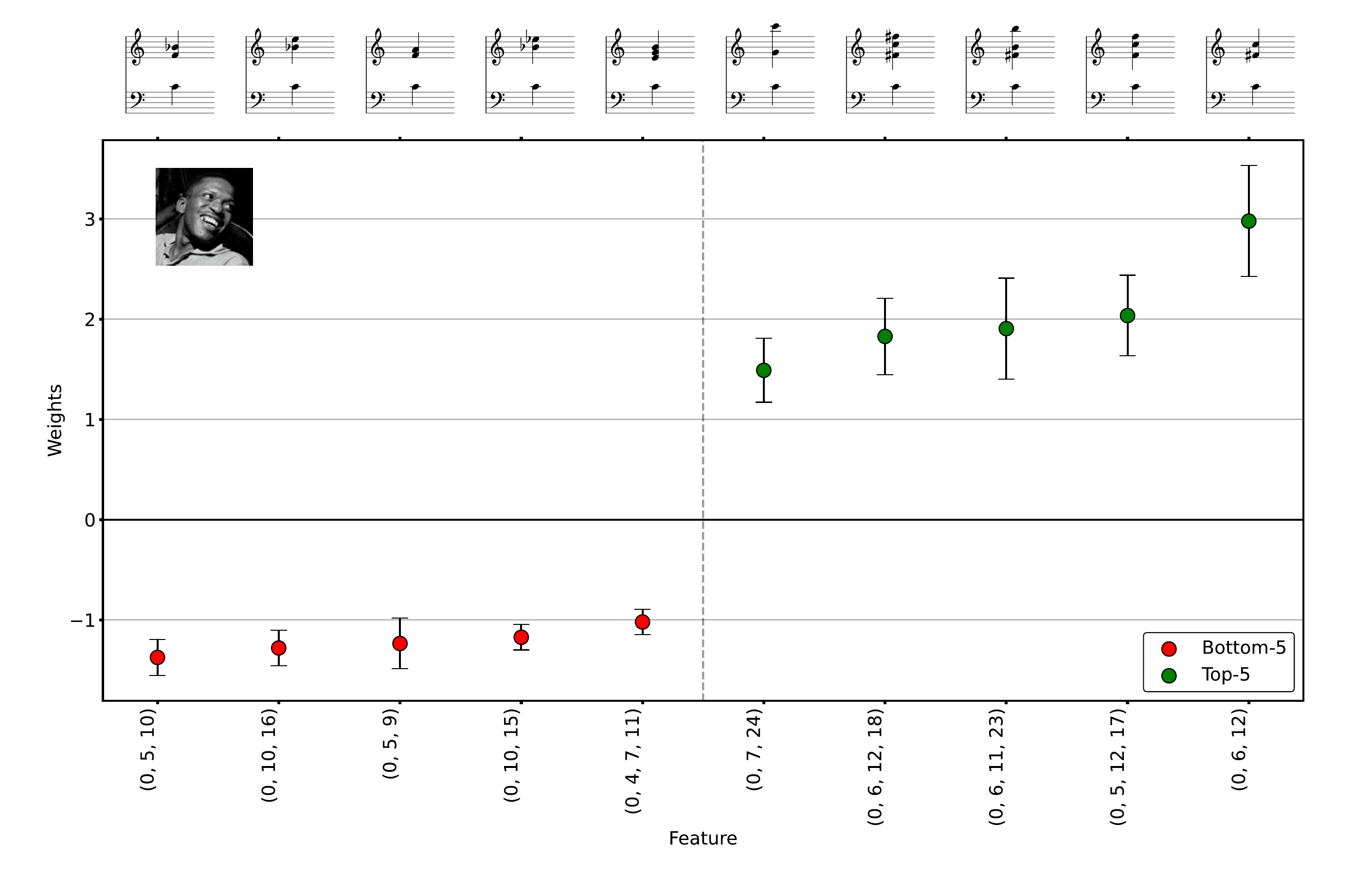}
  \caption{Predictive harmony features, Kenny Drew.}
\label{fig:rsi_sm_drew_harmony}
\end{figure}

\begin{figure}[h!]
  \centering
  \includegraphics[width=1\textwidth]{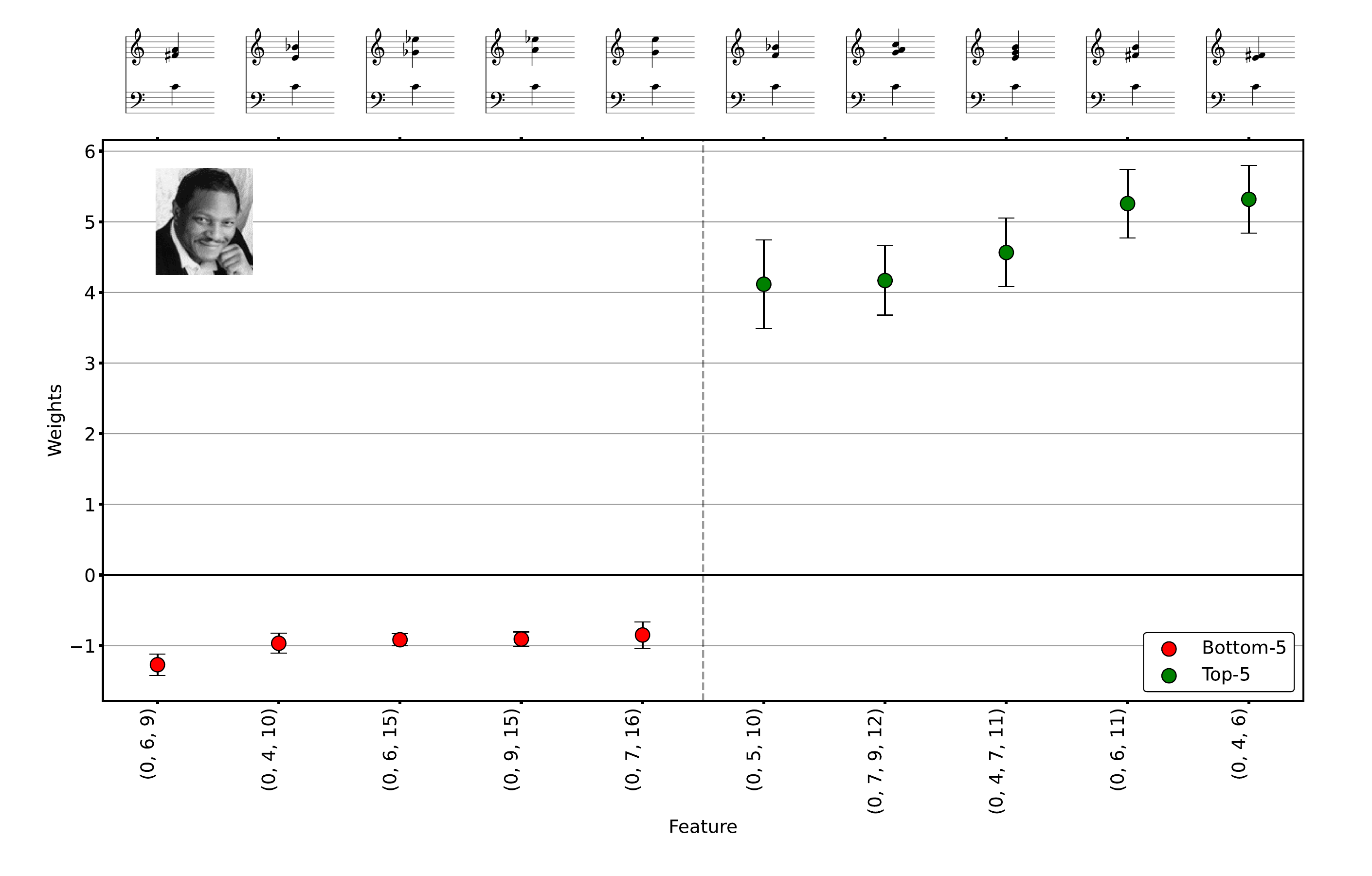}
  \caption{Predictive harmony features, McCoy Tyner.}
\label{fig:rsi_sm_tyner_harmony}
\end{figure}

\begin{figure}[h!]
  \centering
  \includegraphics[width=1\textwidth]{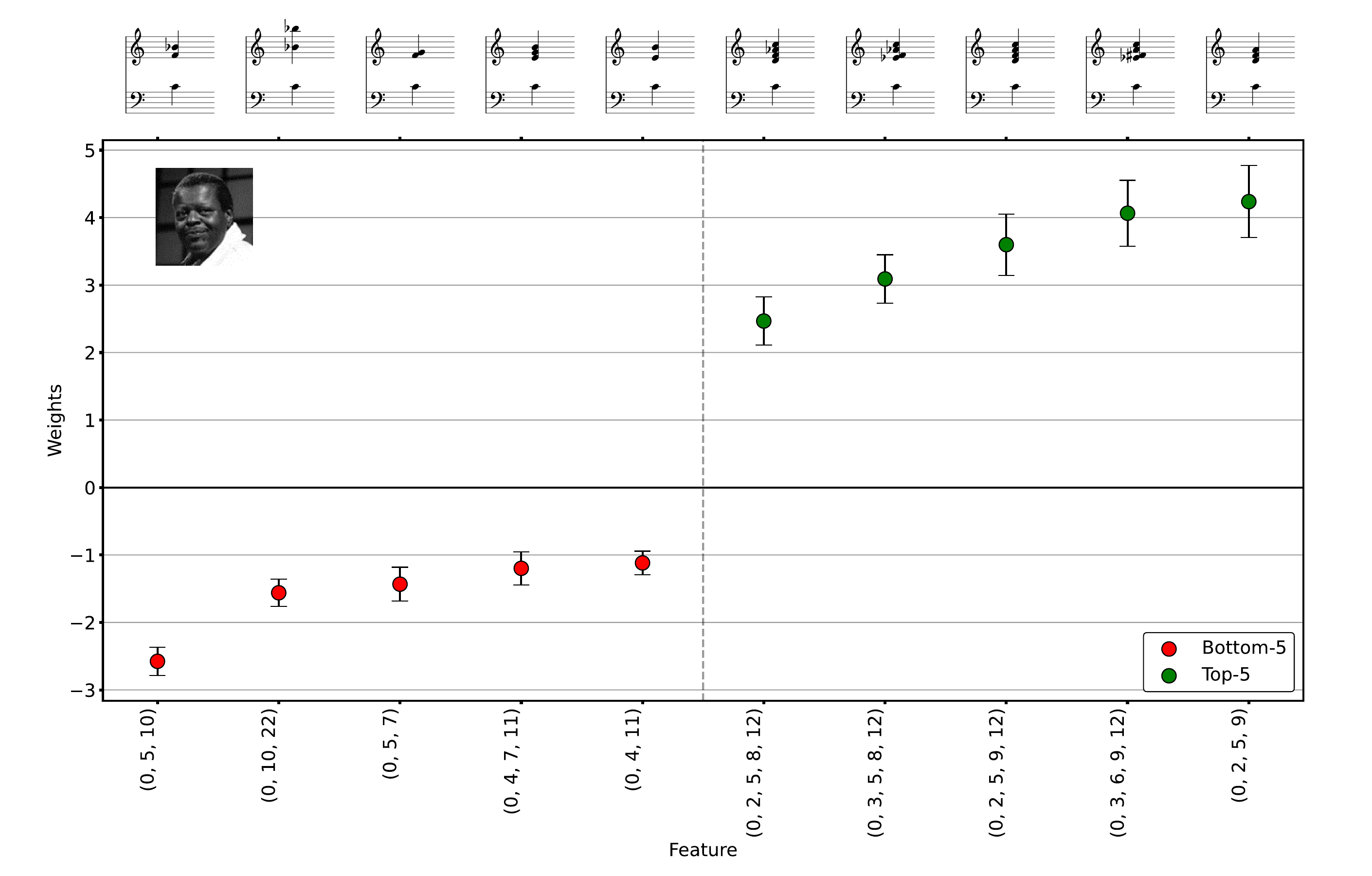}
  \caption{Predictive harmony features, Oscar Peterson.}
\label{fig:rsi_sm_peterson_harmony}
\end{figure}

\begin{figure}[h!]
  \centering
  \includegraphics[width=1\textwidth]{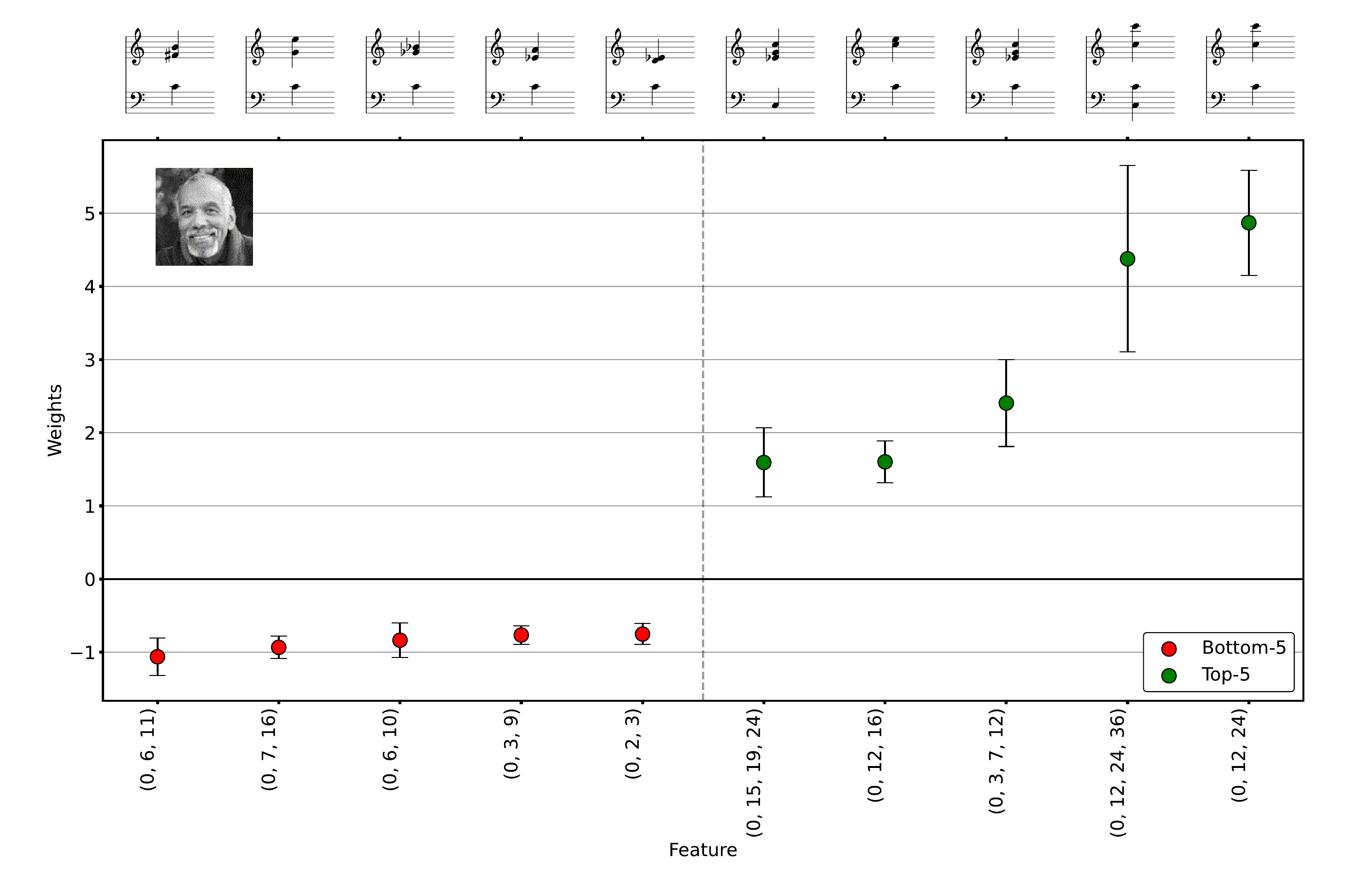}
  \caption{Predictive harmony features, Stanley Cowell.}
\label{fig:rsi_sm_cowell_harmony}
\end{figure}

\begin{figure}[h!]
  \centering
  \includegraphics[width=1\textwidth]{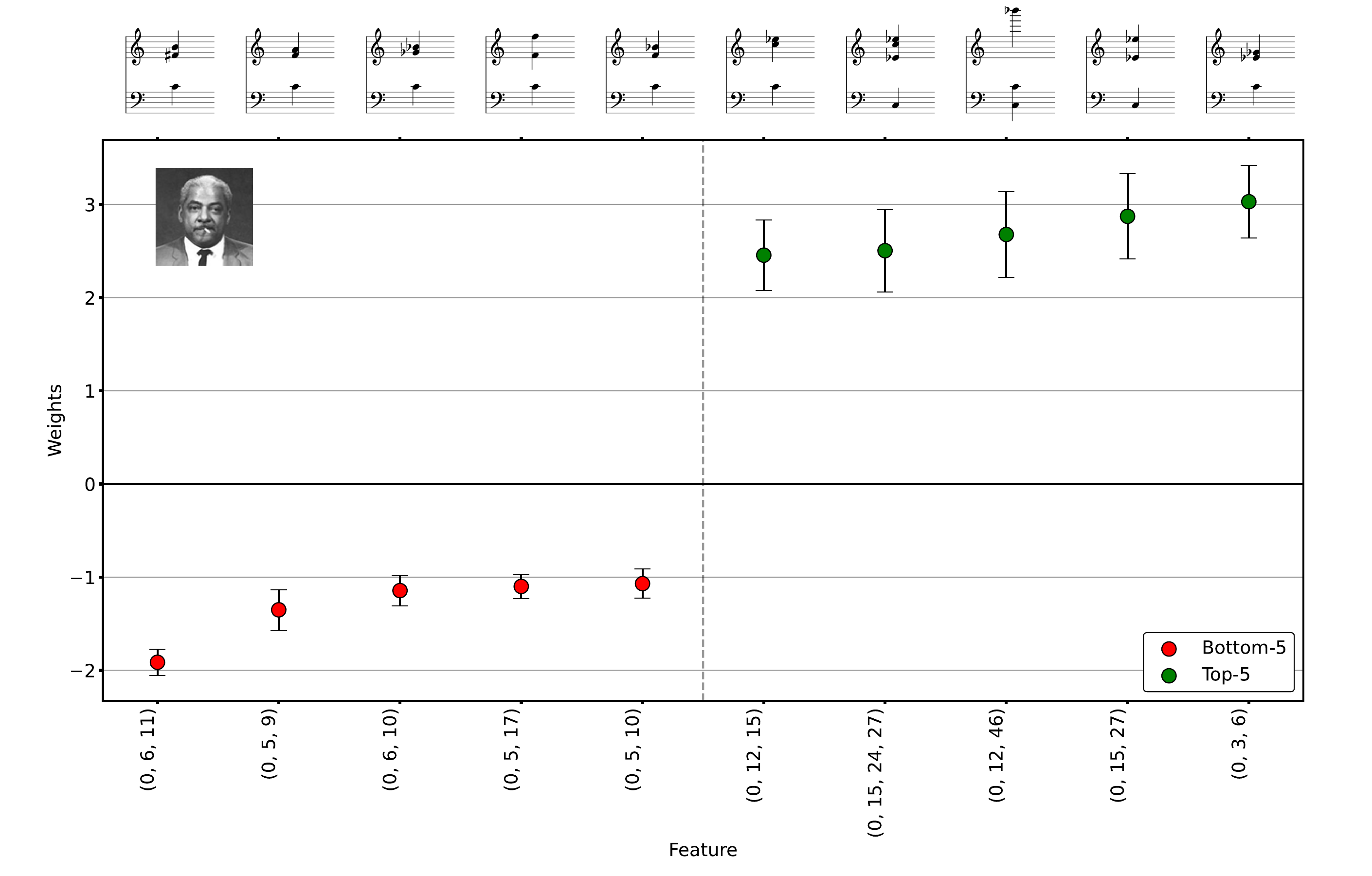}
  \caption{Predictive harmony features, Teddy Wilson.}
\label{fig:rsi_sm_wilson_harmony}
\end{figure}

\begin{figure}[h!]
  \centering
  \includegraphics[width=1\textwidth]{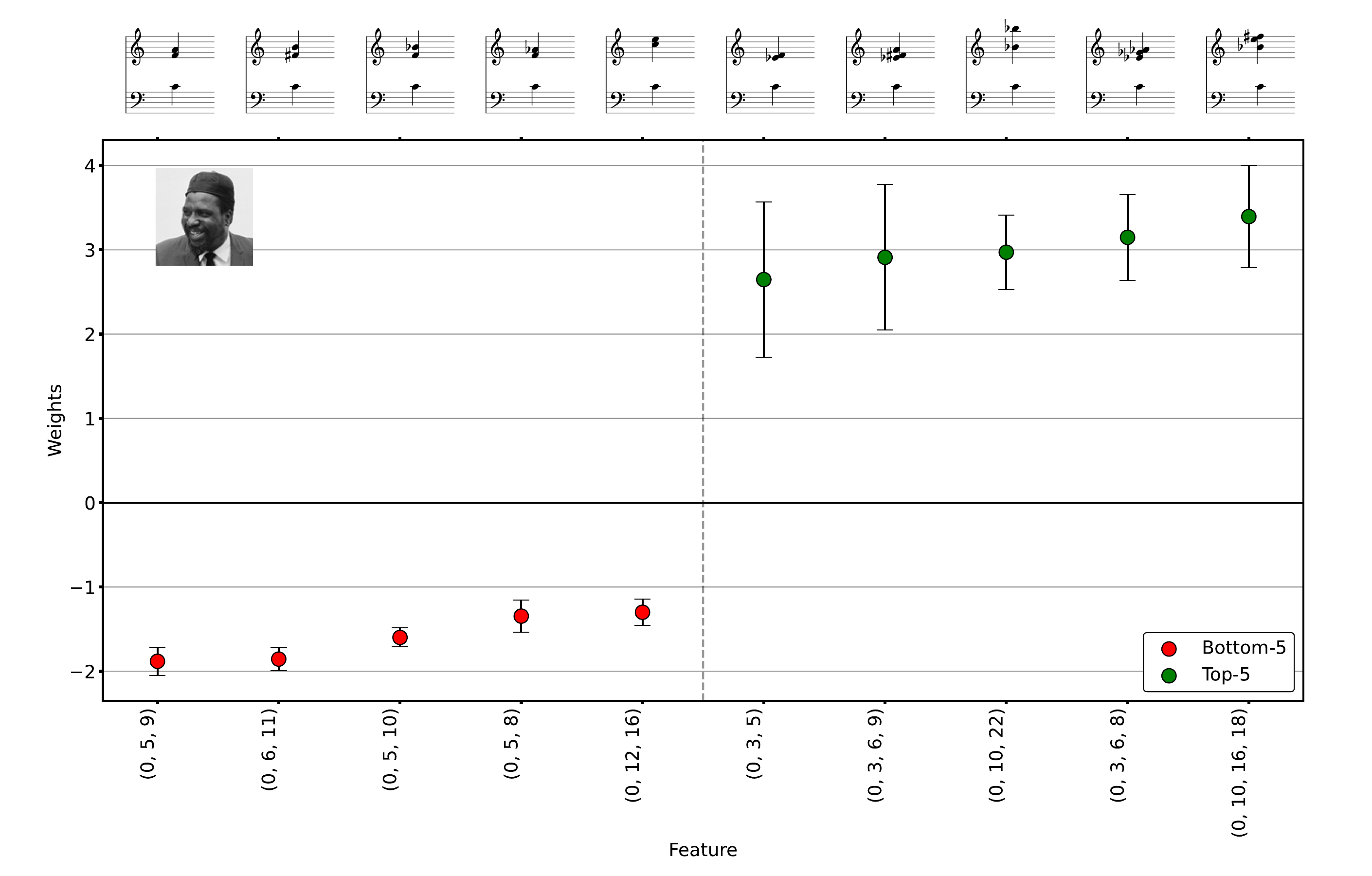}
  \caption{Predictive harmony features, Thelonious Monk.}
\label{fig:rsi_sm_monk_harmony}
\end{figure}

\begin{figure}[h!]
  \centering
  \includegraphics[width=1\textwidth]{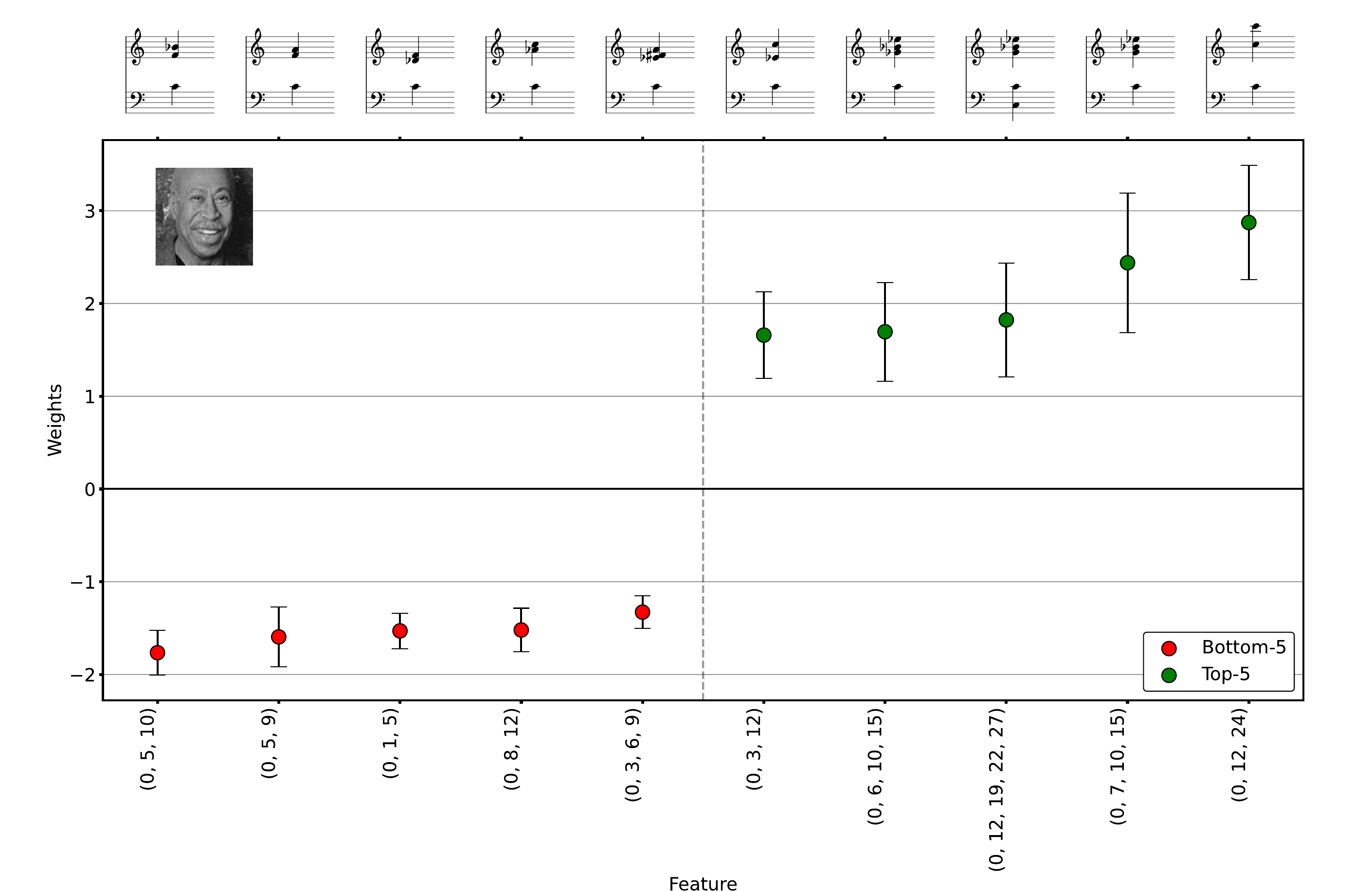}
  \caption{Predictive harmony features, Tommy Flanagan.}
\label{fig:rsi_sm_flanagan_harmony}
\end{figure}

\begin{figure}[h!]
  \centering
  \includegraphics[width=1\textwidth]{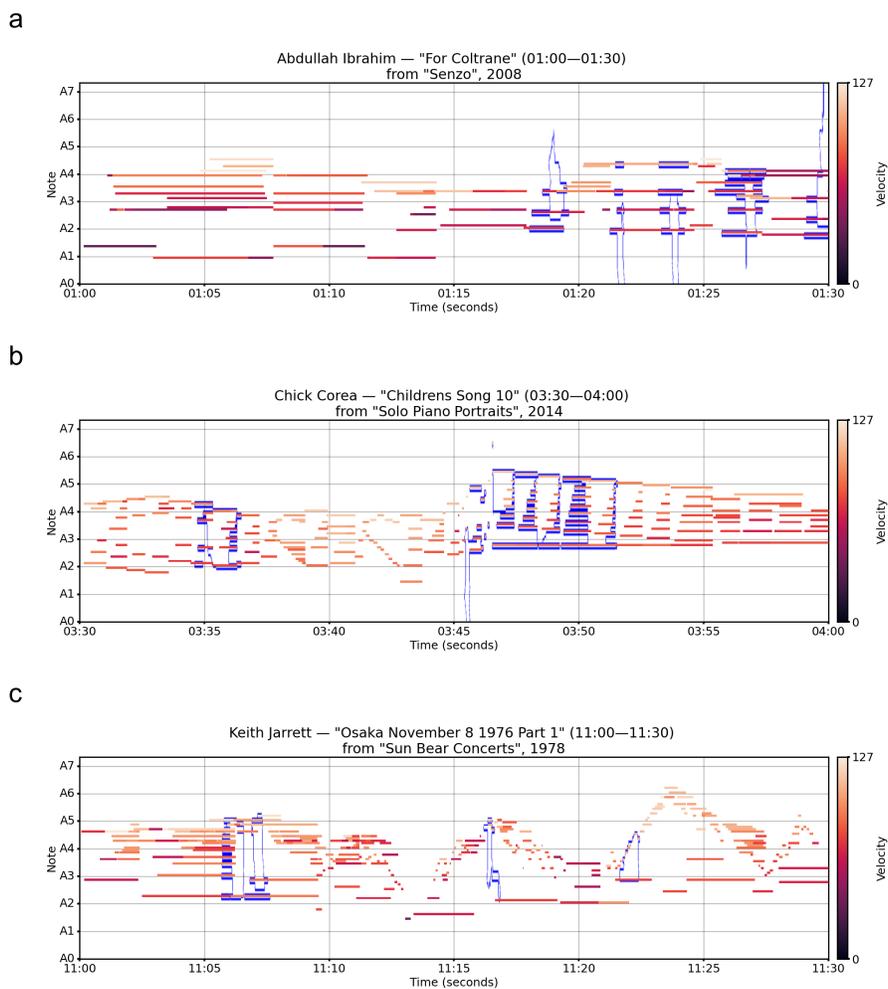}
  \caption{Masked piano rolls generated with LIME. Each panel shows a single 30-second clip from a transcription of a performance by (a) Abdullah Ibrahim, (b) Chick Corea, and (c) Keith Jarrett, taken from the held-out test split. Highlighted in blue are the top five areas of the piano roll that positively contribute to predicting the target label according to LIME.}
\label{fig:rsi_sm_lime_plots}
\end{figure}

\begin{figure}[h!]
  \centering
  \includegraphics[width=1\textwidth]{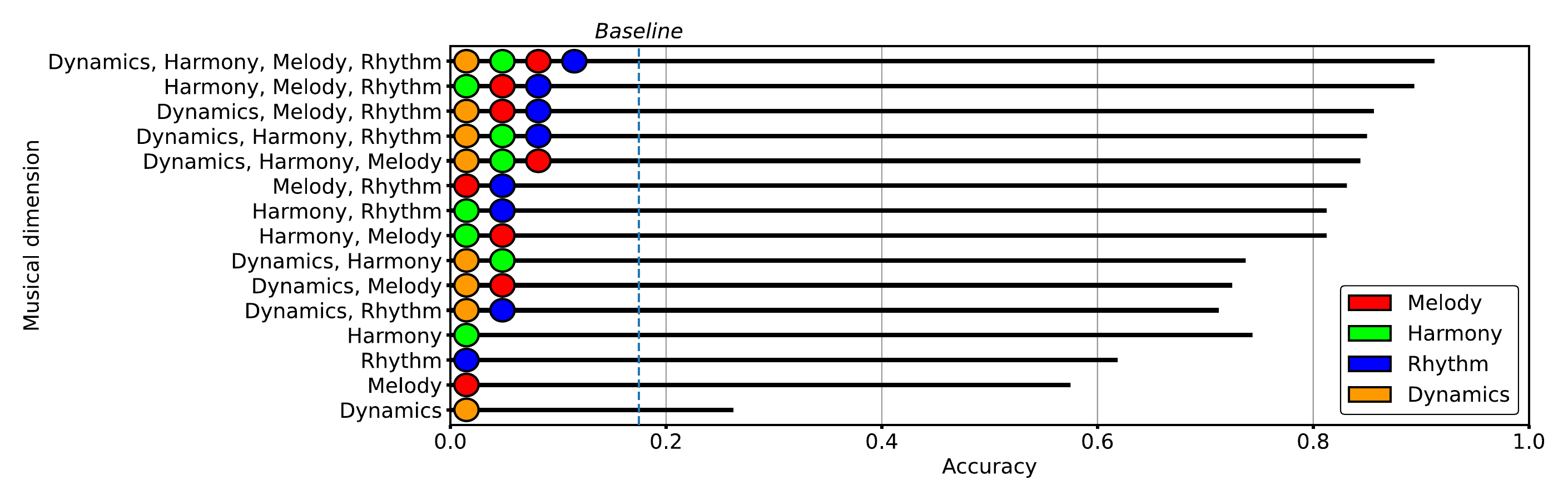}
  \caption{Multi-input model evaluation. The ``lollipop'' plot shows the accuracy obtained from predicting using all combinations of sub-networks, with the colored ``heads'' of each lollipop indicating the particular networks used to make the predictions. The dotted ``baseline'' score is the accuracy obtained from a model that simply predicts the majority class for every recording.}
\label{fig:rsi_sm_factorised_lollipop}
\end{figure}

\begin{figure}[ht!]
  \centering
  \includegraphics[width=1\textwidth]{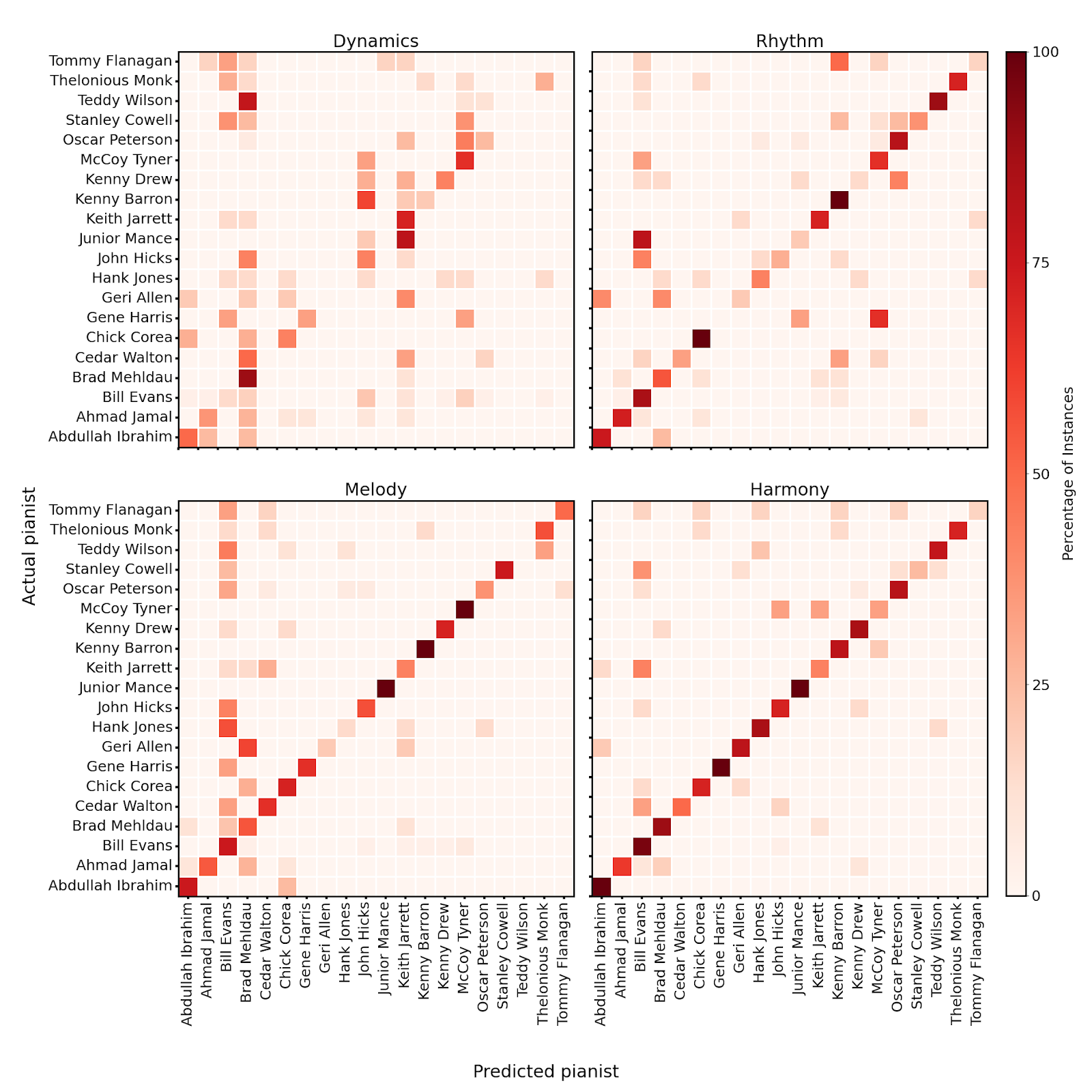}
  \caption{Class accuracy across musical dimensions. Each facet contains a heatmap showing the probability that, when given a recording and the output from a sub-network, the model will identify a particular pianist. The proportion of hits is shown for each facet along the diagonal; all other values are misses. Lighter and darker colours indicate lower and higher predictive probability, respectively.}
\label{fig:rsi_sm_factorised_heatmap}
\end{figure}

\begin{figure}[h!]
  \centering
  \includegraphics[width=1\textwidth]{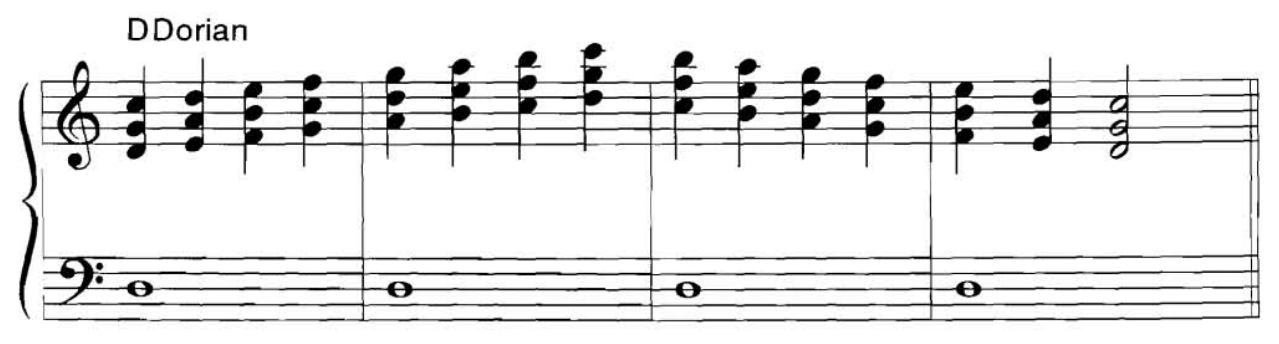}
  \caption{CAV examples. The musical notation shows an example exercise taken from the ``Modal Fourthy Voicings'' chapter in \citet[p. 52]{Haerle1994}.}
\label{fig:rsi_sm_haerle_example}
\end{figure}

\begin{figure}[h!]
  \centering
  \includegraphics[width=1\textwidth]{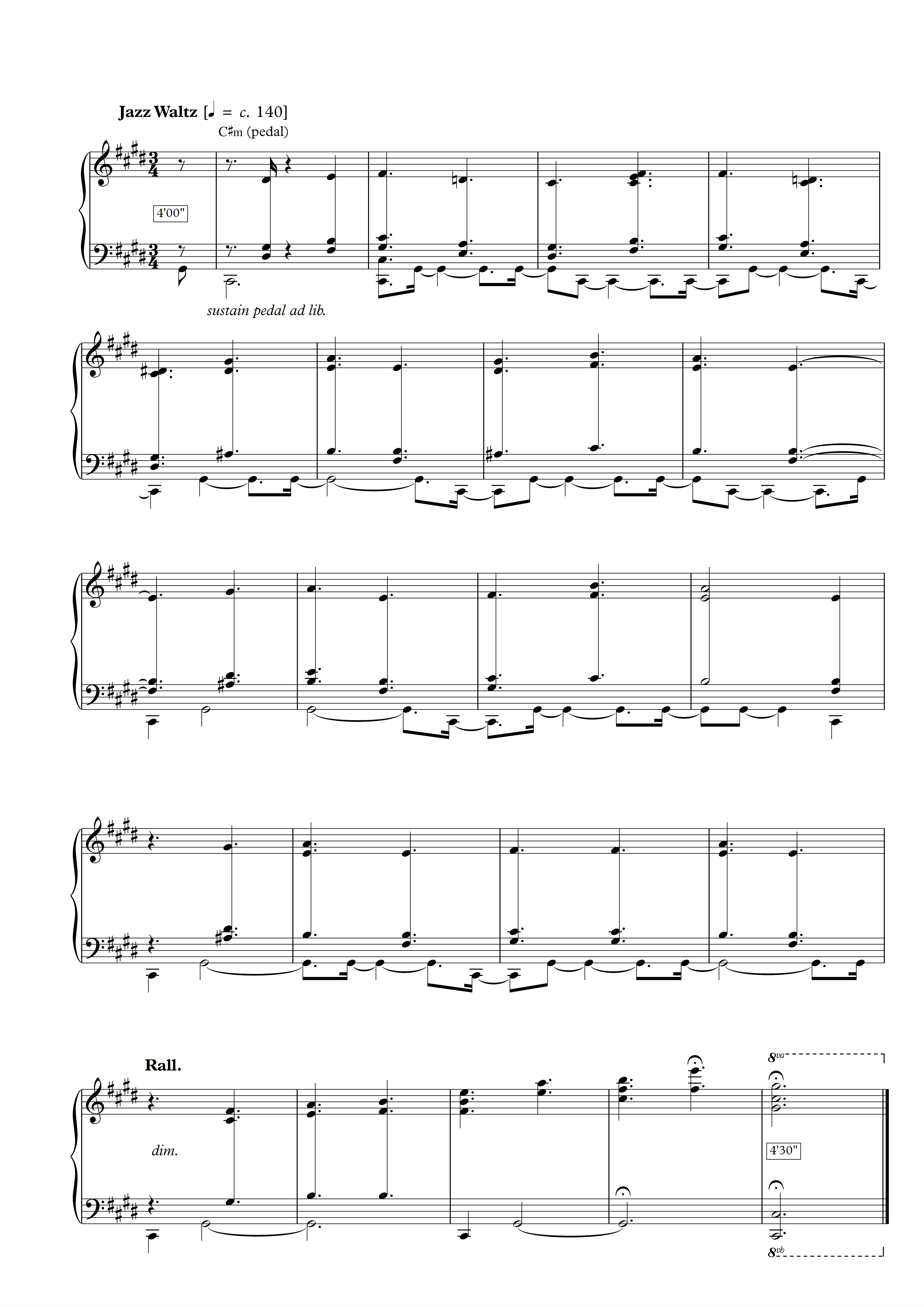}
  \caption{``Tivoli'' (1991) transcription. The musical notation shows a transcription of the final 30 seconds from an unaccompanied performance by McCoy Tyner (``Tivoli'' from ``Soliloquy'', recorded in 1991, originally composed by Dexter Gordon). Transcription created by the first author.}
\label{fig:rsi_sm_tivoli_transcription}
\end{figure}

\begin{figure}[]
  \centering
  \includegraphics[width=1\textwidth]{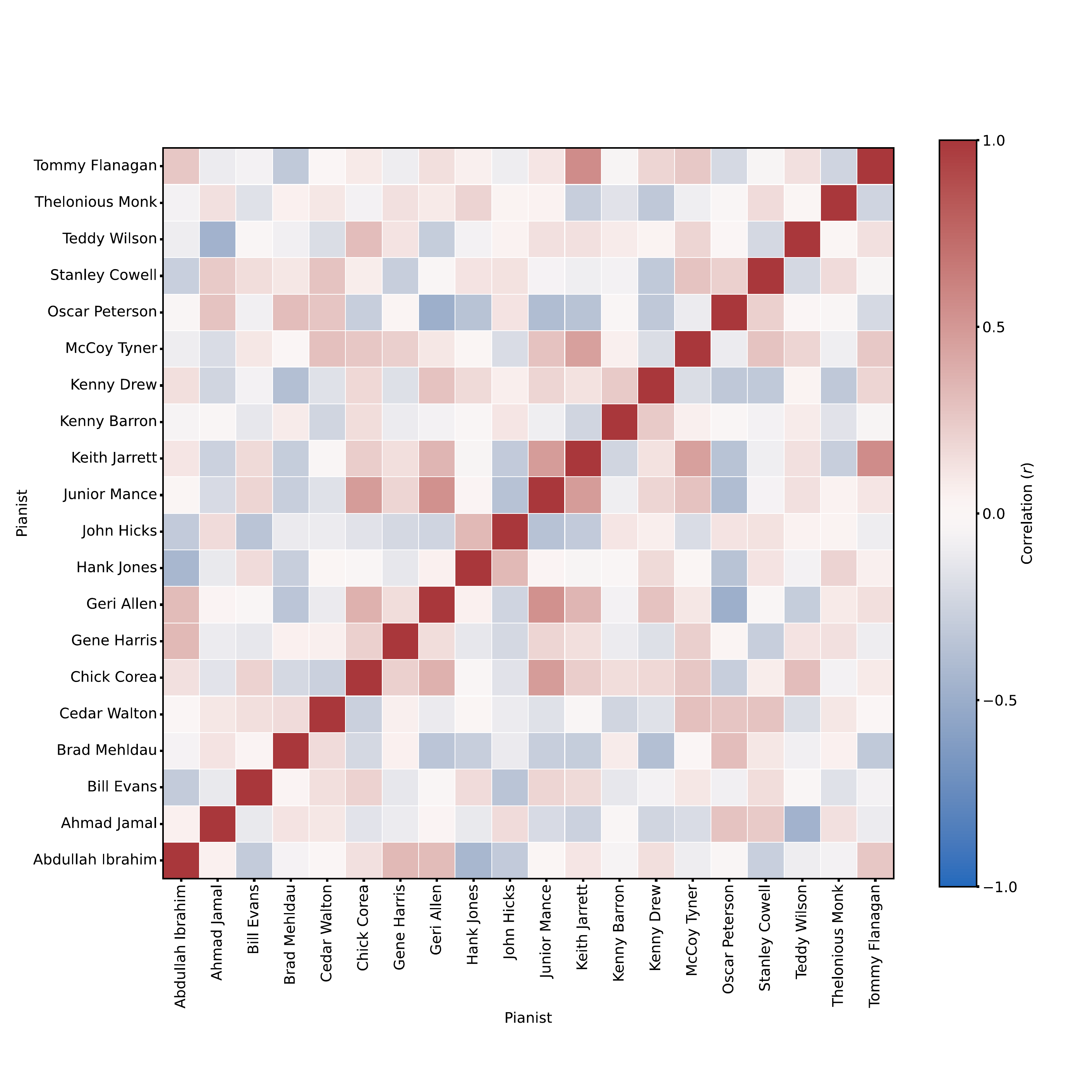}
  \caption{Pairwise correlations between pianists. The colour of each cell shows the strength of the linear relationships (Pearson's $r$) between sign-count ratios obtained between performers along the $x$- and $y$-axis. Darker reds and blues indicate stronger positive and negative associations, respectively. Sign-count ratios are collected for all concepts ($N = 20$) and across all iterations ($N = 10$) of the testing process, such that the total number of values considered for each pianist in every correlation is 200.}
\label{fig:rsi_sm_pianist_correlations}
\end{figure}

\begin{figure}[]
  \centering
  \includegraphics[width=1\textwidth]{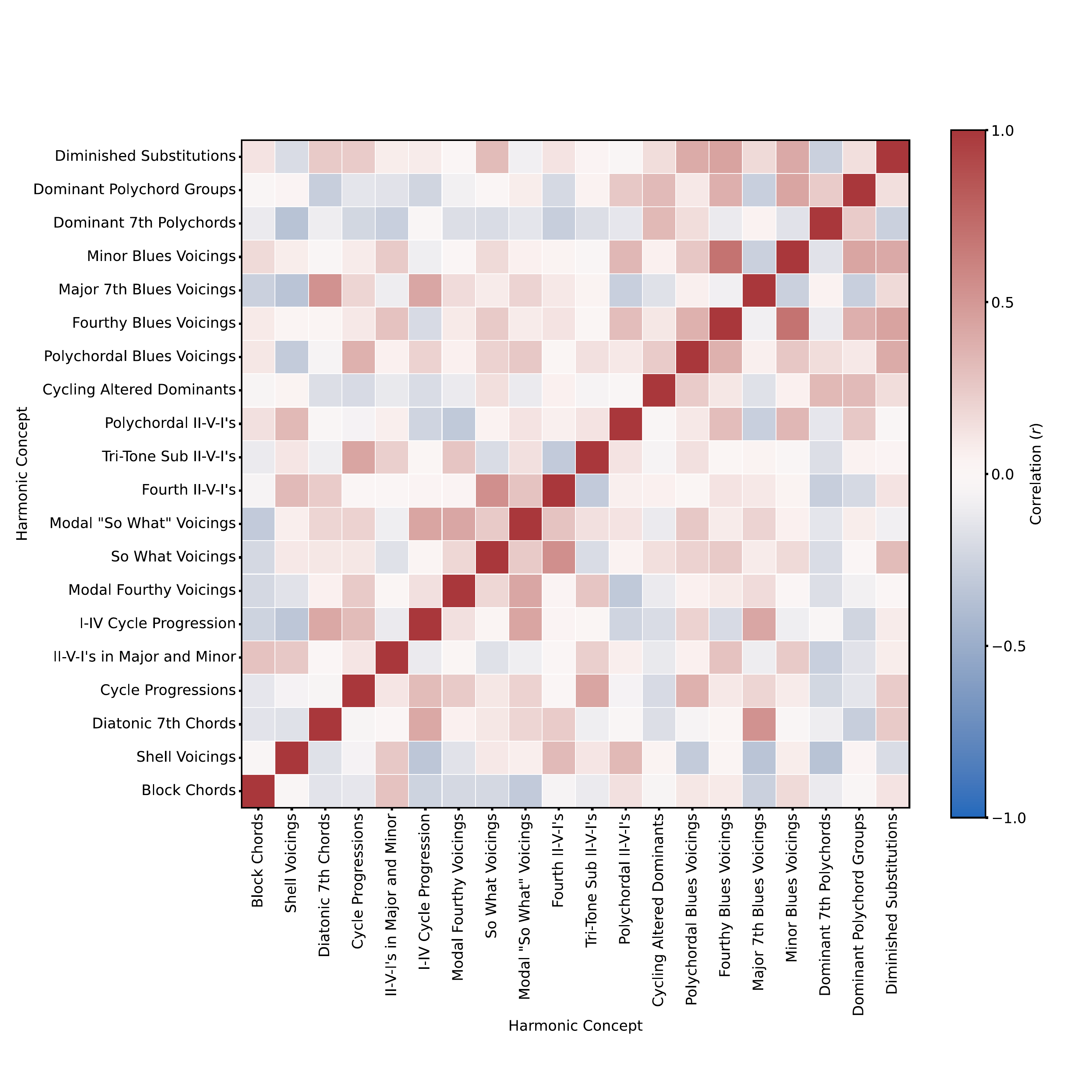}
  \caption{Pairwise correlations between concepts. Sign-count ratios are collected for all performers ($N = 20$) and across all iterations, such that the total number of values considered for each concept in every correlation is 200.}
\label{fig:rsi_sm_cav_correlations}
\end{figure}
\FloatBarrier\clearpage

\subsection{Supplementary Tables}

\begin{table}[h]
    \centering
    \caption{Optimised parameters (RF). These parameters resulted from the randomised search used to select parameters for the random forest classifier. Note that, when values for a parameter are sampled from either a uniform distribution $\mathcal{U}$ or logarithmic distribution $\mathcal{L}$ (with base 10), these are given in the range (start, stop).}
    \label{tab:rsi_sm_rf_optimised_parameters}
    \begin{tabular}{l c p{4cm} c}
        \toprule
        \textbf{Parameter} & \textbf{Values} & \textbf{Description} & \textbf{Result} \\
        \midrule
        \texttt{n\_estimators} & $\sim \mathcal{U}(10, 401)$ & Number of trees & 265 \\
        \texttt{max\_depth} & $\sim \mathcal{U}(1, 41)$ & Maximum depth of each tree & 26 \\
        \texttt{max\_features} & $\sim \mathcal{U}(0.0001, 1)$ & Proportion of features for best split & 0.148 \\
        \texttt{bootstrap} & \{True, False\} & Whether or not to use bootstrap samples & False \\
        \texttt{min\_samples\_leaf} & $\sim \mathcal{U}(1, 11)$ & Minimum samples per leaf & 3 \\
        \texttt{min\_samples\_split} & $\sim \mathcal{U}(2, 11)$ & Minimum samples to split a node & 7 \\
        \bottomrule
    \end{tabular}
\end{table}

\begin{table}[h]
    \centering
    \caption{Optimised parameters (SVM).}
    \label{tab:rsi_sm_svm_optimised_parameters}
    \begin{tabular}{l c p{4cm} c}
        \toprule
        \textbf{Parameter} & \textbf{Values} & \textbf{Description} & \textbf{Result} \\
        \midrule
        \texttt{C} & $\sim \mathcal{L}(0.001, 1000)$ & Regularization parameter: smaller values imply stronger regularization & 1.359 \\
        \texttt{class\_weight} & \{None, balanced\} & If balanced, adjusts weights inversely proportional to class frequencies & None \\
        \bottomrule
    \end{tabular}
\end{table}

\begin{table}[h]
    \centering
    \caption{Optimised parameters (LR).}
    \label{tab:rsi_sm_lr_optimised_parameters}
    \begin{tabular}{l c p{4cm} c}
        \toprule
        \textbf{Parameter} & \textbf{Values} & \textbf{Description} & \textbf{Result} \\
        \midrule
        \texttt{C} & $\sim \mathcal{L}(0.001, 1000)$ & Identical to SVM & 37.017 \\
        \texttt{class\_weight} & \{None, balanced\} & Identical to SVM & balanced \\
        \texttt{penalty} & \{None, $\text{L}_2$\} & Penalty term to use & $\text{L}_2$ \\
        \bottomrule
    \end{tabular}
\end{table}

\begin{table}[h]
    \centering
    \caption{Results for models trained using handcrafted features on held-out test data. Note that scores for the SVM at higher than top-1 accuracy cannot be provided due to limitations in the \texttt{scikit-learn} implementation of this model.}
    \label{tab:rsi_sm_handcrafted_results}
    \begin{tabular}{l c c c c}
        \toprule
        \textbf{Architecture} & \textbf{Top-1} & \textbf{Top-2} & \textbf{Top-3} & \textbf{Top-5} \\
        \midrule
        Random Forest (RF) & 0.454 & 0.626 & 0.706 & 0.785 \\
        Support Vector Machine (SVM) & 0.687 & -- & -- & -- \\
        Logistic Regression (LR) & \textbf{0.767} & \textbf{0.859} & \textbf{0.896} & \textbf{0.939} \\
        \bottomrule
    \end{tabular}
\end{table}

\begin{table}[h]
    \centering
    \caption{Results for representation learning models. Clip accuracy is the accuracy of ``raw'' predictions made from each input clip. Track accuracy is the accuracy of predictions obtained from averaging the estimated class probabilities across all clips from the same initial recording (see section \ref{sec:rsi_dataset})}
    \label{tab:rsi_sm_representation_learning_results}
    \begin{tabular}{l c c c}
        \toprule
        \textbf{Architecture} & \textbf{Data Augmentation} & \multicolumn{2}{c}{\textbf{Accuracy}} \\
        &  & \textbf{Clip} & \textbf{Track} \\
        \midrule
        CRNN & N & 0.594 & 0.769 \\
        CRNN & Y & 0.681 & 0.825 \\
        ResNet-50 & N & 0.772 & 0.875 \\
        ResNet-50 & Y & \textbf{0.805} & \textbf{0.944} \\
        \bottomrule
    \end{tabular}
\end{table}

\newpage

\begin{sidewaystable}[ht]
    \centering
    \caption{Concept dataset. A description of the twenty chapters in \citet{Haerle1994} that make up the concept dataset. The text is adapted and summarised from the didactic descriptions given by Haerle that precede each chapter.}
    \label{tab:rsi_sm_haerle_concepts}
    \begin{tabular}{lcp{12cm}}
        \toprule
         \textbf{Name} 
&\textbf{Number} & \textbf{Description} \\
        \midrule
         Block Chords 
&1 & Includes ``block''-type, closed chords that are built on stacked thirds and outline common voicings for major, dominant, and minor, and diminished sevenths, as well as the same chords with suspended fourths. \\
         Shell Voicings 
&2 & Includes ``shell''-type, open chord voicings that outline extensions beyond the seventh (e.g., sixthes, ninths, flattened fifths). \\
         Diatonic 7th Chords 
&3 & Includes diatonic seventh chords walked sequentially through major and minor tonalities. \\
         Cycle Progressions 
&4 & Connects different inversions of voicings in an idiomatic way as part of typical progressions. Common progressions represented in this concept are I-IV in major and minor, dominant 7th cycles, I-V in major and V-I in major and minor. \\
         II-V-I's in Major and Minor 
&5 & A logical extension of the Cycle Progressions concept which moves on to playing complete II-V-I cadences in major and minor keys. \\
         I-IV Cycle Progression 
&6 & A common cycle progression which moves from the key center to the IV chord which, in turn, becomes a new key center. \\
         Modal Fourthy Voicings 
&7 & Fourthy voicings which are moved through a dorian mode. \\
         ``So What'' Voicings 
&8 & A particular fourthy structure probably best known from its use in the Miles Davis composition ``So What''. Presented in both minor (dorian) and major (lydian) modes. \\
         Modal ``So What'' Voicings 
&9 & Fourthy voicings which are moved through a dorian mode. The ``So What'' voicing is used and, rather than keeping its pure structure, it changes slightly as it is moved through the dorian scale. \\
         Fourthy II-V-I's 
&10 & II-V-I progressions using the ``So What'' voicing. The structure is built on the root of the II chord and on the 3rd of the I chord. Since the voicing doesn't clearly fit or imply a dominant chord, the V chord simply involves chromatin side-slipping up or down to the I chord. \\
         Tri-Tone Sub II-V-I's 
&11 & II-V-I progressions using both the normal II and V chords and the II-V located a tri-tone away. The result is a deception as though the cadence was suddenly modulating to a distant key. \\
         Polychordal II-V-I's 
&12 & II-V-I progressions involving polychordal structures in which the left hand plays a conventional inversion and the right hand plays some kind of triadic structure to create extensions and/or alterations of the harmony. \\
    \end{tabular}
\end{sidewaystable}

\begin{sidewaystable}[ht]
    \centering
    \begin{tabular}{lcp{12cm}}
         Cycling Altered Dominants 
&13 & Progressions involving polychordal structures in which the left hand plays a dominant structure and the right hand plays some kind of triadic structure to create extensions and/or alterations of the harmony. \\
         Polychordal Blues Voicings 
&14 & Twelve-bar blues progressions involving polychordal structures in which the left hand plays a conventional inversion and the right hand plays some kind of triadic structure to fill out a two-hand voicing of the harmony. \\
         Fourthy Blues Voicings 
&15 & Blues progressions involving fourthy structures in which the left hand plays a conventional inversion and the right hand plays a structure of two perfect fourths to fill out a two-hand voicing of the harmony. \\
         Major 7th Blues Voicings 
&16 & Blues progressions incorporating voicings from other concepts, including II-V-I's in Major and Minor (concept 5), I-IV Cycle Progression (6), and Tri-Tone Sub II-V-I's (11). \\
         Minor Blues Voicings 
&17 & Minor blues progressions involving fourthy structures in which the left hand plays a conventional inversion and the right hand plays a structure of two perfect fourths to fill out a two-hand voicing of the harmony. \\
         Dominant 7th Polychords 
&18 & Dominant seventh chords where the left hand plays a conventional inversion and the right hand plays some additional construction (e.g., involving flattening and sharpening of the fifth and ninth scale degrees). \\
         Dominant Polychord Groups 
&19 & Utilizes five of the polychord formulas from the previous concept (Dominant 7th Polychords) in groups with each other against a conventional inversion in the left hand to create melodic motion. \\
         Diminished Substitutions &20 & Utilizes voicings that are derived from the half-whole diminished scale that relates to a dominant seventh chord. Since the same scale relates to four different dominant seventh chords, it makes this device possible. \\
        \bottomrule
    \end{tabular}
\end{sidewaystable}

\end{document}